\documentclass[11pt,draftcls,onecolumn]{IEEEtran}
\ifCLASSINFOpdf
\else
\fi
\usepackage{amsmath}
\usepackage{amsthm}
\usepackage{amssymb}
\usepackage{subfigure}
\usepackage{epsfig}
\usepackage{graphicx}
\usepackage{balance}
\usepackage{epstopdf}
\usepackage{multicol}
\usepackage{multirow}
\usepackage{threeparttable}
\usepackage{booktabs}
\usepackage{cite}
\usepackage{cases}
\usepackage[usenames,dvipsnames]{color}
\usepackage{color}
\makeatletter
\renewcommand{\maketag@@@}[1]{\hbox{\m@th\normalsize\normalfont#1}}
\makeatother

\begin{document}
\title{Convolution Type of Metaplectic Cohen's Distribution Time-Frequency Analysis Theory, Method and Technology}

\author{Manjun~Cui, Zhichao~Zhang,~\IEEEmembership{Member,~IEEE}, Jie~Han, Yunjie~Chen,~\IEEEmembership{Member,~IEEE}, and Chunzheng~Cao
\thanks{This work was supported in part by the National Natural Science Foundation of China under Grant 61901223; in part by the Jiangsu Planned Projects for Postdoctoral Research Funds under Grant 2021K205B; and in part by the Postgraduate Research \& Practice Innovation Program of Jiangsu Province under Grant KYCX24\_1405. \emph{(Corresponding author: Zhichao~Zhang.)}}
\thanks{Manjun~Cui, Zhichao~Zhang, Yunjie~Chen, and Chunzheng~Cao are with the School of Mathematics and Statistics, the Center for Applied Mathematics of Jiangsu Province, and the Jiangsu International Joint Laboratory on System Modeling and Data Analysis, Nanjing University of Information Science and Technology, Nanjing 210044, China (e-mail: cmj1109@163.com; zzc910731@163.com; priestcyj@nuist.edu.cn; caochunzheng@nuist.edu.cn).}
\thanks{Jie~Han is with the School of Remote Sensing and Geomatics Engineering, Nanjing University of Information Science and Technology, Nanjing 210044, China (e-mail: 003645@nuist.edu.cn).}}

\markboth{}
{Shell \MakeLowercase{\textit{et al.}}: Bare Demo of IEEEtran.cls for Journals}

\maketitle

\begin{abstract}
The conventional Cohen's distribution can't meet the requirement of additive noises jamming signals high-performance denoising under the condition of low signal-to-noise ratio, it is necessary to integrate the metaplectic transform for non-stationary signal fractional domain time-frequency analysis. In this paper, we blend time-frequency operators and coordinate operator fractionizations to formulate the definition of the metaplectic Wigner distribution, based on which we integrate the generalized metaplectic convolution to address the unified representation issue of the convolution type of metaplectic Cohen's distribution (CMCD), whose special cases and essential properties are also derived. We blend Wiener filter principle and fractional domain filter mechanism of the metaplectic transform to design the least-squares adaptive filter method in the metaplectic Wigner distribution domain, giving birth to the least-squares adaptive filter-based CMCD whose kernel function can be adjusted with the input signal automatically to achieve the minimum mean-square error (MSE) denoising in Wigner distribution domain. We discuss the optimal symplectic matrices selection strategy of the proposed adaptive CMCD through the minimum MSE minimization modeling and solving. Some examples are also carried out to demonstrate that the proposed filtering method outperforms some state-of-the-arts including Wiener filter and fixed kernel functions-based or adaptive Cohen's distribution in noise suppression.
\end{abstract}


\IEEEpeerreviewmaketitle

\section{Introduction}\label{sec:1}
\indent\IEEEPARstart{T}{he} conventional time-frequency distributions, such as Wigner distribution \cite{Wig32,Sad21}, Choi-Williams distribution \cite{Cho89,The21} and Cohen's distribution \cite{Coh66,Her23}, can't meet the requirement of characterization and analysis of non-stationary signal refined features under complex sound, light, heat, electricity and magnetism environments. Integrating them with fractional domain time-frequency transforms, including fractional Fourier transform \cite{Oza01,Zha24}, linear canonical transform \cite{Xu13,Wei22} and metaplectic transform \cite{Bas16,CorarX24}, makes contribution to the enhancement of refined information features characterization degree of freedom and analysis capability. It therefore becomes one of the research hotspots in the field of fractional domain signal processing \cite{Shi17}, playing an important role in exploring fundamental theories, applied methods and technical principles of non-stationary signal fractional domain time-frequency analysis \cite{Li20}. This is of great scientific significance for addressing practical problems encountered in radar, communications, sonar, biomedical and vibration engineering, and has bright application prospects in seismic exploration, electronic countermeasures, deep-sea detection, spectral imaging and ultrasonic inspection \cite{Tao22}.\\
\indent Cohen's distribution, also known as the bi-linear kernel function time-frequency distribution \cite{Coh89}, is one of the most representative time-frequency analysis tools of the conventional time-frequency distributions \cite{Gro00,Bog19,Bog20}. It includes particular cases many classical bi-linear time-frequency distributions, such as Wigner distribution, Choi-Williams distribution, Kirkwood-Rihaczek distribution \cite{Kir33,Rih68,Pra03}, Born-Jordan distribution \cite{CorACHA18}, Zhao-Atlas-Marks distribution \cite{Zha90,Akd21}, Margenau-Hill distribution \cite{Mar61} and Page distribution \cite{Pag52}. Indeed, it can be regarded as a unified bi-linear time-frequency distribution \cite{Coh95}. Metaplectic transform, also known as the $N$-dimensional nonseparable linear canonical transform, is a fractional domain time-frequency transform parameterized by the symplectic matrix belonging to the symplectic group $Sp(N,\mathbb{R})$ \cite{Bas16}. Many celebrated integral transforms, including Fourier transform, fractional Fourier transform, linear canonical transform, Fresnel transform, Lorentz transform and Laplace transform, are all special cases of the metaplectic transform, which plays a crucial role in fractional domain time-frequency transforms \cite{Bas16}. Therefore, the theory, method and technique of the deep integration of Cohen's distribution and metaplectic transform become the focus and hot topic of the deep integration research of the conventional time-frequency distributions and fractional domain time-frequency transforms. These studies exhibit the characteristics of generality, comprehensiveness, universality and practicability, and are key and difficult problems in the field of non-stationary signal fractional domain time-frequency analysis.\\
\indent It is well-known that the Cohen's distribution can be expressed as the conventional convolution of the Wigner distribution and the kernel function \cite{Bog10}. Namely, the Cohen's distribution is a result of smoothing the Wigner distribution by using the kernel function. The main purpose of the smoothing filter is not just only to suppress cross-terms of the Wigner distribution, but to achieve non-stationary signal filter denoising in Wigner distribution domain, and accordingly extract the target signal from background noises or eliminate the impact of noises through the unique reconstruction property of Wigner distribution. Therefore, it seems feasible to achieve the deep integration of Cohen's distribution and metaplectic transform by generalizing the Wigner distribution and convolution found in the convolution type of Cohen's distribution into metaplectic transform domains. This provides a convolution type of metaplectic transform domains Cohen's distribution-based solution for additive noises jamming signals high-performance denoising under the condition of low signal-to-noise ratio (SNR).\\
\indent To sum up, this paper embarks on the requirement of high-performance denoising faced by additive noises jamming signals fractional domain time-frequency analysis under the condition of low SNR to use the metaplectic transform to fractionize the Wigner and convolution operators, construct the organic integration theory of the convolution type of Cohen's distribution and metaplectic transform, and design and develop the convolution type of metaplectic transform domains Cohen's distribution-based adaptive optimal filter denoising method and technique. This is in line with the development tendency of non-stationary signal fractional domain time-frequency analysis theory, method and technology, which is meaningful to both academic and engineering application.\\
\indent It is still been empty of study in the convolution type of metaplectic transform domains Cohen's distribution time-frequency analysis theory, method and technology. However, there are a number of relative research achievements regarding to Wigner distributions in metaplectic transform domains and convolutions in metaplectic transform domains.\\
\indent Inspired by the idea of fractionizing time-frequency operators (i.e., the identity operator and the partial Fourier operator) found in the Wigner distribution, in 2023 Zhang \emph{et al.} \cite{ZhaTIT23} proposed the high-dimensional closed-form instantaneous cross-correlation function Wigner distribution (CICFWD) by replacing the identity and partial Fourier operators with the metaplectic and partial metaplectic operators, respectively, and established its Heisenberg's uncertainty principles. Motivated by the idea of fractionizing the conventional coordinate operator found in the Wigner distribution, in 2020 Cordero \emph{et al.} \cite{CorAA20,Bay20} formulated the matrix $\mathcal{M}\in GL(2N,\mathbb{R})$-Wigner distribution by replacing the conventional coordinate operator with the matrix coordinate operator, and deduced its essential properties. From 2022 to 2024, Cordero \emph{et al.} \cite{CorACHA22,CorarX22,CorJFA23,CorJMPA23,CorACHA24} replaced the specific partial metaplectic operator composites of the matrix coordinate and partial Fourier operators with the general partial metaplectic operator to extend further the matrix $\mathcal{M}\in GL(2N,\mathbb{R})$-Wigner distribution to the symplectic matrix $\mathcal{M}\in Sp(2N,\mathbb{R})$-Wigner distribution, and investigated its applications in characterizations of modulation spaces, time-frequency spaces, Cohen's distribution, pseudo-differential operators and Schr{\"o}dinger equations. In 2023, Giacchi \cite{GiaarX23} obtained Lieb's uncertainty principle of the matrix $\mathcal{M}\in GL(2N,\mathbb{R})$-Wigner distribution. In 2023 and 2024, Zhang \emph{et al.} \cite{ZhaSP23,ZhaTIT24} rewritten the conventional coordinate operator as a specific symplectic coordinate operator, and replaced it with the general symplectic coordinate operator to generalize the Wigner distribution and basis function Wigner distribution to the symplectic Wigner distribution and basis function symplectic Wigner distribution, respectively. Zhang \emph{et al.} \cite{ZhaSP23,ZhaTIT24} also explored Heisenberg's uncertainty principles of the symplectic Wigner distribution and basis function symplectic Wigner distribution, based on which the time-frequency super-resolution optimization model was established, and its optimal solution was derived for the one-dimensional case.\\
\indent In brief, the high-dimensional CICFWD and symplectic Wigner distribution can be seen as Wigner distributions in metaplectic transform domains with the strongest generalization capability in terms of time-frequency operators fractionization and coordinate operator fractionization, respectively. This indicates that integrating time-frequency operators fractionization and coordinate operator fractionization is expected to yield a unified fractional domain Wigner distribution, that is, the metaplectic Wigner distribution.\\
\indent According to the idea of extending convolution theorems in linear canonical transform domains, in 2021 Shah \emph{et al.} \cite{Sha21} generalized the convolution theorem \cite{Wei11,ShiTSP12,ShiSCIS12} in the product form of linear canonical transform and scale Fourier transform into metaplectic transform domains, giving birth to the I-type metaplectic convolution theorem. They \cite{Sha21} also employed the I-type metaplectic convolution theorem in deducing the lattice-based multi-channel sampling theorem and Shannon's sampling theorem in metaplectic transform domains, and explored their applications in signal reconstruction and image super-resolution reconstruction. In 2022, Shah \emph{et al.} \cite{Sha22} generalized the convolution theorem \cite{Den06,Wei12,Goe13} in the product form of two linear canonical transforms and chirp signal into metaplectic transform domains, giving birth to the II-type metaplectic convolution theorem. They \cite{Sha22} also applied the II-type metaplectic convolution theorem to deduce the classical sampling theorem for band-limited signals in metaplectic transform domains, and explained the multiplicative filter designing method in metaplectic transform domains. In 2023, Tantary \emph{et al.} \cite{Tan23} used the II-type metaplectic convolution theorem to deduce Papoulis' sampling theorem for band-limited signals in metaplectic transform domains. During the same year, Zhao \emph{et al.} \cite{Zha23} generalized the convolution theorem \cite{Wei09} in the product form of two linear canonical transforms and the convolution theorem \cite{ZhaOpt16} in the product form of two scale linear canonical transforms into metaplectic transform domains, giving birth to the III-type and IV-type metaplectic convolution theorems, respectively. They \cite{Zha23} also employed these theorems in designing the multiplicative filter method in metaplectic transform domains, and discussed its application in image denoising. In 2024, Cui \emph{et al.} \cite{Cui24} extended the generalized linear canonical convolution theorem \cite{Shi14} that includes particular cases the I-type, II-type and IV-type linear canonical convolution theorems into metaplectic transform domains, generating the generalized metaplectic convolution theorem, designing the least-squares adaptive filter method in metaplectic transform domains, and improving the mean-square error (MSE), structural similarity index measure and peak SNR (PSNR) performance indexes of image denoising.\\
\indent In brief, the generalized metaplectic convolution includes particular cases the I-type, II-type and IV-type metaplectic convolutions. It is none other than the convolution in metaplectic transform domains with the strongest generalization capability in terms of convolution operator fractionization and can be regarded as a unified fractional domain convolution. In addition, the denoising performance of the least-squares adaptive filter on the basis of the generalized metaplectic convolution outperforms that of the least-squares adaptive filter in the Fourier transform domain (i.e., Wiener filter \cite{Har23}) and the multiplicative filter in metaplectic transform domains. However, the inherent mechanism behind the denoising performance improvement triggered by symplectic matrices found in the generalized metaplectic convolution is still unknown.\\
\indent Considering that there is an urgent requirement on high-performance denoising faced by additive noises jamming signals fractional domain time-frequency analysis under the condition of low SNR, the purpose of this paper is to establish time-frequency analysis theory, method and technology systems of the convolution type of metaplectic Cohen's distribution (CMCD). This boosts comprehensively the development of the ordinary Cohen's distribution time-frequency analysis fundamental theories, applied methods and technical principles in both scale and depth, enriching the fractional domain time-frequency distribution theoretical connotation, the fractional domain filter denoising methodology and the fractional domain parameter selection technical solution. This also provides applied mathematical theory, method and technology for solving practical problems encountered in fractional domain signal processing. The main contributions of this paper are summarized as follows:
\begin{itemize}

    \item This paper blends Wigner operator fractionization and convolution operator fractionization to construct the organic integration theory of the convolution type of Cohen's distribution and metaplectic transform, according to the metaplectic Wigner distribution and generalized metaplectic convolution. It addresses the unified representation issue of the CMCD, and clarifies the mathematical formula of the metaplectic Cohen's distribution with the strongest generalization capability.

    \item This paper blends Wiener filter principle and fractional domain filter mechanism of the metaplectic transform to design the least-squares adaptive filter method in the metaplectic Wigner distribution domain, in analogy with Wiener filter method and the least-squares adaptive filter method in metaplectic transform domains. It discloses the influencing mechanism of the kernel function on the filter denoising effect, and generates the adaptive kernel function formula of the metaplectic Cohen's distribution with the minimum MSE in Wigner distribution domain.

    \item This paper blends fractional domain parameters adjustment mechanism and matrix optimization theory to develop the minimization technique of the minimum MSE in Wigner distribution domain, on the basis of the proposed filter method. It reveals the inherent mechanism behind the high-performance denoising triggered by symplectic matrices, and formulates the optimal symplectic matrices selection strategy of the metaplectic Cohen's distribution that minimizes the minimum MSE function.

\end{itemize}

\indent The remainder of this paper is structured as follows. In Section~\ref{sec:2}, we collect some preparatory works. In Section~\ref{sec:3}, we construct fundamental theories of the CMCD, including the mathematical definition, relations to the classical time-frequency analysis tools, and essential properties. In Section~\ref{sec:4}, we design the CMCD-based adaptive filter method for additive noises jamming signals. In Section~\ref{sec:5}, we introduce some examples to validate the effectiveness, reliability and feasibility of the proposed method. In Section~\ref{sec:6}, we provide a theoretical analysis for the optimal symplectic matrices selection strategy. In Section~\ref{sec:7}, we draw a conclusion. All the technical proofs of our theoretical results are relegated to the appendix parts.
\section{Preliminaries}\label{sec:2}
\indent In this section, we recall some necessary background and notation on the Cohen's distribution, metaplectic transform, high-dimensional CICFWD, symplectic Wigner distribution and generalized metaplectic convolution.
\subsection{Cohen's distribution}\label{sec:2.1}
\indent\emph{Definition 1:} Let a function $f\in L^2(\mathbb{R}^N)$ and a kernel function $\Pi$. The Cohen's distribution of the function $f$ is defined as the conventional convolution of the Wigner distribution and the kernel function, that is \cite{Bog10}
\begin{equation}\label{eq2.1}
\mathrm{C}f(\mathbf{x},\mathbf{w})=\left(\mathrm{W}f\ast\Pi\right)(\mathbf{x},\mathbf{w}),
\end{equation}
where $\ast$ denotes the conventional convolution operator, and the Wigner distribution of the function $f$ is given by \cite{ZhaTIT23}
\begin{equation}\label{eq2.2}
\mathrm{W}f(\mathbf{x},\mathbf{w})=\mathcal{F}_{\mathbf{y},2}\mathfrak{T}_{\mathcal{P}}\left(\mathfrak{I}f\otimes\overline{\mathfrak{I}f}\right)(\mathbf{x},\mathbf{w}),
\end{equation}
where the identity operator $\mathfrak{I}$, the tensor product $\otimes$, the conventional coordinate operator $\mathfrak{T}_{\mathcal{P}}$ with $\mathcal{P}=\begin{pmatrix}\mathbf{I}_N&\mathbf{I}_N\\\frac{\mathbf{I}_N}{2}&-\frac{\mathbf{I}_N}{2}\end{pmatrix}$, and the partial Fourier operator $\mathcal{F}_{\mathbf{y},2}$ with respect to the second variables $\mathbf{y}$ are given by $\mathfrak{I}f:=f$, $(f\otimes\overline{g})(\mathbf{x},\mathbf{y}):=f(\mathbf{x})\overline{g(\mathbf{y})}$, $\mathfrak{T}_{\mathcal{P}}h(\mathbf{x},\mathbf{y}):=\sqrt{|\mathrm{det}(\mathcal{P})|}h((\mathbf{x},\mathbf{y})\mathcal{P})=h\left(\mathbf{x}+\frac{\mathbf{y}}{2},\mathbf{x}-\frac{\mathbf{y}}{2}\right)$, and $\mathcal{F}_{\mathbf{y},2}h(\mathbf{x},\mathbf{w}):=\left\langle h(\mathbf{x},\mathbf{y}),\mathrm{e}^{2\pi\mathrm{i}\mathbf{y}\mathbf{w}^{\mathrm{T}}}\right\rangle_{\mathbf{y}}$, respectively, and where the inner product $\langle,\rangle_{\mathbf{y}}$ with respect to the variables $\mathbf{y}\in\mathbb{R}^N$ is given by $\langle\circ,\diamond\rangle_{\mathbf{y}}:=\int_{\mathbb{R}^N}\circ(\mathbf{y})\overline{\diamond^{\mathrm{T}}(\mathbf{y})}\mathrm{d}\mathbf{y}$. Here, $\mathrm{det}(\cdot)$ denotes the determinant operator for matrices, $\mathbf{I}_N$ denotes the $N\times N$ identity matrix, and the superscripts $\mathrm{T}$ and --- denote the transpose operator and complex conjugate operator, respectively.\\
\indent The convolution type of Cohen's distribution, shown in Eq.~\eqref{eq2.1}, can be rewritten as an integral form \cite{Coh89,Coh95}
\begin{equation}\label{eq2.3}
\mathrm{C}_f(\mathbf{x},\mathbf{w})=\iiint_{\mathbb{R}^{N\times N\times N}}f\left(\mathbf{y}+\frac{\mathbf{z}}{2}\right)\overline{f\left(\mathbf{y}-\frac{\mathbf{z}}{2}\right)}\phi(\mathbf{v},\mathbf{z})\mathrm{e}^{-2\pi\mathrm{i}\left(\mathbf{v}\mathbf{x}^\mathrm{T}+\mathbf{z}\mathbf{w}^\mathrm{T}-\mathbf{y}\mathbf{v}^\mathrm{T}\right)}\mathrm{d}\mathbf{y}\mathrm{d}\mathbf{z}\mathrm{d}\mathbf{v}.
\end{equation}
Here, the kernel functions $\phi(\mathbf{v},\mathbf{z})$ and $\Pi(\mathbf{x},\mathbf{w})$ compose a Fourier transform pair, i.e.,
\begin{align}\label{eq2.4}
\Pi(\mathbf{x},\mathbf{w})=&\mathcal{F}[\phi](\mathbf{x},\mathbf{w})\nonumber\\
=&\left\langle \phi(\mathbf{v},\mathbf{z}),\mathrm{e}^{2\pi\mathrm{i}\left(\mathbf{v}\mathbf{x}^\mathrm{T}+\mathbf{z}\mathbf{w}^\mathrm{T}\right)}\right\rangle_{(\mathbf{v},\mathbf{z})}\nonumber\\
=&\iint_{\mathbb{R}^{N\times N}}\phi(\mathbf{v},\mathbf{z})\mathrm{e}^{-2\pi\mathrm{i}\left(\mathbf{v}\mathbf{x}^\mathrm{T}+\mathbf{z}\mathbf{w}^\mathrm{T}\right)}\mathrm{d}\mathbf{v}\mathrm{d}\mathbf{z},
\end{align}
where $\mathcal{F}$ denotes the Fourier operator. The Wigner distribution, Choi-Williams distribution, Kirkwood-Rihaczek distribution, Born-Jordan distribution, Zhao-Atlas-Marks distribution, Margenau-Hill distribution and Page distribution are special cases of the Cohen's distribution corresponding to $\phi(\mathbf{v},\mathbf{z})=1$, $\phi(\mathbf{v},\mathbf{z})=\mathrm{e}^{-\frac{\lVert\mathbf{v}\rVert^2\lVert\mathbf{z}\rVert^2}{\varsigma}}$, $\phi(\mathbf{v},\mathbf{z})=\mathrm{e}^{\pi\mathrm{i}\mathbf{v}\mathbf{z}^\mathrm{T}}$, $\phi(\mathbf{v},\mathbf{z})=\frac{\sin\left(\pi\mathbf{v}\mathbf{z}^\mathrm{T}\right)}{\pi\mathbf{v}\mathbf{z}^\mathrm{T}}$, $\phi(\mathbf{v},\mathbf{z})=g(\mathbf{z})\left\|\mathbf{z}\right\|_1\frac{\sin\left(2\pi\kappa\mathbf{v}\mathbf{z}^\mathrm{T}\right)}{2\pi\kappa\mathbf{v}\mathbf{z}^\mathrm{T}}$, $\phi(\mathbf{v},\mathbf{z})=\cos\left(\pi\mathbf{v}\mathbf{z}^\mathrm{T}\right)$ and $\phi(\mathbf{v},\mathbf{z})=\mathrm{e}^{2\pi\mathrm{i}\mathbf{v}\left\|\mathbf{z}\right\|_1}$, respectively. Here, $\left\|\cdot\right\|_1$ and $\lVert\cdot\rVert$ denote the $1$-norm and 2-norm for vectors, respectively. Moreover, the Cohen's distribution exhibits some useful and important properties, including marginal distribution, energy conservation, unique reconstruction, Moyal formula, complex conjugate symmetry, time reversal symmetry, scaling property, time translation property, and frequency modulation property.\\
\indent See Table~\ref{tab:1} for a summary of some special cases of the Cohen's distribution.
\begin{table}[htbp]
\centering
\caption{\label{tab:1}Some special cases of the Cohen's distribution}
\footnotesize
\begin{tabular}{cc}
\specialrule{0.1em}{4pt}{4pt}
$\phi(\mathbf{v},\mathbf{z})$ & Cohen's distribution\\
\specialrule{0.1em}{4pt}{4pt}
$1$ & Wigner distribution\\
\specialrule{0em}{4pt}{4pt}
$\mathrm{e}^{-\frac{\lVert\mathbf{v}\rVert^2\lVert\mathbf{z}\rVert^2}{\varsigma}}$ & Choi-Williams distribution\\
\specialrule{0em}{4pt}{4pt}
$\mathrm{e}^{\pi\mathrm{i}\mathbf{v}\mathbf{z}^\mathrm{T}}$ & Kirkwood-Rihaczek distribution\\
\specialrule{0em}{4pt}{4pt}
$\frac{\sin\left(\pi\mathbf{v}\mathbf{z}^\mathrm{T}\right)}{\pi\mathbf{v}\mathbf{z}^\mathrm{T}}$ & Born-Jordan distribution\\
\specialrule{0em}{4pt}{4pt}
$g(\mathbf{z})\left\|\mathbf{z}\right\|_1\frac{\sin\left(2\pi\kappa\mathbf{v}\mathbf{z}^\mathrm{T}\right)}{2\pi\kappa\mathbf{v}\mathbf{z}^\mathrm{T}}$ & Zhao-Atlas-Marks distribution\\
\specialrule{0em}{4pt}{4pt}
$\cos\left(\pi\mathbf{v}\mathbf{z}^\mathrm{T}\right)$ & Margenau-Hill distribution\\
\specialrule{0em}{4pt}{4pt}
$\mathrm{e}^{2\pi\mathrm{i}\mathbf{v}\left\|\mathbf{z}\right\|_1}$ & Page distribution\\
\specialrule{0.1em}{4pt}{4pt}
\end{tabular}
\end{table}
\subsection{Metaplectic transform}\label{sec:2.2}
\indent\emph{Definition 2:} The family of symplectic matrices is known as the symplectic group, defined by \cite{ZhaTIT23,Mor02,Cor15,Wan23}
\begin{equation}\label{eq2.5}
Sp(N,\mathbb{R})=\left\{\mathcal{M}\in GL(2N,\mathbb{R}):\mathcal{M}\mathcal{J}\mathcal{M}^{\mathrm{T}}=\mathcal{J}\right\}.
\end{equation}
(this definition is not the same as that introduced in \cite{CorACHA22,GiaarX23,Fol89,Won98,dGos06,Cor13}, but they are equivalent by interchanging the symplectic matrix $\mathcal{M}$ and its transpose $\mathcal{M}^{\mathrm{T}}$), where $GL(2N,\mathbb{R})$ denotes the linear group of $2N\times 2N$ invertible matrices and $\mathcal{J}:=\begin{pmatrix}\mathbf{0}_N&\mathbf{I}_N\\-\mathbf{I}_N&\mathbf{0}_N\end{pmatrix}$, (here $\mathbf{0}_N$ denotes the $N\times N$ null matrix).\\
\indent\emph{Definition 3:} Let a function $f\in L^2(\mathbb{R}^N)$, a function $h\in L^2(\mathbb{R}^{2N})$ and a symplectic matrix $\mathcal{M}=\begin{pmatrix}\mathbf{A}&\mathbf{B}\\\mathbf{C}&\mathbf{D}\end{pmatrix}\in Sp(N,\mathbb{R})$. The metaplectic transform of the function $f$ and the partial metaplectic transform of the function $h$ with respect to the second variables $\mathbf{y}$ associated with the symplectic matrix $\mathcal{M}$ are defined as \cite{Bas16}
\begin{align}\label{eq2.6}
\mu(\mathcal{M})f(\mathbf{u})=\left\{
\begin{array}{ll}
\left\langle f(\mathbf{x}),\overline{\mathcal{K}_{\mathcal{M}}(\mathbf{u},\mathbf{x})}\right\rangle_{\mathbf{x}},&\mathrm{det}(\mathbf{B})\neq0\\[8pt]
\sqrt{\mathrm{det}(\mathbf{D})}\mathrm{e}^{\pi\mathrm{i}\mathbf{u}\mathbf{C}\mathbf{D}^{\mathrm{T}}\mathbf{u}^{\mathrm{T}}}f(\mathbf{u}\mathbf{D}),&\mathrm{det}(\mathbf{B})=0
\end{array}
\right.
\end{align}
and
\begin{align}\label{eq2.7}
\mu_{\mathbf{y},2}(\mathcal{M})h(\mathbf{x},\mathbf{u})=\left\{
\begin{array}{ll}
\left\langle h(\mathbf{x},\mathbf{y}),\overline{\mathcal{K}_{\mathcal{M}}(\mathbf{u},\mathbf{y})}\right\rangle_{\mathbf{y}},&\mathrm{det}(\mathbf{B})\neq0\\[8pt]
\sqrt{\mathrm{det}(\mathbf{D})}\mathrm{e}^{\pi\mathrm{i}\mathbf{u}\mathbf{C}\mathbf{D}^{\mathrm{T}}\mathbf{u}^{\mathrm{T}}}h(\mathbf{x},\mathbf{u}\mathbf{D}),&\mathrm{det}(\mathbf{B})=0
\end{array}
\right.,
\end{align}
respectively, where the basis function $\mathcal{K}_{\mathcal{M}}$ in the metaplectic transform domain is given by
\begin{equation}\label{eq2.8}
\mathcal{K}_{\mathcal{M}}(\natural,\sharp)=\frac{1}{\sqrt{-\mathrm{det}(\mathbf{B})}}\mathrm{e}^{\pi\mathrm{i}\left(\natural\mathbf{D}\mathbf{B}^{-1}\natural^{\mathrm{T}}+\sharp\mathbf{B}^{-1}\mathbf{A}\sharp^{\mathrm{T}}\right)-2\pi\mathrm{i}\sharp\mathbf{B}^{-1}\natural^{\mathrm{T}}}.
\end{equation}
\indent Note that the identity operator is a special case of the metaplectic operator corresponding to $\mathcal{M}=\mathbf{I}_{2N}$, i.e., $\mu(\mathbf{I}_{2N})=\mathfrak{I}$. Also, the Fourier operator and partial Fourier operator $\mathcal{F}_{\mathbf{y},2}$ with respect to the second variables $\mathbf{y}$ are special cases of the metaplectic operator and partial metaplectic operator with respect to the second variables $\mathbf{y}$ corresponding to $\mathcal{M}=\mathcal{J}$, respectively, i.e., $\mu(\mathcal{J})=\mathcal{F}$ and $\mu_{\mathbf{y},2}(\mathcal{J})=\mathcal{F}_{\mathbf{y},2}$. Moreover, let $\mathcal{I}:=\begin{pmatrix}\mathbf{0}_N&\mathbf{I}_N\\\mathbf{I}_N&\mathbf{0}_N\end{pmatrix}$, the conventional coordinate operator $\mathfrak{T}_{\mathcal{P}}$ is a special case of the symplectic coordinate operator $\mathfrak{T}_{\mathcal{M}\mathcal{I}}$ given by
\begin{equation}\label{eq2.9}
\mathfrak{T}_{\mathcal{M}\mathcal{I}}h(\mathbf{x},\mathbf{y}):=\sqrt{|\mathrm{det}(\mathcal{M}\mathcal{I})|}h((\mathbf{x},\mathbf{y})\mathcal{M}\mathcal{I}),
\end{equation}
corresponding to $\mathcal{M}=\mathcal{P}\mathcal{I}$.\\
\indent See Table~\ref{tab:2} for a summary of some special cases of the metaplectic operator, partial metaplectic operator and symplectic coordinate operator.
\begin{table}[htbp]
\centering
\caption{\label{tab:2}Some special cases of the metaplectic operator, partial metaplectic operator and symplectic coordinate operator}
\footnotesize
\begin{tabular}{cc}
\specialrule{0.1em}{4pt}{4pt}
$\mathcal{M}=\begin{pmatrix}\mathbf{A}&\mathbf{B}\\\mathbf{C}&\mathbf{D}\end{pmatrix}$ & Metaplectic operator\\
\specialrule{0.1em}{4pt}{4pt}
$\mathbf{I}_{2N}$ & Identity operator\\
\specialrule{0em}{4pt}{4pt}
$\mathcal{J}$ & Fourier operator\\
\specialrule{0.1em}{4pt}{4pt}
$\mathcal{M}=\begin{pmatrix}\mathbf{A}&\mathbf{B}\\\mathbf{C}&\mathbf{D}\end{pmatrix}$ & Partial metaplectic operator\\
\specialrule{0.1em}{4pt}{4pt}
$\mathcal{J}$ & Partial Fourier operator\\
\specialrule{0.1em}{4pt}{4pt}
$\mathcal{M}=\begin{pmatrix}\mathbf{A}&\mathbf{B}\\\mathbf{C}&\mathbf{D}\end{pmatrix}$ & Symplectic coordinate operator\\
\specialrule{0.1em}{4pt}{4pt}
$\mathcal{P}\mathcal{I}$ & Conventional coordinate operator\\
\specialrule{0.1em}{4pt}{4pt}
\end{tabular}
\end{table}
\subsection{High-dimensional CICFWD}\label{sec:2.3}
\indent Thanks to $\mu(\mathbf{I}_{2N})=\mathfrak{I}$ and $\mu_{\mathbf{y},2}(\mathcal{J})=\mathcal{F}_{\mathbf{y},2}$, the Wigner distribution can be rewritten in terms of the metaplectic operator and partial metaplectic operator as
\begin{equation}\label{eq2.10}
\mathrm{W}f(\mathbf{x},\mathbf{w})=\mu_{\mathbf{y},2}(\mathcal{J})\mathfrak{T}_{\mathcal{P}}\left(\mu(\mathbf{I}_{2N})f\otimes\overline{\mu(\mathbf{I}_{2N})f}\right)(\mathbf{x},\mathbf{w}).
\end{equation}
Fixed three $2N\times 2N$ symplectic matrices $\mathcal{M},\mathcal{M}_1,\mathcal{M}_2\in Sp(N,\mathbb{R})$, replacing the specific symplectic matrices $\mathcal{J},\mathbf{I}_{2N},\mathbf{I}_{2N}$ found in Eq.~\eqref{eq2.9} with the general forms $\mathcal{M},\mathcal{M}_1,\mathcal{M}_2$ yields the high-dimensional CICFWD.\\
\indent\emph{Definition 4:} Let a function $f\in L^2(\mathbb{R}^N)$, let $\mu_{\mathbf{y},2}(\mathcal{M})$ be the partial metaplectic operator with respect to the second variables $\mathbf{y}$ associated with the symplectic matrix $\mathcal{M}=\begin{pmatrix}\mathbf{A}&\mathbf{B}\\\mathbf{C}&\mathbf{D}\end{pmatrix}\in Sp(N,\mathbb{R})$ with $\mathrm{det}(\mathbf{B})\neq0$, and let $\mu(\mathcal{M}_1),\mu(\mathcal{M}_2)$ be two metaplectic operators associated with the symplectic matrices $\mathcal{M}_1=\begin{pmatrix}\mathbf{A}_1&\mathbf{B}_1\\\mathbf{C}_1&\mathbf{D}_1\end{pmatrix},\mathcal{M}_2=\begin{pmatrix}\mathbf{A}_2&\mathbf{B}_2\\\mathbf{C}_2&\mathbf{D}_2\end{pmatrix}\in Sp(N,\mathbb{R})$ with $\mathrm{det}(\mathbf{B}_1)\neq0,\mathrm{det}(\mathbf{B}_2)\neq0$, respectively. The high-dimensional CICFWD of the function $f$ associated with the symplectic matrices $\mathcal{M},\mathcal{M}_1,\mathcal{M}_2$ is defined as \cite{ZhaTIT23}
\begin{equation}\label{eq2.11}
\mathrm{W}(\mathcal{M},\mathcal{M}_1,\mathcal{M}_2)f(\mathbf{x},\mathbf{u})=\mu_{\mathbf{y},2}(\mathcal{M})\mathfrak{T}_{\mathcal{P}}\left(\mu(\mathcal{M}_1)f\otimes\overline{\mu(\mathcal{M}_2)f}\right)(\mathbf{x},\mathbf{u}).
\end{equation}
\indent When $N=1$, the high-dimensional CICFWD in the case of $\mathcal{M}=\begin{pmatrix}a&b\\c&d\end{pmatrix},\mathcal{M}_1=\begin{pmatrix}a_1&b_1\\c_1&d_1\end{pmatrix},\mathcal{M}_2=\begin{pmatrix}a_2&b_2\\c_2&d_2\end{pmatrix}$ reduces to the CICFWD
\begin{equation}\label{eq2.12}
\mathrm{W}\left(\begin{pmatrix}a&b\\c&d\end{pmatrix},\begin{pmatrix}a_1&b_1\\c_1&d_1\end{pmatrix},\begin{pmatrix}a_2&b_2\\c_2&d_2\end{pmatrix}\right)f=\mathcal{L}_{y,2}\begin{pmatrix}a&b\\c&d\end{pmatrix}\mathfrak{T}_{\mathcal{P}}\left(\mathcal{L}\begin{pmatrix}a_1&b_1\\c_1&d_1\end{pmatrix}f\otimes\overline{\mathcal{L}\begin{pmatrix}a_2&b_2\\c_2&d_2\end{pmatrix}f}\right),
\end{equation}
where $\mathcal{L}$ and $\mathcal{L}_{y,2}$ denote the linear canonical operator and partial linear canonical operator with respect to the second variable $y$, respectively. In addition to the Wigner distribution and CICFWD, the high-dimensional CICFWD includes also particular cases the $N$-dimensional nonseparable affine characteristic Wigner distribution \cite{Pei01} corresponding to $\mathcal{M}=\mathcal{J},\mathcal{M}_1=\mathcal{M}_2$, the $N$-dimensional nonseparable basis function Wigner distribution \cite{Bai12} corresponding to $\mathcal{M},\mathcal{M}_1=\mathcal{M}_2=\mathbf{I}_{2N}$, the $N$-dimensional nonseparable convolution representation Wigner distribution \cite{ZhaSP16} corresponding to $\mathcal{M}=\begin{pmatrix}\frac{\mathbf{D}_1}{4}&-\mathbf{B}_1\\-\mathbf{C}_1&4\mathbf{A}_1\end{pmatrix},\mathcal{M}_1=\begin{pmatrix}\mathbf{A}_1&\mathbf{B}_1\\\mathbf{C}_1&\mathbf{D}_1\end{pmatrix},\mathcal{M}_2=\begin{pmatrix}-\mathbf{A}_1&\mathbf{B}_1\\\mathbf{C}_1&-\mathbf{D}_1\end{pmatrix}$, and the $N$-dimensional nonseparable instantaneous cross-correlation function Wigner distribution \cite{ZhaSP15} corresponding to $\mathcal{M},\mathcal{M}_1,\mathcal{M}_2=\mathbf{I}_{2N}$, respectively.\\
\indent It should be emphasized here that the high-dimensional CICFWD differs essentially from the matrix $\mathcal{M}\in GL(2N,\mathbb{R})$-Wigner distribution and the symplectic matrix $\mathcal{M}\in Sp(2N,\mathbb{R})$-Wigner distribution, due to the different numbers and types of symplectic matrices embedding mechanisms.\\
\indent See Table~\ref{tab:3} for a summary of some special cases of the high-dimensional CICFWD.
\begin{table}[htbp]
\centering
\caption{\label{tab:3}Some special cases of the high-dimensional CICFWD}
\footnotesize
\begin{tabular}{cccc}
\specialrule{0.1em}{4pt}{4pt}
$\mathcal{M}=\begin{pmatrix}\mathbf{A}&\mathbf{B}\\\mathbf{C}&\mathbf{D}\end{pmatrix}$ & $\mathcal{M}_1=\begin{pmatrix}\mathbf{A}_1&\mathbf{B}_1\\\mathbf{C}_1&\mathbf{D}_1\end{pmatrix}$ & $\mathcal{M}_2=\begin{pmatrix}\mathbf{A}_2&\mathbf{B}_2\\\mathbf{C}_2&\mathbf{D}_2\end{pmatrix}$ & High-dimensional CICFWD\\
\specialrule{0.1em}{4pt}{4pt}
$\mathcal{J}$ & $\mathbf{I}_{2N}$ & $\mathbf{I}_{2N}$ & Wigner distribution\\
\specialrule{0em}{4pt}{4pt}
$\begin{pmatrix}a&b\\c&d\end{pmatrix}$  & $\begin{pmatrix}a_1&b_1\\c_1&d_1\end{pmatrix}$ & $\begin{pmatrix}a_2&b_2\\c_2&d_2\end{pmatrix}$ & CICFWD\\
\specialrule{0em}{4pt}{4pt}
\multirow{2}{*}{$\mathcal{J}$} & \multirow{2}{*}{$\mathcal{M}_1$} & \multirow{2}{*}{$\mathcal{M}_1$} & $N$-dimensional nonseparable affine\\
 & & & characteristic Wigner distribution\\
\specialrule{0em}{4pt}{4pt}
\multirow{2}{*}{$\mathcal{M}$} & \multirow{2}{*}{$\mathbf{I}_{2N}$} & \multirow{2}{*}{$\mathbf{I}_{2N}$} & $N$-dimensional nonseparable basis\\
 & & & function Wigner distribution\\
\multirow{2}{*}{$\begin{pmatrix}\frac{\mathbf{D}_1}{4}&-\mathbf{B}_1\\-\mathbf{C}_1&4\mathbf{A}_1\end{pmatrix}$} & \multirow{2}{*}{$\mathcal{M}_1$} & \multirow{2}{*}{$\begin{pmatrix}-\mathbf{A}_1&\mathbf{B}_1\\\mathbf{C}_1&-\mathbf{D}_1\end{pmatrix}$} & $N$-dimensional nonseparable convolution\\
 & & & representation Wigner distribution\\
\multirow{2}{*}{$\mathcal{M}$} & \multirow{2}{*}{$\mathcal{M}_1$} & \multirow{2}{*}{$\mathbf{I}_{2N}$} & $N$-dimensional nonseparable instantaneous\\
 & & & cross-correlation function Wigner distribution\\
\specialrule{0.1em}{4pt}{4pt}
\end{tabular}
\end{table}
\subsection{Symplectic Wigner distribution}\label{sec:2.4}
\indent Owing to Eq.~\eqref{eq2.9}, the Wigner distribution can be rewritten in terms of the symplectic coordinate operator as
\begin{equation}\label{eq2.13}
\mathrm{W}f(\mathbf{x},\mathbf{w})=\mathcal{F}_{\mathbf{y},2}\mathfrak{T}_{(\mathcal{P}\mathcal{I})\mathcal{I}}\left(\mathfrak{I}f\otimes\overline{\mathfrak{I}f}\right)(\mathbf{x},\mathbf{w}),
\end{equation}
where $\mathcal{P}\mathcal{I}\in Sp(N,\mathbb{R})$. Fixed a $2N\times 2N$ symplectic matrix $\mathcal{M}_3\in Sp(N,\mathbb{R})$, replacing the specific symplectic matrix $\mathcal{P}\mathcal{I}$ found in Eq.~\eqref{eq2.13} with the general form $\mathcal{M}_3$ gives the symplectic Wigner distribution.\\
\indent\emph{Definition~5:} Let a function $f\in L^2(\mathbb{R}^N)$ and a symplectic matrix $\mathcal{M}_3=\begin{pmatrix}\mathbf{A}_3&\mathbf{B}_3\\\mathbf{C}_3&\mathbf{D}_3\end{pmatrix}\in Sp(N,\mathbb{R})$. The symplectic Wigner distribution of the function $f$ associated with the symplectic matrix $\mathcal{M}_3$ is defined as \cite{ZhaTIT24}
\begin{equation}\label{eq2.14}
\mathrm{W}(\mathcal{M}_3)f(\mathbf{x},\mathbf{w})=\mathcal{F}_{\mathbf{y},2}\mathfrak{T}_{\mathcal{M}_3\mathcal{I}}\left(\mathfrak{I}f\otimes\overline{\mathfrak{I}f}\right)(\mathbf{x},\mathbf{w}).
\end{equation}
\indent In addition to the Wigner distribution, the symplectic Wigner distribution includes also particular cases the $\tau$-Wigner distribution \cite{Bog10,Bog13,Bog14,Cor18,Cor19,Cor20,Lue19,Lue20,DEl19,Guo22,Vuo22} corresponding to $\mathcal{M}_3=\begin{pmatrix}\mathbf{I}_N&\mathbf{I}_N\\-(1-\tau)\mathbf{I}_N&\tau\mathbf{I}_N\end{pmatrix}$ and the short-time Fourier transform \cite{Gab46,Dau90} corresponding to $\mathcal{M}_3=\begin{pmatrix}\mathbf{I}_N&\mathbf{0}_N\\-\mathbf{I}_N&\mathbf{I}_N\end{pmatrix}$.\\
\indent It should be noted here that the symplectic Wigner distribution is a special case of the symplectic matrix $\mathcal{M}\in Sp(2N,\mathbb{R})$-Wigner distribution, just like the matrix $\mathcal{M}\in GL(2N,\mathbb{R})$-Wigner distribution. According to the relationship between $GL(2N,\mathbb{R})$ and $Sp(N,\mathbb{R})$, shown in Eq.~\eqref{eq2.5}, the main difference and connection between the matrix $\mathcal{M}\in GL(2N,\mathbb{R})$-Wigner distribution and the symplectic Wigner distribution are that the matrix $\mathcal{M}$ embedded in the former is a $2N\times 2N$ invertible matrix while satisfying also $\mathcal{M}\mathcal{J}\mathcal{M}^{\mathrm{T}}=\mathcal{J}$ for the latter.\\
\indent See Table~\ref{tab:4} for a summary of some special cases of the symplectic Wigner distribution.
\begin{table}[htbp]
\centering
\caption{\label{tab:4}Some special cases of the symplectic Wigner distribution}
\footnotesize
\begin{tabular}{cc}
\specialrule{0.1em}{4pt}{4pt}
$\mathcal{M}_3=\begin{pmatrix}\mathbf{A}_3&\mathbf{B}_3\\\mathbf{C}_3&\mathbf{D}_3\end{pmatrix}$ & Symplectic Wigner distribution\\
\specialrule{0.1em}{4pt}{4pt}
$\mathcal{P}\mathcal{I}$ & Wigner distribution\\
\specialrule{0em}{4pt}{4pt}
$\begin{pmatrix}\mathbf{I}_N&\mathbf{I}_N\\-(1-\tau)\mathbf{I}_N&\tau\mathbf{I}_N\end{pmatrix}$ & $\tau$-Wigner distribution\\
\specialrule{0em}{4pt}{4pt}
$\begin{pmatrix}\mathbf{I}_N&\mathbf{0}_N\\-\mathbf{I}_N&\mathbf{I}_N\end{pmatrix}$ & Short-time Fourier transform\\
\specialrule{0.1em}{4pt}{4pt}
\end{tabular}
\end{table}
\subsection{Generalized metaplectic convolution}\label{sec:2.5}
\indent\emph{Definition~6:} Let two functions $f,g\in L^2(\mathbb{R}^N)$ and three symplectic matrices $\mathcal{M}_j=\begin{pmatrix}\mathbf{A}_j&\mathbf{B}_j\\\mathbf{C}_j&\mathbf{D}_j\end{pmatrix}\in Sp(N,\mathbb{R})$ with $\mathrm{det}(\mathbf{B}_j)\neq0$, $j=4,5,6$. The generalized metaplectic convolution operator $\Theta_{\mathcal{M}_4,\mathcal{M}_5,\mathcal{M}_6}$ is defined as \cite{Cui24}
\begin{equation}\label{eq2.15}
\left(f\Theta_{\mathcal{M}_4,\mathcal{M}_5,\mathcal{M}_6}g\right)(\mathbf{x})=\mathrm{e}^{-\pi\mathrm{i}\mathbf{x}\mathbf{B}_6^{-1}\mathbf{A}_6\mathbf{x}^{\mathrm{T}}}\left(f(\mathbf{x})\mathrm{e}^{\pi\mathrm{i}\mathbf{x}\mathbf{B}_4^{-1}\mathbf{A}_4\mathbf{x}^{\mathrm{T}}}\right)\ast\left(g(\mathbf{x})\mathrm{e}^{\pi\mathrm{i}\mathbf{x}\mathbf{B}_5^{-1}\mathbf{A}_5\mathbf{x}^{\mathrm{T}}}\right).
\end{equation}
\indent\emph{Generalized metaplectic convolution theorem:} Let two functions $f,g\in L^2(\mathbb{R}^N)$, and let $\mu(\mathcal{M}_4),\mu(\mathcal{M}_5),\mu(\mathcal{M}_6)$ be three metaplectic operators associated with the symplectic matrices $\mathcal{M}_4=\begin{pmatrix}\mathbf{A}_4&\mathbf{B}_4\\\mathbf{C}_4&\mathbf{D}_4\end{pmatrix},\mathcal{M}_5=\begin{pmatrix}\mathbf{A}_5&\mathbf{B}_5\\\mathbf{C}_5&\mathbf{D}_5\end{pmatrix},\mathcal{M}_6=\begin{pmatrix}\mathbf{A}_6&\mathbf{B}_6\\\mathbf{C}_6&\mathbf{D}_6\end{pmatrix}\in Sp(N,\mathbb{R})$ with $\mathrm{det}(\mathbf{B}_4)\neq0,\mathrm{det}(\mathbf{B}_5)\neq0,\mathrm{det}(\mathbf{B}_6)\neq0$, respectively. The metaplectic transform of the function $f\Theta_{\mathcal{M}_4,\mathcal{M}_5,\mathcal{M}_6}g$ associated with the symplectic matrix $\mathcal{M}_6$ reads \cite{Cui24}
\begin{equation}\label{eq2.16}
\mu(\mathcal{M}_6)\left(f\Theta_{\mathcal{M}_4,\mathcal{M}_5,\mathcal{M}_6}g\right)(\mathbf{u})=\epsilon_{\mathcal{M}_4,\mathcal{M}_5,\mathcal{M}_6}(\mathbf{u})\mu(\mathcal{M}_4)f\left(\mathbf{u}\left(\mathbf{B}_6^{-1}\right)^{\mathrm{T}}\mathbf{B}_4^{\mathrm{T}}\right)\mu(\mathcal{M}_5)g\left(\mathbf{u}\left(\mathbf{B}_6^{-1}\right)^{\mathrm{T}}\mathbf{B}_5^{\mathrm{T}}\right),
\end{equation}
where
\begin{equation}\label{eq2.17}
\epsilon_{\mathcal{M}_4,\mathcal{M}_5,\mathcal{M}_6}(\mathbf{u})=\sqrt{-\frac{\mathrm{det}(\mathbf{B}_4)\mathrm{det}(\mathbf{B}_5)}{\mathrm{det}(\mathbf{B}_6)}}\mathrm{e}^{\pi\mathrm{i}\left[\mathbf{u}\mathbf{D}_6\mathbf{B}_6^{-1}\mathbf{u}^{\mathrm{T}}-\mathbf{u}\left(\mathbf{B}_6^{-1}\right)^{\mathrm{T}}\mathbf{B}_4^{\mathrm{T}}\mathbf{D}_4\mathbf{B}_6^{-1}\mathbf{u}^{\mathrm{T}}-\mathbf{u}\left(\mathbf{B}_6^{-1}\right)^{\mathrm{T}}\mathbf{B}_5^{\mathrm{T}}\mathbf{D}_5\mathbf{B}_6^{-1}\mathbf{u}^{\mathrm{T}}\right]}.
\end{equation}
\indent The generalized metaplectic convolution includes particular cases the conventional convolution corresponding to $\mathcal{M}_4=\mathcal{M}_5=\mathcal{M}_6=\mathcal{J}$, the I-type metaplectic convolution corresponding to $\mathcal{M}_4=\mathcal{M}_6,\mathcal{M}_5=\mathcal{J}$, the II-type metaplectic convolution corresponding to $\mathcal{M}_4=\mathcal{M}_5=\mathcal{M}_6$, and the IV-type metaplectic convolution corresponding to $\mathcal{M}_4=\mathcal{M}_5,\mathcal{M}_6=\begin{pmatrix}\frac{\mathbf{A}_4}{\sqrt{2}}&\sqrt{2}\mathbf{B}_4\\\frac{\mathbf{C}_4}{\sqrt{2}}&\sqrt{2}\mathbf{D}_4\end{pmatrix}$.\\
\indent See Table~\ref{tab:5} for a summary of some special cases of the generalized metaplectic convolution.
\begin{table}[htbp]
\centering
\caption{\label{tab:5}Some special cases of the generalized metaplectic convolution}
\footnotesize
\begin{tabular}{cccc}
\specialrule{0.1em}{4pt}{4pt}
$\mathcal{M}_4=\begin{pmatrix}\mathbf{A}_4&\mathbf{B}_4\\\mathbf{C}_4&\mathbf{D}_4\end{pmatrix}$ & $\mathcal{M}_5=\begin{pmatrix}\mathbf{A}_5&\mathbf{B}_5\\\mathbf{C}_5&\mathbf{D}_5\end{pmatrix}$ & $\mathcal{M}_6=\begin{pmatrix}\mathbf{A}_6&\mathbf{B}_6\\\mathbf{C}_6&\mathbf{D}_6\end{pmatrix}$ & Generalized metaplectic convolution\\
\specialrule{0.1em}{4pt}{4pt}
$\mathcal{J}$ & $\mathcal{J}$ & $\mathcal{J}$ & Conventional convolution\\
\specialrule{0em}{4pt}{4pt}
$\mathcal{M}_4$  & $\mathcal{J}$ & $\mathcal{M}_4$ & I-type metaplectic convolution\\
\specialrule{0em}{4pt}{4pt}
$\mathcal{M}_4$  & $\mathcal{M}_4$ & $\mathcal{M}_4$ & II-type metaplectic convolution\\
\specialrule{0em}{4pt}{4pt}
$\mathcal{M}_4$  & $\mathcal{M}_4$ & $\begin{pmatrix}\frac{\mathbf{A}_4}{\sqrt{2}}&\sqrt{2}\mathbf{B}_4\\\frac{\mathbf{C}_4}{\sqrt{2}}&\sqrt{2}\mathbf{D}_4\end{pmatrix}$ & IV-type metaplectic convolution\\
\specialrule{0.1em}{4pt}{4pt}
\end{tabular}
\end{table}
\section{Convolution type of metaplectic Cohen's distribution (CMCD)}\label{sec:3}
\indent In this section, we first integrate the high-dimensional CICFWD and symplectic Wigner distribution to propose the definition of the metaplectic Wigner distribution. Then, we combine it with the generalized metaplectic convolution to formulate the definition of the CMCD. We also disclose its relations to many celebrated time-frequency analysis tools and deduce its essential properties.
\subsection{Joint time-frequency operators and coordinate operator fractionizations}\label{sec:3.1}
\indent Recall that the high-dimensional CICFWD and symplectic Wigner distribution, given respectively by Eqs.~\eqref{eq2.11} and \eqref{eq2.14}, are Wigner distributions in metaplectic transform domains with the strongest generalization capability in terms of time-frequency operators fractionization and coordinate operator fractionization, respectively. Combining them yields the metaplectic Wigner distribution, which can be regarded as a unified fractional domain Wigner distribution.\\
\indent\emph{Definition~7:} Let a function $f\in L^2(\mathbb{R}^N)$ and a symplectic matrix $\mathcal{M}_3=\begin{pmatrix}\mathbf{A}_3&\mathbf{B}_3\\\mathbf{C}_3&\mathbf{D}_3\end{pmatrix}\in Sp(N,\mathbb{R})$, let $\mu_{\mathbf{y},2}(\mathcal{M})$ be the partial metaplectic operator with respect to the second variables $\mathbf{y}$ associated with the symplectic matrix $\mathcal{M}=\begin{pmatrix}\mathbf{A}&\mathbf{B}\\\mathbf{C}&\mathbf{D}\end{pmatrix}\in Sp(N,\mathbb{R})$ with $\mathrm{det}(\mathbf{B})\neq0$, and let $\mu(\mathcal{M}_1),\mu(\mathcal{M}_2)$ be two metaplectic operators associated with the symplectic matrices $\mathcal{M}_1=\begin{pmatrix}\mathbf{A}_1&\mathbf{B}_1\\\mathbf{C}_1&\mathbf{D}_1\end{pmatrix},\mathcal{M}_2=\begin{pmatrix}\mathbf{A}_2&\mathbf{B}_2\\\mathbf{C}_2&\mathbf{D}_2\end{pmatrix}\in Sp(N,\mathbb{R})$ with $\mathrm{det}(\mathbf{B}_1)\neq0,\mathrm{det}(\mathbf{B}_2)\neq0$, respectively. The metaplectic Wigner distribution of the function $f$ associated with the symplectic matrices $\mathcal{M},\mathcal{M}_1,\mathcal{M}_2,\mathcal{M}_3$ is defined as
\begin{equation}\label{eq3.1}
\mathrm{W}(\mathcal{M},\mathcal{M}_1,\mathcal{M}_2,\mathcal{M}_3)f(\mathbf{x},\mathbf{u})=\mu_{\mathbf{y},2}(\mathcal{M})\mathfrak{T}_{\mathcal{M}_3\mathcal{I}}\left(\mu(\mathcal{M}_1)f\otimes\overline{\mu(\mathcal{M}_2)f}\right)(\mathbf{x},\mathbf{u}).
\end{equation}
\indent Three points deserve to be underlined regarding the relationships between the metaplectic Wigner distribution and the existing Wigner distributions in metaplectic transform domains.
\begin{itemize}

    \item The metaplectic Wigner distribution is none other than an organic integration of the high-dimensional CICFWD and symplectic Wigner distribution, which are two special cases corresponding to $\mathcal{M}_3=\mathcal{P}\mathcal{I}$ and $\mathcal{M}=\mathcal{J},\mathcal{M}_1=\mathcal{M}_2=\mathbf{I}_{2N}$, respectively. Therefore, it can be considered as the Wigner distribution in metaplectic transform domains with the strongest generalization capability under the combined action of time-frequency operators and coordinate operator fractionizations.

    \item The special cases of the high-dimensional CICFWD and symplectic Wigner distribution are also those of the metaplectic Wigner distribution, including the Wigner distribution, the CICFWD, the $N$-dimensional nonseparable affine characteristic Wigner distribution, the $N$-dimensional nonseparable basis function Wigner distribution, the $N$-dimensional nonseparable convolution representation Wigner distribution, the $N$-dimensional nonseparable instantaneous cross-correlation function Wigner distribution, the $\tau$-Wigner distribution, and the short-time Fourier transform.

    \item Because of the different numbers and types of symplectic matrices embedding mechanisms, the metaplectic Wigner distribution differs essentially from the matrix $\mathcal{M}\in GL(2N,\mathbb{R})$-Wigner distribution and the symplectic matrix $\mathcal{M}\in Sp(2N,\mathbb{R})$-Wigner distribution.

\end{itemize}

\indent See Table~\ref{tab:6} for a summary of some special cases of the metaplectic Wigner distribution.
\begin{table}[htbp]
\centering
\caption{\label{tab:6}Some special cases of the metaplectic Wigner distribution}
\tiny
\begin{tabular}{ccccc}
\specialrule{0.1em}{4pt}{4pt}
$\mathcal{M}=\begin{pmatrix}\mathbf{A}&\mathbf{B}\\\mathbf{C}&\mathbf{D}\end{pmatrix}$ & $\mathcal{M}_1=\begin{pmatrix}\mathbf{A}_1&\mathbf{B}_1\\\mathbf{C}_1&\mathbf{D}_1\end{pmatrix}$ & $\mathcal{M}_2=\begin{pmatrix}\mathbf{A}_2&\mathbf{B}_2\\\mathbf{C}_2&\mathbf{D}_2\end{pmatrix}$ & $\mathcal{M}_3=\begin{pmatrix}\mathbf{A}_3&\mathbf{B}_3\\\mathbf{C}_3&\mathbf{D}_3\end{pmatrix}$ & Metaplectic Wigner distribution\\
\specialrule{0.1em}{4pt}{4pt}
$\mathcal{M}$ & $\mathcal{M}_1$ & $\mathcal{M}_2$ & $\mathcal{P}\mathcal{I}$ & High-dimensional CICFWD\\
\specialrule{0em}{4pt}{4pt}
$\mathcal{J}$ & $\mathbf{I}_{2N}$ & $\mathbf{I}_{2N}$ & $\mathcal{M}_3$ & Symplectic Wigner distribution\\
\specialrule{0em}{4pt}{4pt}
$\mathcal{J}$ & $\mathbf{I}_{2N}$ & $\mathbf{I}_{2N}$ & $\mathcal{P}\mathcal{I}$ & Wigner distribution\\
\specialrule{0em}{4pt}{4pt}
$\begin{pmatrix}a&b\\c&d\end{pmatrix}$  & $\begin{pmatrix}a_1&b_1\\c_1&d_1\end{pmatrix}$ & $\begin{pmatrix}a_2&b_2\\c_2&d_2\end{pmatrix}$ & $\mathcal{P}\mathcal{I}$ & CICFWD\\
\specialrule{0em}{4pt}{4pt}
\multirow{2}{*}{$\mathcal{J}$} & \multirow{2}{*}{$\mathcal{M}_1$} & \multirow{2}{*}{$\mathcal{M}_1$} & \multirow{2}{*}{$\mathcal{P}\mathcal{I}$} & $N$-dimensional nonseparable affine\\
 & & & & characteristic Wigner distribution\\
\specialrule{0em}{4pt}{4pt}
\multirow{2}{*}{$\mathcal{M}$} & \multirow{2}{*}{$\mathbf{I}_{2N}$} & \multirow{2}{*}{$\mathbf{I}_{2N}$} & \multirow{2}{*}{$\mathcal{P}\mathcal{I}$} & $N$-dimensional nonseparable basis\\
 & & & & function Wigner distribution\\
\multirow{2}{*}{$\begin{pmatrix}\frac{\mathbf{D}_1}{4}&-\mathbf{B}_1\\-\mathbf{C}_1&4\mathbf{A}_1\end{pmatrix}$} & \multirow{2}{*}{$\mathcal{M}_1$} & \multirow{2}{*}{$\begin{pmatrix}-\mathbf{A}_1&\mathbf{B}_1\\\mathbf{C}_1&-\mathbf{D}_1\end{pmatrix}$} & \multirow{2}{*}{$\mathcal{P}\mathcal{I}$} & $N$-dimensional nonseparable convolution\\
 & & & & representation Wigner distribution\\
\multirow{2}{*}{$\mathcal{M}$} & \multirow{2}{*}{$\mathcal{M}_1$} & \multirow{2}{*}{$\mathbf{I}_{2N}$} & \multirow{2}{*}{$\mathcal{P}\mathcal{I}$} & $N$-dimensional nonseparable instantaneous\\
 & & & & cross-correlation function Wigner distribution\\
\specialrule{0em}{4pt}{4pt}
$\mathcal{J}$ & $\mathbf{I}_{2N}$ & $\mathbf{I}_{2N}$ & $\begin{pmatrix}\mathbf{I}_N&\mathbf{I}_N\\-(1-\tau)\mathbf{I}_N&\tau\mathbf{I}_N\end{pmatrix}$ & $\tau$-Wigner distribution\\
\specialrule{0em}{4pt}{4pt}
$\mathcal{J}$ & $\mathbf{I}_{2N}$ & $\mathbf{I}_{2N}$ & $\begin{pmatrix}\mathbf{I}_N&\mathbf{0}_N\\-\mathbf{I}_N&\mathbf{I}_N\end{pmatrix}$ & Short-time Fourier transform\\
\specialrule{0.1em}{4pt}{4pt}
\end{tabular}
\end{table}
\subsection{Joint Wigner operator and convolution operator fractionizations}\label{sec:3.2}
\indent Recall that the generalized metaplectic convolution, given by Eq.~\eqref{eq2.15}, is the convolution in metaplectic transform domains with the strongest generalization capability in terms of convolution operator fractionization. Replacing the Wigner operator $\mathrm{W}$ and the convolution operator $\ast$ found in the ordinary Cohen's distribution with the metaplectic Wigner operator $\mathrm{W}(\mathcal{M},\mathcal{M}_1,\mathcal{M}_2,\mathcal{M}_3)$ ($\mathcal{M},\mathcal{M}_1,\mathcal{M}_2,\mathcal{M}_3\in Sp(N,\mathbb{R})$) and the generalized metaplectic convolution operator $\Theta_{\mathcal{M}_4,\mathcal{M}_5,\mathcal{M}_6}$ ($\mathcal{M}_4,\mathcal{M}_5,\mathcal{M}_6\in Sp(2N,\mathbb{R})$), respectively, gives the CMCD.\\
\indent\emph{Definition~8:} Let a function $f\in L^2(\mathbb{R}^N)$ and a kernel function $\Pi$, let $\mathrm{W}(\mathcal{M},\mathcal{M}_1,\mathcal{M}_2,\mathcal{M}_3)$ be the metaplectic Wigner operator associated with the symplectic matrices $\mathcal{M}=\begin{pmatrix}\mathbf{A}&\mathbf{B}\\\mathbf{C}&\mathbf{D}\end{pmatrix},\mathcal{M}_1=\begin{pmatrix}\mathbf{A}_1&\mathbf{B}_1\\\mathbf{C}_1&\mathbf{D}_1\end{pmatrix},\mathcal{M}_2=\begin{pmatrix}\mathbf{A}_2&\mathbf{B}_2\\\mathbf{C}_2&\mathbf{D}_2\end{pmatrix},\mathcal{M}_3=\begin{pmatrix}\mathbf{A}_3&\mathbf{B}_3\\\mathbf{C}_3&\mathbf{D}_3\end{pmatrix}\in Sp(N,\mathbb{R})$ with $\mathrm{det}(\mathbf{B})\neq0,\mathrm{det}(\mathbf{B}_1)\neq0,\mathrm{det}(\mathbf{B}_2)\neq0$, and let $\Theta_{\mathcal{M}_4,\mathcal{M}_5,\mathcal{M}_6}$ be the generalized metaplectic convolution operator associated with the symplectic matrices $\mathcal{M}_4=\begin{pmatrix}\mathbf{A}_4&\mathbf{B}_4\\\mathbf{C}_4&\mathbf{D}_4\end{pmatrix},\mathcal{M}_5=\begin{pmatrix}\mathbf{A}_5&\mathbf{B}_5\\\mathbf{C}_5&\mathbf{D}_5\end{pmatrix},\mathcal{M}_6=\begin{pmatrix}\mathbf{A}_6&\mathbf{B}_6\\\mathbf{C}_6&\mathbf{D}_6\end{pmatrix}\in Sp(2N,\mathbb{R})$ with $\mathrm{det}(\mathbf{B}_4)\neq0,\mathrm{det}(\mathbf{B}_5)\neq0,\mathrm{det}(\mathbf{B}_6)\neq0$. The CMCD of the function $f$ associated with the symplectic matrices $\mathcal{M},\mathcal{M}_1,\mathcal{M}_2,\mathcal{M}_3,\mathcal{M}_4,\mathcal{M}_5,\mathcal{M}_6$ is defined as
\begin{equation}\label{eq3.2}
\mathrm{C}(\mathcal{M},\mathcal{M}_1,\mathcal{M}_2,\mathcal{M}_3,\mathcal{M}_4,\mathcal{M}_5,\mathcal{M}_6)f(\mathbf{x},\mathbf{u})=\left(\mathrm{W}(\mathcal{M},\mathcal{M}_1,\mathcal{M}_2,\mathcal{M}_3)f\Theta_{\mathcal{M}_4,\mathcal{M}_5,\mathcal{M}_6}\Pi\right)(\mathbf{x},\mathbf{u}).
\end{equation}
\indent The CMCD can be considered as the convolution type of metaplectic transform domains Cohen's distribution with the strongest generalization capability under the combined action of Wigner operator and convolution operator fractionizations.
\subsection{Relations to the classical time-frequency analysis tools}\label{sec:3.3}
\indent When $\mathcal{M}=\mathcal{J},\mathcal{M}_1=\mathcal{M}_2=\mathbf{I}_{2N},\mathcal{M}_3=\mathcal{P}\mathcal{I}$, the CMCD reduces to
\begin{equation}\label{eq3.3}
\left(\mathrm{W}f\Theta_{\mathcal{M}_4,\mathcal{M}_5,\mathcal{M}_6}\Pi\right)(\mathbf{x},\mathbf{u}),
\end{equation}
which denotes the generalized metaplectic convolution-based Cohen's distribution.\\
\indent When $\mathcal{M}_4=\mathcal{M}_5=\mathcal{M}_6=\begin{pmatrix}\mathbf{0}_{2N}&\mathbf{I}_{2N}\\-\mathbf{I}_{2N}&\mathbf{0}_{2N}\end{pmatrix}$, the CMCD reduces to
\begin{equation}\label{eq3.4}
\left(\mathrm{W}(\mathcal{M},\mathcal{M}_1,\mathcal{M}_2,\mathcal{M}_3)f\ast\Pi\right)(\mathbf{x},\mathbf{u}),
\end{equation}
which denotes the metaplectic Wigner distribution-based Cohen's distribution.\\
\indent When $\mathcal{M}=\mathcal{J},\mathcal{M}_1=\mathcal{M}_2=\mathbf{I}_{2N},\mathcal{M}_3=\mathcal{P}\mathcal{I},\mathcal{M}_4=\mathcal{M}_5=\mathcal{M}_6=\begin{pmatrix}\mathbf{0}_{2N}&\mathbf{I}_{2N}\\-\mathbf{I}_{2N}&\mathbf{0}_{2N}\end{pmatrix}$, the CMCD reduces to the Cohen's distribution. The special cases of the Cohen's distribution are also those of the CMCD, including the Wigner distribution, the Choi-Williams distribution, the Kirkwood-Rihaczek distribution, the Born-Jordan distribution, the Zhao-Atlas-Marks distribution, the Margenau-Hill distribution, and the Page distribution.\\
\indent When $\mathcal{M}_4=\mathcal{M}_6$ and $\Pi=\delta$, which denotes the Dirac delta operator, the CMCD reduces to the metaplectic Wigner distribution. The special cases of the metaplectic Wigner distribution are also those of the CMCD, including the Wigner distribution, the CICFWD, the $N$-dimensional nonseparable affine characteristic Wigner distribution, the $N$-dimensional nonseparable basis function Wigner distribution, the $N$-dimensional nonseparable convolution representation Wigner distribution, the $N$-dimensional nonseparable instantaneous cross-correlation function Wigner distribution, the $\tau$-Wigner distribution, and the short-time Fourier transform.\\
\indent When $\mathcal{M}_4=\mathcal{M}_6,\mathcal{M}_5=\begin{pmatrix}\mathbf{0}_{2N}&\mathbf{I}_{2N}\\-\mathbf{I}_{2N}&\mathbf{0}_{2N}\end{pmatrix}$, the CMCD reduces to
\begin{equation}\label{eq3.5}
\left(\mathrm{W}(\mathcal{M},\mathcal{M}_1,\mathcal{M}_2,\mathcal{M}_3)f\Theta_{\mathcal{M}_4,\mathcal{J},\mathcal{M}_4}\Pi\right)(\mathbf{x},\mathbf{u})\triangleq\mathrm{C}_{\mathrm{I}}(\mathcal{M},\mathcal{M}_1,\mathcal{M}_2,\mathcal{M}_3,\mathcal{M}_4)f(\mathbf{x},\mathbf{u}),
\end{equation}
which denotes joint metaplectic Wigner and I-type metaplectic convolution operators Cohen's distribution.\\
\indent When $\mathcal{M}_4=\mathcal{M}_5=\mathcal{M}_6$, the CMCD reduces to
\begin{equation}\label{eq3.6}
\left(\mathrm{W}(\mathcal{M},\mathcal{M}_1,\mathcal{M}_2,\mathcal{M}_3)f\Theta_{\mathcal{M}_4,\mathcal{M}_4,\mathcal{M}_4}\Pi\right)(\mathbf{x},\mathbf{u})\triangleq\mathrm{C}_{\mathrm{II}}(\mathcal{M},\mathcal{M}_1,\mathcal{M}_2,\mathcal{M}_3,\mathcal{M}_4)f(\mathbf{x},\mathbf{u}),
\end{equation}
which denotes joint metaplectic Wigner and II-type metaplectic convolution operators Cohen's distribution.\\
\indent When $\mathcal{M}_4=\mathcal{M}_5,\mathcal{M}_6=\begin{pmatrix}\frac{\mathbf{A}_4}{\sqrt{2}}&\sqrt{2}\mathbf{B}_4\\\frac{\mathbf{C}_4}{\sqrt{2}}&\sqrt{2}\mathbf{D}_4\end{pmatrix}$, the CMCD reduces to
\begin{equation}\label{eq3.7}
\left(\mathrm{W}(\mathcal{M},\mathcal{M}_1,\mathcal{M}_2,\mathcal{M}_3)f\Theta_{\mathcal{M}_4,\mathcal{M}_4,\begin{pmatrix}\frac{\mathbf{A}_4}{\sqrt{2}}&\sqrt{2}\mathbf{B}_4\\\frac{\mathbf{C}_4}{\sqrt{2}}&\sqrt{2}\mathbf{D}_4\end{pmatrix}}\Pi\right)(\mathbf{x},\mathbf{u})\triangleq\mathrm{C}_{\mathrm{IV}}(\mathcal{M},\mathcal{M}_1,\mathcal{M}_2,\mathcal{M}_3,\mathcal{M}_4)f(\mathbf{x},\mathbf{u}),
\end{equation}
which denotes joint metaplectic Wigner and IV-type metaplectic convolution operators Cohen's distribution.\\
\indent See Table~\ref{tab:7} for a summary of some special cases of the CMCD.
\begin{table}[htbp]
\centering
\caption{\label{tab:7}Some special cases of the CMCD}
\tiny
\begin{tabular}{ccccccccc}
\specialrule{0.1em}{4pt}{4pt}
$\mathcal{M}$ & $\mathcal{M}_1$ & $\mathcal{M}_2$ & $\mathcal{M}_3$ & $\mathcal{M}_4$ & $\mathcal{M}_5$ & $\mathcal{M}_6$ & $\Pi$ & CMCD\\
\specialrule{0.1em}{4pt}{4pt}
\multirow{2}{*}{$\mathcal{J}$} & \multirow{2}{*}{$\mathbf{I}_{2N}$} & \multirow{2}{*}{$\mathbf{I}_{2N}$} & \multirow{2}{*}{$\mathcal{P}\mathcal{I}$} & \multirow{2}{*}{$\mathcal{M}_4$} & \multirow{2}{*}{$\mathcal{M}_5$} & \multirow{2}{*}{$\mathcal{M}_6$} & \multirow{2}{*}{$\Pi$} & Generalized metaplectic convolution-based\\
 & & & & & & & & Cohen's distribution\\
\specialrule{0em}{4pt}{4pt}
\multirow{2}{*}{$\mathcal{M}$} & \multirow{2}{*}{$\mathcal{M}_1$} & \multirow{2}{*}{$\mathcal{M}_2$} & \multirow{2}{*}{$\mathcal{M}_3$} & \multirow{2}{*}{$\begin{pmatrix}\mathbf{0}_{2N}&\mathbf{I}_{2N}\\-\mathbf{I}_{2N}&\mathbf{0}_{2N}\end{pmatrix}$} & \multirow{2}{*}{$\begin{pmatrix}\mathbf{0}_{2N}&\mathbf{I}_{2N}\\-\mathbf{I}_{2N}&\mathbf{0}_{2N}\end{pmatrix}$} & \multirow{2}{*}{$\begin{pmatrix}\mathbf{0}_{2N}&\mathbf{I}_{2N}\\-\mathbf{I}_{2N}&\mathbf{0}_{2N}\end{pmatrix}$} & \multirow{2}{*}{$\Pi$} & Metaplectic Wigner distribution-based\\
 & & & & & & & & Cohen's distribution\\
\specialrule{0em}{4pt}{4pt}
$\mathcal{J}$ & $\mathbf{I}_{2N}$ & $\mathbf{I}_{2N}$ & $\mathcal{P}\mathcal{I}$ & $\begin{pmatrix}\mathbf{0}_{2N}&\mathbf{I}_{2N}\\-\mathbf{I}_{2N}&\mathbf{0}_{2N}\end{pmatrix}$ & $\begin{pmatrix}\mathbf{0}_{2N}&\mathbf{I}_{2N}\\-\mathbf{I}_{2N}&\mathbf{0}_{2N}\end{pmatrix}$ & $\begin{pmatrix}\mathbf{0}_{2N}&\mathbf{I}_{2N}\\-\mathbf{I}_{2N}&\mathbf{0}_{2N}\end{pmatrix}$ & $\Pi$ & Cohen's distribution\\
\specialrule{0em}{4pt}{4pt}
$\mathcal{M}$ & $\mathcal{M}_1$ & $\mathcal{M}_2$ & $\mathcal{M}_3$ & $\mathcal{M}_4$ & $\mathcal{M}_5$ & $\mathcal{M}_4$ & $\delta$ & Metaplectic Wigner distribution\\
\specialrule{0em}{4pt}{4pt}
\multirow{2}{*}{$\mathcal{M}$} & \multirow{2}{*}{$\mathcal{M}_1$} & \multirow{2}{*}{$\mathcal{M}_2$} & \multirow{2}{*}{$\mathcal{M}_3$} & \multirow{2}{*}{$\mathcal{M}_4$} & \multirow{2}{*}{$\begin{pmatrix}\mathbf{0}_{2N}&\mathbf{I}_{2N}\\-\mathbf{I}_{2N}&\mathbf{0}_{2N}\end{pmatrix}$} & \multirow{2}{*}{$\mathcal{M}_4$} & \multirow{2}{*}{$\Pi$} & Joint metaplectic Wigner and I-type metaplectic\\
 & & & & & & & & convolution operators Cohen's distribution\\
\specialrule{0em}{4pt}{4pt}
\multirow{2}{*}{$\mathcal{M}$} & \multirow{2}{*}{$\mathcal{M}_1$} & \multirow{2}{*}{$\mathcal{M}_2$} & \multirow{2}{*}{$\mathcal{M}_3$} & \multirow{2}{*}{$\mathcal{M}_4$} & \multirow{2}{*}{$\mathcal{M}_4$} & \multirow{2}{*}{$\mathcal{M}_4$} & \multirow{2}{*}{$\Pi$} & Joint metaplectic Wigner and II-type metaplectic\\
 & & & & & & & & convolution operators Cohen's distribution\\
\specialrule{0em}{4pt}{4pt}
\multirow{2}{*}{$\mathcal{M}$} & \multirow{2}{*}{$\mathcal{M}_1$} & \multirow{2}{*}{$\mathcal{M}_2$} & \multirow{2}{*}{$\mathcal{M}_3$} & \multirow{2}{*}{$\mathcal{M}_4$} & \multirow{2}{*}{$\mathcal{M}_4$} & \multirow{2}{*}{$\begin{pmatrix}\frac{\mathbf{A}_4}{\sqrt{2}}&\sqrt{2}\mathbf{B}_4\\\frac{\mathbf{C}_4}{\sqrt{2}}&\sqrt{2}\mathbf{D}_4\end{pmatrix}$} & \multirow{2}{*}{$\Pi$} & Joint metaplectic Wigner and IV-type metaplectic\\
 & & & & & & & & convolution operators Cohen's distribution\\
\specialrule{0.1em}{4pt}{4pt}
\end{tabular}
\end{table}
\subsection{Essential properties}\label{sec:3.4}
\indent In this section, we extend some essential properties of the Cohen's distribution to those of the CMCD, including marginal distribution, energy conservation, unique reconstruction, Moyal formula, complex conjugate symmetry, time reversal symmetry, scaling property, time translation property, frequency modulation property, and metaplectic invariance. We also demonstrate that how those general properties get reflected as constraints on the symplectic matrices and the kernel function. For simplicity in describing and proving these properties, we denote $\mathbf{B}_j^{-1}\mathbf{A}_j=\begin{pmatrix}\mathbf{F}_{j1}&\mathbf{F}_{j2}\\\mathbf{F}_{j3}&\mathbf{F}_{j4}\end{pmatrix}$, $j=4,5,6$.\\
\indent\emph{Property~1 (Marginal distribution):} There are four types of marginal distributions for the CMCD, including time, frequency, time delay and frequency shift marginal distributions.\\
\indent\emph{Time marginal distribution:} When $\mathbf{F}_{j2}=\mathbf{F}_{j3}=\mathbf{0}_N$, $j=4,5$, $\mathbf{F}_{44}=-\mathbf{D}\mathbf{B}^{-1}$, $\mathbf{F}_{54}=\mathbf{0}_N$, and the kernel function satisfies $\phi(\mathbf{v},\mathbf{0})=1$, the integration of the CMCD with respect to the variables $\mathbf{u}$ reads
\begin{align}\label{Time marginal distribution} &\int_{\mathbb{R}^N}\mathrm{C}(\mathcal{M},\mathcal{M}_1,\mathcal{M}_2,\mathcal{M}_3,\mathcal{M}_4,\mathcal{M}_5,\mathcal{M}_6)f(\mathbf{x},\mathbf{u})\mathrm{e}^{\pi\mathrm{i}\left(\mathbf{x},\mathbf{u}\right)\mathbf{B}_6^{-1}\mathbf{A}_6\left(\mathbf{x},\mathbf{u}\right)^{\mathrm{T}}}\mathrm{d}\mathbf{u}\nonumber\\ =&\sqrt{-\mathrm{det}(\mathbf{B})}\mathrm{e}^{\pi\mathrm{i}\mathbf{x}\mathbf{F}_{41}\mathbf{x}^{\mathrm{T}}}\mu(\mathcal{M}_1)f\left(\mathbf{x}\mathbf{B}_3\right)\overline{\mu(\mathcal{M}_2)f\left(\mathbf{x}\mathbf{A}_3\right)}.
\end{align}
\indent\emph{Frequency marginal distribution:} When $\mathcal{M}_5=\mathbf{0}_{4N}$, $\mathbf{A}=\mathbf{0}_N$, $\mathbf{F}_{41}=\mathbf{F}_{42}=\mathbf{F}_{43}=\mathbf{0}_N$, $\mathrm{det}(\mathbf{A}_3)\neq0$, and the kernel function satisfies $\phi(\mathbf{0},\mathbf{z})=1$, the integration of the CMCD with respect to the variables $\mathbf{x}$ reads
\begin{align}\label{Frequency marginal distribution} &\int_{\mathbb{R}^N}\mathrm{C}(\mathcal{M},\mathcal{M}_1,\mathcal{M}_2,\mathcal{M}_3,\mathcal{M}_4,\mathcal{M}_5,\mathcal{M}_6)f(\mathbf{x},\mathbf{u})\mathrm{e}^{\pi\mathrm{i}\left(\mathbf{x},\mathbf{u}\right)\mathbf{B}_6^{-1}\mathbf{A}_6\left(\mathbf{x},\mathbf{u}\right)^{\mathrm{T}}}\mathrm{d}\mathbf{x}\nonumber\\ =&\frac{\left|\mathrm{det}(\mathbf{A}_3)\right|\left|\mathrm{det}(\mathbf{D}_3)\right|}{\sqrt{-\mathrm{det}(\mathbf{B})}}\mathrm{e}^{\pi\mathrm{i}\left(\mathbf{u}\mathbf{D}\mathbf{B}^{-1}\mathbf{u}^{\mathrm{T}}+\mathbf{u}\mathbf{F}_{44}\mathbf{u}^{\mathrm{T}}\right)}\mu\left(\begin{pmatrix}\mathbf{0}_N&\mathbf{B}{\left(\mathbf{A}_3^{-1}\right)}^{\mathrm{T}}\\-\left(\mathbf{B}^{-1}\right)^{\mathrm{T}}\mathbf{A}_3&\mathbf{I}_N\end{pmatrix}\mathcal{M}_1\right)f(\mathbf{u})\nonumber\\ &\times\overline{\mu\left(\begin{pmatrix}\mathbf{0}_N&\mathbf{B}{\left(\mathbf{B}_3^{-1}\right)}^{\mathrm{T}}\\-\mathbf{B}_3^{\mathrm{T}}\mathbf{B}^{-1}&\mathbf{B}_3^{-1}\mathbf{A}_3\end{pmatrix}\mathcal{M}_2\right)f(\mathbf{u})},
\end{align}
where $\begin{pmatrix}\mathbf{0}_N&\mathbf{B}{\left(\mathbf{A}_3^{-1}\right)}^{\mathrm{T}}\\-\left(\mathbf{B}^{-1}\right)^{\mathrm{T}}\mathbf{A}_3&\mathbf{I}_N\end{pmatrix},\begin{pmatrix}\mathbf{0}_N&\mathbf{B}{\left(\mathbf{B}_3^{-1}\right)}^{\mathrm{T}}\\-\mathbf{B}_3^{\mathrm{T}}\mathbf{B}^{-1}&\mathbf{B}_3^{-1}\mathbf{A}_3\end{pmatrix}\in Sp(N,\mathbb{R})$.\\
\indent\emph{Time delay marginal distribution:} The CMCD of $\mathbf{x}=\mathbf{0}$ reads
\begin{align}\label{Time delay marginal distribution}	&\mathrm{C}(\mathcal{M},\mathcal{M}_1,\mathcal{M}_2,\mathcal{M}_3,\mathcal{M}_4,\mathcal{M}_5,\mathcal{M}_6)f(\mathbf{0},\mathbf{u})\nonumber\\ =&\mathrm{e}^{-\pi\mathrm{i}\left(\mathbf{0},\mathbf{u}\right)\mathbf{B}_6^{-1}\mathbf{A}_6\left(\mathbf{0},\mathbf{u}\right)^{\mathrm{T}}}\iiint_{\mathbb{R}^{N\times N\times N}}\mu(\mathcal{M}_1)f\left(\mathbf{p}\mathbf{B}_3+\mathbf{t}\mathbf{D}_3\right)\overline{\mu(\mathcal{M}_2)f\left(\mathbf{p}\mathbf{A}_3+\mathbf{t}\mathbf{C}_3\right)}\mathrm{e}^{\pi\mathrm{i}\left(\mathbf{p},\mathbf{q}\right)\mathbf{B}_4^{-1}\mathbf{A}_4\left(\mathbf{p},\mathbf{q}\right)^{\mathrm{T}}}\nonumber\\
&\times\mathrm{e}^{\pi\mathrm{i}\left(-\mathbf{p},\mathbf{u}-\mathbf{q}\right)\mathbf{B}_5^{-1}\mathbf{A}_5\left(-\mathbf{p},\mathbf{u}-\mathbf{q}\right)^{\mathrm{T}}}\mathcal{K}_{\mathcal{M}}(\mathbf{q},\mathbf{t})\iint_{\mathbb{R}^{N\times N}}\phi\left(\mathbf{v},\mathbf{z}\right)\mathrm{e}^{-2\pi\mathrm{i}\left(-\mathbf{v}\mathbf{p}^{\mathrm{T}}+\mathbf{z}(\mathbf{u}-\mathbf{q})^{\mathrm{T}}\right)}\mathrm{d}\mathbf{v}\mathrm{d}\mathbf{z}\mathrm{d}\mathbf{t}\mathrm{d}\mathbf{p}\mathrm{d}\mathbf{q}.
\end{align}
\indent\emph{Frequency shift marginal distribution:} The CMCD of $\mathbf{u}=\mathbf{0}$ reads
\begin{align}\label{Frequency shift marginal distribution} &\mathrm{C}(\mathcal{M},\mathcal{M}_1,\mathcal{M}_2,\mathcal{M}_3,\mathcal{M}_4,\mathcal{M}_5,\mathcal{M}_6)f(\mathbf{x},\mathbf{0})\nonumber\\ =&\mathrm{e}^{-\pi\mathrm{i}\left(\mathbf{x},\mathbf{0}\right)\mathbf{B}_6^{-1}\mathbf{A}_6\left(\mathbf{x},\mathbf{0}\right)^{\mathrm{T}}}\iiint_{\mathbb{R}^{N\times N\times N}}\mu(\mathcal{M}_1)f\left(\mathbf{p}\mathbf{B}_3+\mathbf{t}\mathbf{D}_3\right)\overline{\mu(\mathcal{M}_2)f\left(\mathbf{p}\mathbf{A}_3+\mathbf{t}\mathbf{C}_3\right)}\mathrm{e}^{\pi\mathrm{i}\left(\mathbf{p},\mathbf{q}\right)\mathbf{B}_4^{-1}\mathbf{A}_4\left(\mathbf{p},\mathbf{q}\right)^{\mathrm{T}}}\nonumber\\
&\times\mathrm{e}^{\pi\mathrm{i}\left(\mathbf{x}-\mathbf{p},-\mathbf{q}\right)\mathbf{B}_5^{-1}\mathbf{A}_5\left(\mathbf{x}-\mathbf{p},-\mathbf{q}\right)^{\mathrm{T}}}\mathcal{K}_{\mathcal{M}}(\mathbf{q},\mathbf{t})\iint_{\mathbb{R}^{N\times N}}\phi\left(\mathbf{v},\mathbf{z}\right)\mathrm{e}^{-2\pi\mathrm{i}\left(\mathbf{v}(\mathbf{x-p})^{\mathrm{T}}-\mathbf{z}\mathbf{q}^{\mathrm{T}}\right)}\mathrm{d}\mathbf{v}\mathrm{d}\mathbf{z}\mathrm{d}\mathbf{t}\mathrm{d}\mathbf{p}\mathrm{d}\mathbf{q}.
\end{align}
\indent\emph{Property~2 (Energy conservation):} The energy conservation is closely related to the marginal distribution. There are three equivalent forms regarding to the energy conservation of the CMCD, including time marginal distribution, frequency marginal distribution and time delay or frequency shift marginal distribution based energy conservations.\\
\indent\emph{Time marginal distribution based energy conservation:} Integrating with respect to the variables $\mathbf{x}$ on both sides of Eq.~\eqref{Time marginal distribution} yields
\begin{align}\label{Time marginal distribution based energy conservation} &\int_{\mathbb{R}^N}\mathrm{e}^{-\pi\mathrm{i}\mathbf{x}\mathbf{F}_{41}\mathbf{x}^{\mathrm{T}}}\int_{\mathbb{R}^N}\mathrm{C}(\mathcal{M},\mathcal{M}_1,\mathcal{M}_2,\mathcal{M}_3,\mathcal{M}_4,\mathcal{M}_5,\mathcal{M}_6)f(\mathbf{x},\mathbf{u})\mathrm{e}^{\pi\mathrm{i}\left(\mathbf{x},\mathbf{u}\right)\mathbf{B}_6^{-1}\mathbf{A}_6\left(\mathbf{x},\mathbf{u}\right)^{\mathrm{T}}}\mathrm{d}\mathbf{u}\mathrm{d}\mathbf{x}\nonumber\\ =&\frac{\sqrt{-\mathrm{det}(\mathbf{B})}}{\left|\mathrm{det}(\mathbf{B}_3)\right|}\int_{\mathbb{R}^N}\left|f(\mathbf{x})\right|^2\mathrm{d}\mathbf{x}
\end{align}
for $\mathbf{A}_3=\mathbf{B}_3$ and $\mathcal{M}_1=\mathcal{M}_2$.\\
\indent\emph{Frequency marginal distribution based energy conservation:} Integrating with respect to the variables $\mathbf{u}$ on both sides of Eq.~\eqref{Frequency marginal distribution} yields
\begin{align}\label{Frequency marginal distribution based energy conservation}	&\int_{\mathbb{R}^N}\mathrm{e}^{-\pi\mathrm{i}\left(\mathbf{u}\mathbf{D}\mathbf{B}^{-1}\mathbf{u}^{\mathrm{T}}+\mathbf{u}\mathbf{F}_{44}\mathbf{u}^{\mathrm{T}}\right)}\int_{\mathbb{R}^N}\mathrm{C}(\mathcal{M},\mathcal{M}_1,\mathcal{M}_2,\mathcal{M}_3,\mathcal{M}_4,\mathcal{M}_5,\mathcal{M}_6)f(\mathbf{x},\mathbf{u})\mathrm{e}^{\pi\mathrm{i}\left(\mathbf{x},\mathbf{u}\right)\mathbf{B}_6^{-1}\mathbf{A}_6\left(\mathbf{x},\mathbf{u}\right)^{\mathrm{T}}}\mathrm{d}\mathbf{x}\mathrm{d}\mathbf{u}\nonumber\\ =&\frac{\left|\mathrm{det}(\mathbf{A}_3)\right|\left|\mathrm{det}(\mathbf{D}_3)\right|}{\sqrt{-\mathrm{det}(\mathbf{B})}}\int_{\mathbb{R}^N}\left|f(\mathbf{x})\right|^2\mathrm{d}\mathbf{x},
\end{align}
for $\begin{pmatrix}\mathbf{0}_N&\mathbf{B}{\left(\mathbf{A}_3^{-1}\right)}^{\mathrm{T}}\\-\left(\mathbf{B}^{-1}\right)^{\mathrm{T}}\mathbf{A}_3&\mathbf{I}_N\end{pmatrix}\mathcal{M}_1=\begin{pmatrix}\mathbf{0}_N&\mathbf{B}{\left(\mathbf{B}_3^{-1}\right)}^{\mathrm{T}}\\-\mathbf{B}_3^{\mathrm{T}}\mathbf{B}^{-1}&\mathbf{B}_3^{-1}\mathbf{A}_3\end{pmatrix}\mathcal{M}_2$.\\
\indent\emph{Time delay or frequency shift marginal distribution based energy conservation:} The CMCD of $\mathbf{x}=\mathbf{u}=\mathbf{0}$ reads
\begin{align}\label{Time delay or frequency shift marginal distribution based energy conservation} \mathrm{C}(\mathcal{M},\mathcal{M}_1,\mathcal{M}_2,\mathcal{M}_3,\mathcal{M}_4,\mathcal{M}_5,\mathcal{M}_6)f(\mathbf{0},\mathbf{0})=\frac{1}{\left|\mathrm{det}(\mathbf{C}_3)\right|}\int_{\mathbb{R}^N}\left|f(\mathbf{x})\right|^2\mathrm{d}\mathbf{x}
\end{align}
for $\mathbf{A}=\mathbf{0}_N$, $\mathbf{C}_3=\mathbf{D}_3$, $\mathcal{M}_1=\mathcal{M}_2$, and $\phi\left(\mathbf{z},\mathbf{v}\right)=1$.\\
\indent\emph{Property~3 (Unique reconstruction):} When $\mathbf{A}_3=\mathbf{0}_N$, the function $f(\mathbf{y})$ can be recovered by its CMCD, i.e.,
\begin{align}\label{Reconstruction formula}
f(\mathbf{y})=&\frac{1}{\overline{\mu(\mathcal{M}_2)f(\mathbf{0})}}\frac{\left|\mathrm{det}(\mathbf{B}_3)\right|}{\sqrt{-\mathrm{det}(\mathbf{B})}}\iiiint_{\mathbb{R}^{N\times N\times N\times N}}\mathrm{e}^{-\pi\mathrm{i}\mathbf{u}\mathbf{D}\mathbf{B}^{-1}\mathbf{u}^{\mathrm{T}}}\mathrm{e}^{-\pi\mathrm{i}\left(\mathbf{x},\mathbf{u}\right)\mathbf{B}_4^{-1}\mathbf{A}_4\left(\mathbf{x},\mathbf{u}\right)^{\mathrm{T}}}\mathrm{e}^{2\pi\mathrm{i}\left(\mathbf{p}\mathbf{x}^{\mathrm{T}}+\mathbf{q}\mathbf{u}^{\mathrm{T}}\right)}\nonumber\\
&\times\frac{\mathcal{F}\left[\mathrm{e}^{\pi\mathrm{i}\left(\mathbf{x},\mathbf{u}\right)\mathbf{B}_6^{-1}\mathbf{A}_6\left(\mathbf{x},\mathbf{u}\right)^{\mathrm{T}}}\mathrm{C}(\mathcal{M},\mathcal{M}_1,\mathcal{M}_2,\mathcal{M}_3,\mathcal{M}_4,\mathcal{M}_5,\mathcal{M}_6)f(\mathbf{x},\mathbf{u})\right]\left(\mathbf{p},\mathbf{q}\right)}{\mathcal{F}\left[\mathrm{e}^{\pi\mathrm{i}\left(\mathbf{x},\mathbf{u}\right)\mathbf{B}_5^{-1}\mathbf{A}_5\left(\mathbf{x},\mathbf{u}\right)^{\mathrm{T}}}\Pi(\mathbf{x,u})\right]\left(\mathbf{p},\mathbf{q}\right)}\nonumber\\
&\times\mathcal{K}_{\mathcal{M}_1^{-1}}\left(\mathbf{x}\mathbf{B}_3,\mathbf{y}\right)\mathrm{d}\mathbf{x}\mathrm{d}\mathbf{p}\mathrm{d}\mathbf{q}\mathrm{d}\mathbf{u}.
\end{align}
\indent\emph{Property~4 (Moyal formula):} Moyal formula is an extension of the energy conservation. It involves the integration of the product of CMCDs of two functions $f,g$ over the variables $\mathbf{x}$ and $\mathbf{u}$. When $\left|\mu(\mathcal{M}_{5})\Pi(\mathbf{w})\right|^2=1$ for all $\mathbf{w}\in\mathbb{R}^{2N}$, there is
\begin{align}\label{Moyal formula}
&\iint_{\mathbb{R}^{N\times N}}\mathrm{C}(\mathcal{M},\mathcal{M}_1,\mathcal{M}_2,\mathcal{M}_3,\mathcal{M}_4,\mathcal{M}_5,\mathcal{M}_6)f(\mathbf{x},\mathbf{u})\nonumber\\ &\times\overline{\mathrm{C}(\mathcal{M},\mathcal{M}_1,\mathcal{M}_2,\mathcal{M}_3,\mathcal{M}_4,\mathcal{M}_5,\mathcal{M}_6)g(\mathbf{x},\mathbf{u})}\mathrm{d}\mathbf{x}\mathrm{d}\mathbf{u}\nonumber\\ =&\frac{\left|\mathrm{det}(\mathbf{B}_5)\right|}{\left|\mathrm{det}(\mathbf{B}_3)\right|\left|\mathrm{det}(\mathbf{C}_3)\right|}\left|\int_{\mathbb{R}^N}f(\mathbf{x})\overline{g(\mathbf{x})}\mathrm{d}\mathbf{x}\right|^2.
\end{align}
\indent\emph{Property~5 (Complex conjugate symmetry):} Complex conjugate symmetry reveals the relationship between the CMCD of a complex conjugate function $\overline{f}$ and that of the original function $f$. When the kernel function is a real-valued function, i.e., $\overline{\Pi}=\Pi$, there is
\begin{align}\label{Complex conjugate symmetry} \mathrm{C}(\mathcal{M},\mathcal{M}_1,\mathcal{M}_2,\mathcal{M}_3,\mathcal{M}_4,\mathcal{M}_5,\mathcal{M}_6)\overline{f}(\mathbf{x},\mathbf{u})=\overline{\mathrm{C}(\widehat{\mathcal{M}},\widehat{\mathcal{M}}_1,\widehat{\mathcal{M}}_2,\mathcal{M}_3,\widehat{\mathcal{M}_4},\widehat{\mathcal{M}}_5,\widehat{\mathcal{M}}_6)f(\mathbf{x},\mathbf{u})},
\end{align}
where $\widehat{\mathcal{M}}=\begin{pmatrix}\mathbf{A}&-\mathbf{B}\\-\mathbf{C}&\mathbf{D}\end{pmatrix}$ and $\widehat{\mathcal{M}}_j=\begin{pmatrix}\mathbf{A}_j&-\mathbf{B}_j\\-\mathbf{C}_j&\mathbf{D}_j\end{pmatrix}$, $j=1,2,4,5,6$.\\
\indent\emph{Property~6 (Time reversal symmetry):} Let $\underline{f}(\mathbf{x}):=f(-\mathbf{x})$. Time reversal symmetry reveals the relationship between the CMCD of a time reversal function $\underline{f}$ and that of the original function $f$, that is,
\begin{align}\label{Time reversal symmetry} \mathrm{C}(\mathcal{M},\mathcal{M}_1,\mathcal{M}_2,\mathcal{M}_3,\mathcal{M}_4,\mathcal{M}_5,\mathcal{M}_6)\underline{f}(\mathbf{x},\mathbf{u})=\mathrm{C}(\mathcal{M},-\mathcal{M}_1,-\mathcal{M}_2,\mathcal{M}_3,\mathcal{M}_4,\mathcal{M}_5,\mathcal{M}_6)f(\mathbf{x},\mathbf{u}).
\end{align}
\indent\emph{Property~7 (Scaling property):} Let $\mathbf{S}_{\sigma}=\begin{pmatrix}\frac{\mathbf{I}_N}{\sigma}&\mathbf{0}_N\\\mathbf{0}_N&\sigma\mathbf{I}_N\end{pmatrix}$, and let $\mathcal{S}_\sigma$ denote the scaling operator, i.e., $\mathcal{S}_\sigma f(\mathbf{x}):=f\left(\sigma\mathbf{x}\right)$. The scaling property reveals the relationship between the CMCD of a scaling function $\mathcal{S}_\sigma f$ and that of the original function $f$, that is,
\begin{align} \label{Scaling property} &\mathrm{C}(\mathcal{M},\mathcal{M}_1,\mathcal{M}_2,\mathcal{M}_3,\mathcal{M}_4,\mathcal{M}_5,\mathcal{M}_6)\mathcal{S}_\sigma f(\mathbf{x},\mathbf{u})\nonumber\\ =&\frac{1}{\sigma^{2N}}\mathrm{C}(\mathcal{M},\mathcal{M}_{1}\mathbf{S}_{\sigma},\mathcal{M}_{2}\mathbf{S}_{\sigma},\mathcal{M}_3,\mathcal{M}_4,\mathcal{M}_5,\mathcal{M}_6)f(\mathbf{x},\mathbf{u}).
\end{align}
\indent\emph{Property~8 (Time translation property):} Let $T_{\boldsymbol{\tau}}$ be the translation operator, i.e., $T_{\boldsymbol{\tau}}f(\mathbf{x}):=f(\mathbf{x}-\boldsymbol{\tau})$. The time translation property reveals the relationship between the CMCD of a time translation function $T_{\boldsymbol{\tau}}f$ and that of the original function $f$. When $\mathcal{M}_3^{\mathrm{T}}$ is a symplectic matrix, i.e., $\mathcal{M}_3^{\mathrm{T}}\in Sp(N,\mathbb{R})$, $\mathbf{A}_1=\mathbf{A}_2$, and $\mathbf{B}_1^{\mathrm{T}}\mathbf{C}_1\mathbf{B}_1^{-1}\mathbf{D}_3-\mathbf{B}_2^{\mathrm{T}}\mathbf{C}_2\mathbf{B}_2^{-1}\mathbf{B}_3=\mathbf{0}_N$, there is
\begin{align}\label{Time translation property} &\mathrm{C}(\mathcal{M},\mathcal{M}_1,\mathcal{M}_2,\mathcal{M}_3,\mathcal{M}_4,\mathcal{M}_5,\mathcal{M}_6)T_{\boldsymbol{\tau}}f(\mathbf{x},\mathbf{u})\nonumber\\
=&\bigg[\mathrm{e}^{\pi\mathrm{i}\boldsymbol{\tau}\left(\mathbf{B}_1^{-1}-\mathbf{B}_2^{-1}\right)\mathbf{A}_1\boldsymbol{\tau}^{\mathrm{T}}}\mathrm{e}^{-\pi\mathrm{i}\boldsymbol{\tau}\mathbf{A}_1^{\mathrm{T}}\left(\mathbf{D}_1\mathbf{B}_1^{-1}-\mathbf{D}_2\mathbf{B}_2^{-1}\right)\mathbf{A}_1\boldsymbol{\tau}^{\mathrm{T}}}\mathrm{e}^{-\pi\mathrm{i}\boldsymbol{\tau}\mathbf{A}_1^{\mathrm{T}}\left(\mathbf{A}_3-\mathbf{B}_3\right)^{\mathrm{T}}\mathbf{B}^{\mathrm{T}}\mathbf{C}\mathbf{B}^{-1}\mathbf{A}\left(\mathbf{A}_3-\mathbf{B}_3\right)\mathbf{A}_1\boldsymbol{\tau}^{\mathrm{T}}}\nonumber\\
&\times\mathrm{e}^{2\pi\mathrm{i}\boldsymbol{\tau}\mathbf{B}_1^{\mathrm{T}}\mathbf{C}_1\mathbf{B}_1^{-1}\mathbf{B}_3^{\mathrm{T}}\mathbf{x}^{\mathrm{T}}}\mathrm{e}^{-2\pi\mathrm{i}\boldsymbol{\tau}\mathbf{B}_2^{\mathrm{T}}\mathbf{C}_2\mathbf{B}_2^{-1}\mathbf{A}_3^{\mathrm{T}}\mathbf{x}^{\mathrm{T}}}\mathrm{e}^{2\pi\mathrm{i}\boldsymbol{\tau}\mathbf{A}_1^{\mathrm{T}}\left(\mathbf{A}_3-\mathbf{B}_3\right)^{\mathrm{T}}\mathbf{B}^{\mathrm{T}}\mathbf{C}\mathbf{B}^{-1}\mathbf{u}^{\mathrm{T}}}\mathrm{W}(\mathcal{M},\mathcal{M}_1,\mathcal{M}_2,\mathcal{M}_3)f\nonumber\\
&\left(\mathbf{x}-\boldsymbol{\tau}\mathbf{A}_1^{\mathrm{T}}\left(\mathbf{D}_3-\mathbf{C}_3\right)^{\mathrm{T}},\mathbf{u}-\boldsymbol{\tau}\mathbf{A}_1^{\mathrm{T}}\left(\mathbf{A}_3-\mathbf{B}_3\right)^{\mathrm{T}}\mathbf{A}^{\mathrm{T}}\right)\bigg]\Theta_{\mathcal{M}_4,\mathcal{M}_5,\mathcal{M}_6}\Pi(\mathbf{x},\mathbf{u}).
\end{align}
\indent\emph{Property~9 (Frequency modulation property):} Let $M_{\mathbf{w}}$ be the modulation operator, i.e., $M_{\mathbf{w}}f(\mathbf{x}):=f(\mathbf{x})\mathrm{e}^{2\pi\mathrm{i}\mathbf{x}\mathbf{w}^{\mathrm{T}}}$. The frequency modulation property reveals the relationship between the CMCD of a frequency modulation function $M_{\mathbf{w}}f$ and that of the original function $f$. When $\mathcal{M}_3^{\mathrm{T}}$ is a symplectic matrix, i.e., $\mathcal{M}_3^{\mathrm{T}}\in Sp(N,\mathbb{R})$, and $\mathbf{B}_1=\mathbf{B}_2$, there is
\begin{align}\label{Frequency modulation property} &\mathrm{C}(\mathcal{M},\mathcal{M}_1,\mathcal{M}_2,\mathcal{M}_3,\mathcal{M}_4,\mathcal{M}_5,\mathcal{M}_6)M_{\mathbf{w}}f(\mathbf{x},\mathbf{u})\nonumber\\
=&\bigg[\mathrm{e}^{-\pi\mathrm{i}\mathbf{w}\left(\mathbf{B}_1^{\mathrm{T}}\mathbf{D}_1-\mathbf{B}_2^{\mathrm{T}}\mathbf{D}_2\right)\mathbf{w}^{\mathrm{T}}}\mathrm{e}^{-\pi\mathrm{i}\mathbf{w}\left(\mathbf{C}_3\mathbf{D}_2-\mathbf{D}_3\mathbf{D}_1\right)^{\mathrm{T}}\mathbf{B}^{\mathrm{T}}\mathbf{D}\left(\mathbf{C}_3\mathbf{D}_2-\mathbf{D}_3\mathbf{D}_1\right)\mathbf{w}^{\mathrm{T}}}\nonumber\\
&\times\mathrm{e}^{-\pi\mathrm{i}\mathbf{w}\mathbf{B}_1^{\mathrm{T}}\left(\mathbf{A}_3-\mathbf{B}_3\right)^{\mathrm{T}}\mathbf{B}^{\mathrm{T}}\mathbf{C}\mathbf{B}^{-1}\mathbf{A}\left(\mathbf{A}_3-\mathbf{B}_3\right)\mathbf{B}_1\mathbf{w}^{\mathrm{T}}}\mathrm{e}^{2\pi\mathrm{i}\mathbf{w}\mathbf{B}_1^{\mathrm{T}}\left(\mathbf{A}_3-\mathbf{B}_3\right)^{\mathrm{T}}\mathbf{B}^{\mathrm{T}}\mathbf{C}\left(\mathbf{C}_3\mathbf{D}_2-\mathbf{D}_3\mathbf{D}_1\right)\mathbf{w}^{\mathrm{T}}}\mathrm{e}^{2\pi\mathrm{i}\mathbf{x}\left(\mathbf{B}_3\mathbf{D}_1-\mathbf{A}_3\mathbf{D}_2\right)\mathbf{w}^{\mathrm{T}}}\nonumber\\
&\times\mathrm{e}^{-2\pi\mathrm{i}\mathbf{u}\mathbf{D}\left(\mathbf{C}_3\mathbf{D}_2-\mathbf{D}_3\mathbf{D}_1\right)\mathbf{w}^{\mathrm{T}}}\mathrm{e}^{2\pi\mathrm{i}\mathbf{w}\mathbf{B}_1^{\mathrm{T}}\left(\mathbf{A}_3-\mathbf{B}_3\right)^{\mathrm{T}}\mathbf{B}^{\mathrm{T}}\mathbf{C}\mathbf{B}^{-1}\mathbf{u}^{\mathrm{T}}}\mathrm{W}(\mathcal{M},\mathcal{M}_1,\mathcal{M}_2,\mathcal{M}_3)f\nonumber\\ &\left(\mathbf{x}-\mathbf{w}\mathbf{B}_1^\mathrm{T}(\mathbf{D}_3-\mathbf{C}_3)^{\mathrm{T}},\mathbf{u}+\mathbf{w}\mathbf{D}_2^{\mathrm{T}}\mathbf{C}_3^{\mathrm{T}}\mathbf{B}^{\mathrm{T}}-\mathbf{w}\mathbf{D}_1^{\mathrm{T}}\mathbf{D}_3^{\mathrm{T}}\mathbf{B}^{\mathrm{T}}-\mathbf{w}\mathbf{B}_1^{\mathrm{T}}(\mathbf{A}_3-\mathbf{B}_3)^{\mathrm{T}}\mathbf{A}^{\mathrm{T}}\right)\bigg]\nonumber\\
&\Theta_{\mathcal{M}_4,\mathcal{M}_5,\mathcal{M}_6}\Pi(\mathbf{x},\mathbf{u}).
\end{align}
\indent\emph{Property~10 (Metaplectic invariance):} The metaplectic invariance reveals the equivalence between the CMCD of the metaplectic transform of a function $\mu\left(\mathcal{M}_0\right)f$ and that of the original function $f$, i.e.,
\begin{align}\label{Metaplectic invariance} &\mathrm{C}(\mathcal{M},\mathcal{M}_1,\mathcal{M}_2,\mathcal{M}_3,\mathcal{M}_4,\mathcal{M}_5,\mathcal{M}_6)\mu\left(\mathcal{M}_0\right)f(\mathbf{x},\mathbf{u})\nonumber\\ =&\mathrm{C}(\mathcal{M},\mathcal{M}_1\mathcal{M}_0,\mathcal{M}_2\mathcal{M}_0,\mathcal{M}_3,\mathcal{M}_4,\mathcal{M}_5,\mathcal{M}_6)f(\mathbf{x},\mathbf{u}).
\end{align}
\indent The detailed derivations of the above properties can refer to Appendix~\ref{sec:AppA}. See Table~\ref{tab:8} for a summary of these properties and how they get reflected as constraints on the symplectic matrices and the kernel function.
\begin{table}[htbp]
	\centering
	\caption{\label{tab:8}Some useful and important properties of the CMCD}
	\tiny
	\begin{threeparttable}
		\begin{tabular}{cc}
			\specialrule{0.1em}{4pt}{4pt}
			Property & Expression \\
			\specialrule{0.1em}{4pt}{4pt}
			\multirow{2}{*}{Time marginal distribution\tnote{1}} & $\int_{\mathbb{R}^N}\mathrm{C}(\mathcal{M},\mathcal{M}_1,\mathcal{M}_2,\mathcal{M}_3,\mathcal{M}_4,\mathcal{M}_5,\mathcal{M}_6)f(\mathbf{x},\mathbf{u})\mathrm{e}^{\pi\mathrm{i}\left(\mathbf{x},\mathbf{u}\right)\mathbf{B}_6^{-1}\mathbf{A}_6\left(\mathbf{x},\mathbf{u}\right)^{\mathrm{T}}}\mathrm{d}\mathbf{u}$\\
& $=\sqrt{-\mathrm{det}(\mathbf{B})}\mathrm{e}^{\pi\mathrm{i}\mathbf{x}\mathbf{F}_{41}\mathbf{x}^{\mathrm{T}}}\mu(\mathcal{M}_1)f\left(\mathbf{x}\mathbf{B}_3\right)\overline{\mu(\mathcal{M}_2)f\left(\mathbf{x}\mathbf{A}_3\right)}$\\
			\specialrule{0em}{4pt}{4pt}
			\multirow{4}{*}{Frequency marginal distribution\tnote{2}} & $\int_{\mathbb{R}^N}\mathrm{C}(\mathcal{M},\mathcal{M}_1,\mathcal{M}_2,\mathcal{M}_3,\mathcal{M}_4,\mathcal{M}_5,\mathcal{M}_6)f(\mathbf{x},\mathbf{u})\mathrm{e}^{\pi\mathrm{i}\left(\mathbf{x},\mathbf{u}\right)\mathbf{B}_6^{-1}\mathbf{A}_6\left(\mathbf{x},\mathbf{u}\right)^{\mathrm{T}}}\mathrm{d}\mathbf{x}$\\ & $=\frac{\left|\mathrm{det}(\mathbf{A}_3)\right|\left|\mathrm{det}(\mathbf{D}_3)\right|}{\sqrt{-\mathrm{det}(\mathbf{B})}}\mathrm{e}^{\pi\mathrm{i}\left(\mathbf{u}\mathbf{D}\mathbf{B}^{-1}\mathbf{u}^{\mathrm{T}}+\mathbf{u}\mathbf{F}_{44}\mathbf{u}^{\mathrm{T}}\right)}$\\
&
$\times\mu\left(\begin{pmatrix}\mathbf{0}_N&\mathbf{B}{\left(\mathbf{A}_3^{-1}\right)}^{\mathrm{T}}\\-\left(\mathbf{B}^{-1}\right)^{\mathrm{T}}\mathbf{A}_3&\mathbf{I}_N\end{pmatrix}\mathcal{M}_1\right)f(\mathbf{u})\overline{\mu\left(\begin{pmatrix}\mathbf{0}_N&\mathbf{B}{\left(\mathbf{B}_3^{-1}\right)}^{\mathrm{T}}\\-\mathbf{B}_3^{\mathrm{T}}\mathbf{B}^{-1}&\mathbf{B}_3^{-1}\mathbf{A}_3\end{pmatrix}\mathcal{M}_2\right)f(\mathbf{u})}$\\
			\specialrule{0em}{4pt}{4pt}
			\multirow{3}{*}{Time delay marginal distribution} & $\mathrm{C}(\mathcal{M},\mathcal{M}_1,\mathcal{M}_2,\mathcal{M}_3,\mathcal{M}_4,\mathcal{M}_5,\mathcal{M}_6)f(\mathbf{0},\mathbf{u})$\\
&
$=\mathrm{e}^{-\pi\mathrm{i}\left(\mathbf{0},\mathbf{u}\right)\mathbf{B}_6^{-1}\mathbf{A}_6\left(\mathbf{0},\mathbf{u}\right)^{\mathrm{T}}}\iiint_{\mathbb{R}^{N\times N\times N}}\mu(\mathcal{M}_1)f\left(\mathbf{p}\mathbf{B}_3+\mathbf{t}\mathbf{D}_3\right)\overline{\mu(\mathcal{M}_2)f\left(\mathbf{p}\mathbf{A}_3+\mathbf{t}\mathbf{C}_3\right)}\mathrm{e}^{\pi\mathrm{i}\left(\mathbf{p},\mathbf{q}\right)\mathbf{B}_4^{-1}\mathbf{A}_4\left(\mathbf{p},\mathbf{q}\right)^{\mathrm{T}}}$\\
& $\times\mathrm{e}^{\pi\mathrm{i}\left(-\mathbf{p},\mathbf{u}-\mathbf{q}\right)\mathbf{B}_5^{-1}\mathbf{A}_5\left(-\mathbf{p},\mathbf{u}-\mathbf{q}\right)^{\mathrm{T}}}\mathcal{K}_{\mathcal{M}}(\mathbf{q},\mathbf{t})\iint_{\mathbb{R}^{N\times N}}\phi\left(\mathbf{v},\mathbf{z}\right)\mathrm{e}^{-2\pi\mathrm{i}\left(-\mathbf{v}\mathbf{p}^{\mathrm{T}}+\mathbf{z}(\mathbf{u}-\mathbf{q})^{\mathrm{T}}\right)}\mathrm{d}\mathbf{v}\mathrm{d}\mathbf{z}\mathrm{d}\mathbf{t}\mathrm{d}\mathbf{p}\mathrm{d}\mathbf{q}$\\
			\specialrule{0em}{4pt}{4pt}
			\multirow{3}{*}{Frequency shift marginal distribution} & $\mathrm{C}(\mathcal{M},\mathcal{M}_1,\mathcal{M}_2,\mathcal{M}_3,\mathcal{M}_4,\mathcal{M}_5,\mathcal{M}_6)f(\mathbf{x},\mathbf{0})$\\ & $=\mathrm{e}^{-\pi\mathrm{i}\left(\mathbf{x},\mathbf{0}\right)\mathbf{B}_6^{-1}\mathbf{A}_6\left(\mathbf{x},\mathbf{0}\right)^{\mathrm{T}}}\iiint_{\mathbb{R}^{N\times N\times N}}\mu(\mathcal{M}_1)f\left(\mathbf{p}\mathbf{B}_3+\mathbf{t}\mathbf{D}_3\right)\overline{\mu(\mathcal{M}_2)f\left(\mathbf{p}\mathbf{A}_3+\mathbf{t}\mathbf{C}_3\right)}\mathrm{e}^{\pi\mathrm{i}\left(\mathbf{p},\mathbf{q}\right)\mathbf{B}_4^{-1}\mathbf{A}_4\left(\mathbf{p},\mathbf{q}\right)^{\mathrm{T}}}$\\
& $\times\mathrm{e}^{\pi\mathrm{i}\left(\mathbf{x}-\mathbf{p},-\mathbf{q}\right)\mathbf{B}_5^{-1}\mathbf{A}_5\left(\mathbf{x}-\mathbf{p},-\mathbf{q}\right)^{\mathrm{T}}}\mathcal{K}_{\mathcal{M}}(\mathbf{q},\mathbf{t})\iint_{\mathbb{R}^{N\times N}}\phi\left(\mathbf{v},\mathbf{z}\right)\mathrm{e}^{-2\pi\mathrm{i}\left(\mathbf{v}(\mathbf{x}-\mathbf{p})^{\mathrm{T}}-\mathbf{z}\mathbf{q}^{\mathrm{T}}\right)}\mathrm{d}\mathbf{v}\mathrm{d}\mathbf{z}\mathrm{d}\mathbf{t}\mathrm{d}\mathbf{p}\mathrm{d}\mathbf{q}$\\
			\specialrule{0em}{4pt}{4pt}
			Time marginal distribution & $\int_{\mathbb{R}^N}\mathrm{e}^{-\pi\mathrm{i}\mathbf{x}\mathbf{F}_{41}\mathbf{x}^{\mathrm{T}}}\int_{\mathbb{R}^N}\mathrm{C}(\mathcal{M},\mathcal{M}_1,\mathcal{M}_2,\mathcal{M}_3,\mathcal{M}_4,\mathcal{M}_5,\mathcal{M}_6)f(\mathbf{x},\mathbf{u})\mathrm{e}^{\pi\mathrm{i}\left(\mathbf{x},\mathbf{u}\right)\mathbf{B}_6^{-1}\mathbf{A}_6\left(\mathbf{x},\mathbf{u}\right)^{\mathrm{T}}}\mathrm{d}\mathbf{u}\mathrm{d}\mathbf{x}$\\
			based energy conservation\tnote{3} & $=\frac{\sqrt{-\mathrm{det}(\mathbf{B})}}{\left|\mathrm{det}(\mathbf{B}_3)\right|}\int_{\mathbb{R}^N}\left|f(\mathbf{x})\right|^2\mathrm{d}\mathbf{x}$\\
			\specialrule{0em}{4pt}{4pt}
			Frequency marginal distribution & $\int_{\mathbb{R}^N}\mathrm{e}^{-\pi\mathrm{i}\left(\mathbf{u}\mathbf{D}\mathbf{B}^{-1}\mathbf{u}^{\mathrm{T}}+\mathbf{u}\mathbf{F}_{44}\mathbf{u}^{\mathrm{T}}\right)}\int_{\mathbb{R}^N}\mathrm{C}(\mathcal{M},\mathcal{M}_1,\mathcal{M}_2,\mathcal{M}_3,\mathcal{M}_4,\mathcal{M}_5,\mathcal{M}_6)f(\mathbf{x},\mathbf{u})\mathrm{e}^{\pi\mathrm{i}\left(\mathbf{x},\mathbf{u}\right)\mathbf{B}_6^{-1}\mathbf{A}_6\left(\mathbf{x},\mathbf{u}\right)^{\mathrm{T}}}\mathrm{d}\mathbf{x}\mathrm{d}\mathbf{u}$\\
			based energy conservation\tnote{4} & $=\frac{\left|\mathrm{det}(\mathbf{A}_3)\right|\left|\mathrm{det}(\mathbf{D}_3)\right|}{\sqrt{-\mathrm{det}(\mathbf{B})}}\int_{\mathbb{R}^N}\left|f(\mathbf{x})\right|^2\mathrm{d}\mathbf{x}$\\
			\specialrule{0em}{4pt}{4pt}
			Time delay or frequency shift & \multirow{3}{*}{$\mathrm{C}(\mathcal{M},\mathcal{M}_1,\mathcal{M}_2,\mathcal{M}_3,\mathcal{M}_4,\mathcal{M}_5,\mathcal{M}_6)f(\mathbf{0},\mathbf{0})=\frac{1}{\left|\mathrm{det}(\mathbf{C}_3)\right|}\int_{\mathbb{R}^N}\left|f(\mathbf{x})\right|^2\mathrm{d}\mathbf{x}$}\\
			marginal distribution &\\
			based energy conservation\tnote{5}&\\
			\specialrule{0em}{4pt}{4pt}
			\multirow{4}{*}{Unique reconstruction\tnote{6}} & $f(\mathbf{y})=\frac{1}{\overline{\mu(\mathcal{M}_2)f(\mathbf{0})}}\frac{\left|\mathrm{det}(\mathbf{B}_3)\right|}{\sqrt{-\mathrm{det}(\mathbf{B})}}\iiiint_{\mathbb{R}^{N\times N\times N\times N}}\mathrm{e}^{-\pi\mathrm{i}\mathbf{u}\mathbf{D}\mathbf{B}^{-1}\mathbf{u}^{\mathrm{T}}}\mathrm{e}^{-\pi\mathrm{i}\left(\mathbf{x},\mathbf{u}\right)\mathbf{B}_4^{-1}\mathbf{A}_4\left(\mathbf{x},\mathbf{u}\right)^{\mathrm{T}}}\mathrm{e}^{2\pi\mathrm{i}\left(\mathbf{p}\mathbf{x}^{\mathrm{T}}+\mathbf{q}\mathbf{u}^{\mathrm{T}}\right)}$\\
& $\times\mathcal{K}_{\mathcal{M}_1^{-1}}\left(\mathbf{x}\mathbf{B}_3,\mathbf{y}\right)\frac{\mathcal{F}\left[\mathrm{e}^{\pi\mathrm{i}\left(\mathbf{x},\mathbf{u}\right)\mathbf{B}_6^{-1}\mathbf{A}_6\left(\mathbf{x},\mathbf{u}\right)^{\mathrm{T}}}\mathrm{C}(\mathcal{M},\mathcal{M}_1,\mathcal{M}_2,\mathcal{M}_3,\mathcal{M}_4,\mathcal{M}_5,\mathcal{M}_6)f(\mathbf{x},\mathbf{u})\right]\left(\mathbf{p},\mathbf{q}\right)}{\mathcal{F}\left[\mathrm{e}^{\pi\mathrm{i}\left(\mathbf{x},\mathbf{u}\right)\mathbf{B}_5^{-1}\mathbf{A}_5\left(\mathbf{x},\mathbf{u}\right)^{\mathrm{T}}}\Pi(\mathbf{x,u})\right]\left(\mathbf{p},\mathbf{q}\right)}\mathrm{d}\mathbf{x}\mathrm{d}\mathbf{p}\mathrm{d}\mathbf{q}\mathrm{d}\mathbf{u}$\\
			\specialrule{0em}{4pt}{4pt}
			\multirow{2}{*}{Moyal formula\tnote{7}}&$\iint_{\mathbb{R}^{N\times N}}\mathrm{C}(\mathcal{M},\mathcal{M}_1,\mathcal{M}_2,\mathcal{M}_3,\mathcal{M}_4,\mathcal{M}_5,\mathcal{M}_6)f(\mathbf{x},\mathbf{u})\overline{\mathrm{C}(\mathcal{M},\mathcal{M}_1,\mathcal{M}_2,\mathcal{M}_3,\mathcal{M}_4,\mathcal{M}_5,\mathcal{M}_6)g(\mathbf{x},\mathbf{u})}\mathrm{d}\mathbf{x}\mathrm{d}\mathbf{u}$\\ & $=\frac{\left|\mathrm{det}(\mathbf{B}_5)\right|}{\left|\mathrm{det}(\mathbf{B}_3)\right|\left|\mathrm{det}(\mathbf{C}_3)\right|}\left|\int_{\mathbb{R}^N}f(\mathbf{x})\overline{g(\mathbf{x})}\mathrm{d}\mathbf{x}\right|^2$\\
			\specialrule{0em}{4pt}{4pt}
			Complex conjugate symmetry\tnote{8} & $\mathrm{C}(\mathcal{M},\mathcal{M}_1,\mathcal{M}_2,\mathcal{M}_3,\mathcal{M}_4,\mathcal{M}_5,\mathcal{M}_6)\overline{f}(\mathbf{x},\mathbf{u})=\overline{\mathrm{C}(\widehat{\mathcal{M}},\widehat{\mathcal{M}}_1,\widehat{\mathcal{M}}_2,\mathcal{M}_3,\widehat{\mathcal{M}}_4,\widehat{\mathcal{M}}_5,\widehat{\mathcal{M}}_6)f(\mathbf{x},\mathbf{u})}$\\
			\specialrule{0em}{4pt}{4pt}
			Time reversal symmetry & $\mathrm{C}(\mathcal{M},\mathcal{M}_1,\mathcal{M}_2,\mathcal{M}_3,\mathcal{M}_4,\mathcal{M}_5,\mathcal{M}_6)\underline{f}(\mathbf{x},\mathbf{u})=\mathrm{C}(\mathcal{M},-\mathcal{M}_1,-\mathcal{M}_2,\mathcal{M}_3,\mathcal{M}_4,\mathcal{M}_5,\mathcal{M}_6)f(\mathbf{x},\mathbf{u})$\\
			\specialrule{0em}{4pt}{4pt}
			Scaling property & $\mathrm{C}(\mathcal{M},\mathcal{M}_1,\mathcal{M}_2,\mathcal{M}_3,\mathcal{M}_4,\mathcal{M}_5,\mathcal{M}_6)\mathcal{S}_\sigma f(\mathbf{x},\mathbf{u})=\frac{1}{\sigma^{2N}}\mathrm{C}(\mathcal{M},\mathcal{M}_1\mathbf{S}_{\sigma},\mathcal{M}_2\mathbf{S}_{\sigma},\mathcal{M}_3,\mathcal{M}_4,\mathcal{M}_5,\mathcal{M}_6)f(\mathbf{x},\mathbf{u})$\\
			\specialrule{0em}{4pt}{4pt}
			\multirow{4}{*}{Time translation property\tnote{9}} & $\mathrm{C}(\mathcal{M},\mathcal{M}_1,\mathcal{M}_2,\mathcal{M}_3,\mathcal{M}_4,\mathcal{M}_5,\mathcal{M}_6)T_{\boldsymbol{\tau}}f(\mathbf{x},\mathbf{u})=\bigg[\mathrm{e}^{\pi\mathrm{i}\boldsymbol{\tau}\left(\mathbf{B}_1^{-1}-\mathbf{B}_2^{-1}\right)\mathbf{A}_1\boldsymbol{\tau}^{\mathrm{T}}}\mathrm{e}^{-\pi\mathrm{i}\boldsymbol{\tau}\mathbf{A}_1^{\mathrm{T}}\left(\mathbf{D}_1\mathbf{B}_1^{-1}-\mathbf{D}_2\mathbf{B}_2^{-1}\right)\mathbf{A}_1\boldsymbol{\tau}^{\mathrm{T}}}$\\ & $\times\mathrm{e}^{-\pi\mathrm{i}\boldsymbol{\tau}\mathbf{A}_1^{\mathrm{T}}\left(\mathbf{A}_3-\mathbf{B}_3\right)^{\mathrm{T}}\mathbf{B}^{\mathrm{T}}\mathbf{C}\mathbf{B}^{-1}\mathbf{A}\left(\mathbf{A}_3-\mathbf{B}_3\right)\mathbf{A}_1\boldsymbol{\tau}^{\mathrm{T}}}\mathrm{e}^{2\pi\mathrm{i}\boldsymbol{\tau}\mathbf{B}_1^{\mathrm{T}}\mathbf{C}_1\mathbf{B}_1^{-1}\mathbf{B}_3^{\mathrm{T}}\mathbf{x}^{\mathrm{T}}}\mathrm{e}^{-2\pi\mathrm{i}\boldsymbol{\tau}\mathbf{B}_2^{\mathrm{T}}\mathbf{C}_2\mathbf{B}_2^{-1}\mathbf{A}_3^{\mathrm{T}}\mathbf{x}^{\mathrm{T}}}$\\ &
$\times\mathrm{e}^{2\pi\mathrm{i}\boldsymbol{\tau}\mathbf{A}_1^{\mathrm{T}}\left(\mathbf{A}_3-\mathbf{B}_3\right)^{\mathrm{T}}\mathbf{B}^{\mathrm{T}}\mathbf{C}\mathbf{B}^{-1}\mathbf{u}^{\mathrm{T}}}\mathrm{W}(\mathcal{M},\mathcal{M}_1,\mathcal{M}_2,\mathcal{M}_3)f\left(\mathbf{x}-\boldsymbol{\tau}\mathbf{A}_1^{\mathrm{T}}\left(\mathbf{D}_3-\mathbf{C}_3\right)^{\mathrm{T}},\mathbf{u}-\boldsymbol{\tau}\mathbf{A}_1^{\mathrm{T}}\left(\mathbf{A}_3-\mathbf{B}_3\right)^{\mathrm{T}}\mathbf{A}^{\mathrm{T}}\right)\bigg]$\\ & $\Theta_{\mathcal{M}_4,\mathcal{M}_5,\mathcal{M}_6}\Pi(\mathbf{x},\mathbf{u})$ \\
			\specialrule{0.1em}{4pt}{4pt}
		\end{tabular}
	\end{threeparttable}
\end{table}

\setcounter{table}{7} 

\begin{table}[htbp]
	\centering
	\tiny
	\caption{Some useful and important properties of the CMCD (Continued)}
	\begin{threeparttable}
		\begin{tabular}{cc}
			\specialrule{0.1em}{4pt}{4pt}
			Property & Expression \\
			\specialrule{0.1em}{4pt}{4pt}
			\multirow{7}{*}{Frequency modulation property\tnote{10}} & $\mathrm{C}(\mathcal{M},\mathcal{M}_1,\mathcal{M}_2,\mathcal{M}_3,\mathcal{M}_4,\mathcal{M}_5,\mathcal{M}_6)M_{\mathbf{w}}f(\mathbf{x},\mathbf{u})=\bigg[\mathrm{e}^{-\pi\mathrm{i}\mathbf{w}\left(\mathbf{B}_1^{\mathrm{T}}\mathbf{D}_1-\mathbf{B}_2^{\mathrm{T}}\mathbf{D}_2\right)\mathbf{w}^{\mathrm{T}}}$ \\ & $\times\mathrm{e}^{-\pi\mathrm{i}\mathbf{w}\left(\mathbf{C}_3\mathbf{D}_2-\mathbf{D}_3\mathbf{D}_1\right)^{\mathrm{T}}\mathbf{B}^{\mathrm{T}}\mathbf{D}\left(\mathbf{C}_3\mathbf{D}_2-\mathbf{D}_3\mathbf{D}_1\right)\mathbf{w}^{\mathrm{T}}}\mathrm{e}^{-\pi\mathrm{i}\mathbf{w}\mathbf{B}_1^{\mathrm{T}}\left(\mathbf{A}_3-\mathbf{B}_3\right)^{\mathrm{T}}\mathbf{B}^{\mathrm{T}}\mathbf{C}\mathbf{B}^{-1}\mathbf{A}\left(\mathbf{A}_3-\mathbf{B}_3\right)\mathbf{B}_1\mathbf{w}^{\mathrm{T}}}$ \\ & $\times\mathrm{e}^{2\pi\mathrm{i}\mathbf{w}\mathbf{B}_1^{\mathrm{T}}\left(\mathbf{A}_3-\mathbf{B}_3\right)^{\mathrm{T}}\mathbf{B}^{\mathrm{T}}\mathbf{C}\left(\mathbf{C}_3\mathbf{D}_2-\mathbf{D}_3\mathbf{D}_1\right)\mathbf{w}^{\mathrm{T}}}\mathrm{e}^{2\pi\mathrm{i}\mathbf{x}\left(\mathbf{B}_3\mathbf{D}_1-\mathbf{A}_3\mathbf{D}_2\right)\mathbf{w}^{\mathrm{T}}}$ \\ & $\times\mathrm{e}^{-2\pi\mathrm{i}\mathbf{u}\mathbf{D}\left(\mathbf{C}_3\mathbf{D}_2-\mathbf{D}_3\mathbf{D}_1\right)\mathbf{w}^{\mathrm{T}}}\mathrm{e}^{2\pi\mathrm{i}\mathbf{w}\mathbf{B}_1^{\mathrm{T}}\left(\mathbf{A}_3-\mathbf{B}_3\right)^{\mathrm{T}}\mathbf{B}^{\mathrm{T}}\mathbf{C}\mathbf{B}^{-1}\mathbf{u}^{\mathrm{T}}}$ \\ & $\times\mathrm{W}(\mathcal{M},\mathcal{M}_1,\mathcal{M}_2,\mathcal{M}_3)f\left(\mathbf{x}-\mathbf{w}\mathbf{B}_1^\mathrm{T}(\mathbf{D}_3-\mathbf{C}_3)^{\mathrm{T}},\mathbf{u}+\mathbf{w}\mathbf{D}_2^{\mathrm{T}}\mathbf{C}_3^{\mathrm{T}}\mathbf{B}^{\mathrm{T}}-\mathbf{w}\mathbf{D}_1^{\mathrm{T}}\mathbf{D}_3^{\mathrm{T}}\mathbf{B}^{\mathrm{T}}-\mathbf{w}\mathbf{B}_1^{\mathrm{T}}(\mathbf{A}_3-\mathbf{B}_3)^{\mathrm{T}}\mathbf{A}^{\mathrm{T}}\right)\bigg]$ \\ & $\Theta_{\mathcal{M}_4,\mathcal{M}_5,\mathcal{M}_6}\Pi(\mathbf{x},\mathbf{u})$ \\
			Metaplectic invariance & $\mathrm{C}(\mathcal{M},\mathcal{M}_1,\mathcal{M}_2,\mathcal{M}_3,\mathcal{M}_4,\mathcal{M}_5,\mathcal{M}_6)\mu\left(\mathcal{M}_0\right)f(\mathbf{x},\mathbf{u})=\mathrm{C}(\mathcal{M},\mathcal{M}_1\mathcal{M}_0,\mathcal{M}_2\mathcal{M}_0,\mathcal{M}_3,\mathcal{M}_4,\mathcal{M}_5,\mathcal{M}_6)f(\mathbf{x},\mathbf{u})$\\
			\specialrule{0.1em}{4pt}{4pt}
		\end{tabular}
		\begin{tablenotes}
			\item[1] Constraints on the symplectic matrices and the kernel function: $\mathbf{F}_{j2}=\mathbf{F}_{j3}=\mathbf{0}_N$, $j=4,5$, $\mathbf{F}_{44}=-\mathbf{D}\mathbf{B}^{-1}$, $\mathbf{F}_{54}=\mathbf{0}_N$, and $\phi(\mathbf{v},\mathbf{0})=1$
			\item[2] Constraints on the symplectic matrices and the kernel function: $\mathcal{M}_5=\mathbf{0}_{4N}$, $\mathbf{A}=\mathbf{0}_N$, $\mathbf{F}_{41}=\mathbf{F}_{42}=\mathbf{F}_{43}=\mathbf{0}_N$, $\mathrm{det}(\mathbf{A}_3)\neq0$, $\begin{pmatrix}\mathbf{0}_N&\mathbf{B}{\left(\mathbf{A}_3^{-1}\right)}^{\mathrm{T}}\\-\left(\mathbf{B}^{-1}\right)^{\mathrm{T}}\mathbf{A}_3&\mathbf{I}_N\end{pmatrix},\begin{pmatrix}\mathbf{0}_N&\mathbf{B}{\left(\mathbf{B}_3^{-1}\right)}^{\mathrm{T}}\\-\mathbf{B}_3^{\mathrm{T}}\mathbf{B}^{-1}&\mathbf{B}_3^{-1}\mathbf{A}_3\end{pmatrix}\in Sp(N,\mathbb{R})$, and $\phi(\mathbf{0},\mathbf{z})=1$
			\item[3] Constraints on the symplectic matrices and the kernel function: $\mathbf{A}_3=\mathbf{B}_3$, $\mathcal{M}_1=\mathcal{M}_2$, $\mathbf{F}_{j2}=\mathbf{F}_{j3}=\mathbf{0}_N$, $j=4,5$, $\mathbf{F}_{44}=-\mathbf{D}\mathbf{B}^{-1}$, $\mathbf{F}_{54}=\mathbf{0}_N$, and $\phi(\mathbf{v},\mathbf{0})=1$
			\item[4] Constraints on the symplectic matrices and the kernel function: $\begin{pmatrix}\mathbf{0}_N&\mathbf{B}{\left(\mathbf{A}_3^{-1}\right)}^{\mathrm{T}}\\-\left(\mathbf{B}^{-1}\right)^{\mathrm{T}}\mathbf{A}_3&\mathbf{I}_N\end{pmatrix}\mathcal{M}_1=\begin{pmatrix}\mathbf{0}_N&\mathbf{B}{\left(\mathbf{B}_3^{-1}\right)}^{\mathrm{T}}\\-\mathbf{B}_3^{\mathrm{T}}\mathbf{B}^{-1}&\mathbf{B}_3^{-1}\mathbf{A}_3\end{pmatrix}\mathcal{M}_2$, $\mathcal{M}_5=\mathbf{0}_{4N}$, $\mathbf{A}=\mathbf{0}_N$, $\mathbf{F}_{41}=\mathbf{F}_{42}=\mathbf{F}_{43}=\mathbf{0}_N$, $\mathrm{det}(\mathbf{A}_3)\neq0$, and $\phi(\mathbf{0},\mathbf{z})=1$
			\item[5] Constraints on the symplectic matrices and the kernel function: $\mathbf{A}=\mathbf{0}_N$, $\mathbf{C}_3=\mathbf{D}_3$, $\mathcal{M}_1=\mathcal{M}_2$, and $\phi\left(\mathbf{z,v}\right)=1$
			\item[6] Constraints on the symplectic matrices: $\mathbf{A}_3=\mathbf{0}_N$
			\item[7] Constraints on the kernel function:
			$\left|\mu(\mathcal{M}_5)\Pi(\mathbf{w})\right|^2=1$ for all $\mathbf{w}\in\mathbb{R}^{2N}$
			\item[8] Constraints on the kernel function: $\overline{\Pi}=\Pi$
			\item[9] Constraints on the symplectic matrices: $\mathcal{M}_3^{\mathrm{T}}\in Sp(N,\mathbb{R})$, $\mathbf{A}_1=\mathbf{A}_2$, and $\mathbf{B}_1^{\mathrm{T}}\mathbf{C}_1\mathbf{B}_1^{-1}\mathbf{D}_3-\mathbf{B}_2^{\mathrm{T}}\mathbf{C}_2\mathbf{B}_2^{-1}\mathbf{B}_3=\mathbf{0}_N$
			\item[10] Constraints on the symplectic matrices: $\mathcal{M}_3^{\mathrm{T}}\in Sp(N,\mathbb{R})$ and $\mathbf{B}_1=\mathbf{B}_2$
		\end{tablenotes}
	\end{threeparttable}
\end{table}
\section{Least-squares adaptive filter-based CMCD}\label{sec:4}
\indent For a given noise polluted signal $g(\mathbf{x})=f(\mathbf{x})+n(\mathbf{x})$, where $f(\mathbf{x})$ and $n(\mathbf{x})$ denote the pure signal and the additive noise, respectively, the common tactic of filters is to restore the pure signal as accurately as possible, namely, find the estimate $\widehat{f}(\mathbf{x})$ as close as possible to the ideal $f(\mathbf{x})$. Thanks to the convolution nature of the CMCD and the unique reconstruction property of Wigner distribution, this is equivalent to design an adaptive filter $H(\mathbf{x},\mathbf{u})$ in the metaplectic Wigner distribution domain, which can find the estimate
\begin{align}\label{eq4.1} \mathrm{W}\widehat{f}(\mathbf{x},\mathbf{u})=\left(\mathrm{W}(\mathcal{M},\mathcal{M}_1,\mathcal{M}_2,\mathcal{M}_3)g\Theta_{\mathcal{M}_4,\mathcal{M}_5,\mathcal{M}_6}H\right)(\mathbf{x},\mathbf{u})
\end{align}
as close as possible to the ideal $\mathrm{W}f(\mathbf{x,u})$. According to Wiener filter principle, a natural criterion to characterize the estimation accuracy is the MSE criterion
\begin{align}\label{eq4.2} \sigma_{\mathrm{MSE}}^2\overset{\mathrm{def}}{=}\mathbb{E}\left\{\left|\mathrm{W}f(\mathbf{x},\mathbf{u})-\mathrm{W}\widehat{f}(\mathbf{x},\mathbf{u})\right|^2\right\},
\end{align}
where $\mathbb{E}(\cdot)$ denotes the mathematical expectation operator.\\
\indent For simplicity, let $\mathbf{z}=(\mathbf{x},\mathbf{u})\in\mathbb{R}^{2N}$ and $\mathop{\widetilde{\mathrm{W}}}\limits^jf(\mathbf{z})=\mathrm{e}^{\pi\mathrm{i}\mathbf{z}\mathbf{B}_j^{-1}\mathbf{A}_j\mathbf{z}^{\mathrm{T}}}\mathrm{W}f(\mathbf{z})$, $j=4,5,6$, then Eqs.~\eqref{eq4.1} and \eqref{eq4.2} become
\begin{align}\label{eq4.3}
\mathrm{W}\widehat{f}(\mathbf{z})&=\left(\mathrm{W}(\mathcal{M},\mathcal{M}_1,\mathcal{M}_2,\mathcal{M}_3)g\Theta_{\mathcal{M}_4,\mathcal{M}_5,\mathcal{M}_6}H\right)(\mathbf{z})\nonumber\\ &=\mathrm{e}^{-\pi\mathrm{i}\mathbf{z}\mathbf{B}_6^{-1}\mathbf{A}_6\mathbf{z}^{\mathrm{T}}}\int_{\mathbb{R}^{2N}}\mathop{\widetilde{\mathrm{W}}}\limits^4(\mathcal{M},\mathcal{M}_1,\mathcal{M}_2,\mathcal{M}_3)g(\mathbf{k})\mathop{\widetilde{H}}\limits^5(\mathbf{z}-\mathbf{k})\mathrm{d}\mathbf{k}
\end{align}
and
\begin{align}\label{eq4.4}	\sigma_{\mathrm{MSE}}^2=\mathbb{E}\left\{\left|\mathrm{W}f(\mathbf{z})-\mathrm{W}\widehat{f}(\mathbf{z})\right|^2\right\},
\end{align}
respectively. Now, our goal is to design an adaptive optimal filter $H_{\mathrm{opt}}(\mathbf{z})$ in the metaplectic Wigner distribution domain to minimize the MSE given by Eq.~\eqref{eq4.4}, or equivalently,
\begin{align}\label{eq4.5}
H_{\mathrm{opt}}(\mathbf{z})=\mathop{\arg\min}\limits_{H(\mathbf{z})}\sigma_{\mathrm{MSE}}^2.
\end{align}
By using the orthogonal principle \cite{Pro85}, the chirp-stationary assumption and the conventional convolution and correlation theorems to establish, simplify and solve the Wiener-Hopf equation, respectively, the least-squares adaptive filter transfer function in the metaplectic Wigner distribution domain reads
\begin{align}\label{eq4.6}
&\mu\left(\mathcal{M}_5\right)H_{\mathrm{opt}}(\mathbf{w})\nonumber\\
=&\frac{\sqrt{-\mathrm{det}(\mathbf{B}_6)}}{\sqrt{-\mathrm{det}(\mathbf{B}_4)}\sqrt{-\mathrm{det}(\mathbf{B}_5)}}\mathrm{e}^{\pi\mathrm{i}\mathbf{w}\left(\mathbf{B}_5^{-1}\right)^{\mathrm{T}}\mathbf{B}_4^{\mathrm{T}}\mathbf{D}_4\mathbf{B}_5^{-1}\mathbf{w}^{\mathrm{T}}}
\mathrm{e}^{\pi\mathrm{i}\mathbf{w}\mathbf{D}_5\mathbf{B}_5^{-1}\mathbf{w}^{\mathrm{T}}}\mathrm{e}^{-\pi\mathrm{i}\mathbf{w}\left(\mathbf{B}_5^{-1}\right)^{\mathrm{T}}\mathbf{B}_6^{\mathrm{T}}\mathbf{D}_6\mathbf{B}_5^{-1}\mathbf{w}^{\mathrm{T}}}\nonumber\\
&\times\frac{\mu\left(\mathcal{M}_6\right)\mathrm{W}f\left(\mathbf{w}\left(\mathbf{B}_5^{-1}\right)^{\mathrm{T}}\mathbf{B}_6^{\mathrm{T}}\right)}{\mu\left(\mathcal{M}_4\right)\mathrm{W}(\mathcal{M},\mathcal{M}_1,\mathcal{M}_2,\mathcal{M}_3)g\left(\mathbf{w}\left(\mathbf{B}_5^{-1}\right)^{\mathrm{T}}\mathbf{B}_4^{\mathrm{T}}\right)}.
\end{align}
Taking the inverse metaplectic transform on both sides of the above equation yields the least-squares adaptive filter in the metaplectic Wigner distribution domain
\begin{align}\label{eq4.7}
H_{\mathrm{opt}}(\mathbf{z})=&\frac{\sqrt{-\mathrm{det}(\mathbf{B}_6)}}{\sqrt{-\mathrm{det}(\mathbf{B}_4)}\sqrt{-\mathrm{det}(\mathbf{B}_5)}}\int_{\mathbb{R}^{2N}}
\mathrm{e}^{\pi\mathrm{i}\mathbf{w}\left(\mathbf{B}_5^{-1}\right)^{\mathrm{T}}\mathbf{B}_4^{\mathrm{T}}\mathbf{D}_4\mathbf{B}_5^{-1}\mathbf{w}^{\mathrm{T}}}
\mathrm{e}^{\pi\mathrm{i}\mathbf{w}\mathbf{D}_5\mathbf{B}_5^{-1}\mathbf{w}^{\mathrm{T}}}\mathrm{e}^{-\pi\mathrm{i}\mathbf{w}\left(\mathbf{B}_5^{-1}\right)^{\mathrm{T}}\mathbf{B}_6^{\mathrm{T}}\mathbf{D}_6\mathbf{B}_5^{-1}\mathbf{w}^{\mathrm{T}}}\nonumber\\
&\times\frac{\mu\left(\mathcal{M}_6\right)\mathrm{W}f\left(\mathbf{w}\left(\mathbf{B}_5^{-1}\right)^{\mathrm{T}}\mathbf{B}_6^{\mathrm{T}}\right)}{\mu\left(\mathcal{M}_4\right)\mathrm{W}(\mathcal{M},\mathcal{M}_1,\mathcal{M}_2,\mathcal{M}_3)g\left(\mathbf{w}\left(\mathbf{B}_5^{-1}\right)^{\mathrm{T}}\mathbf{B}_4^{\mathrm{T}}\right)}\mathcal{K}_{\mathcal{M}_5^{-1}}\left(\mathbf{z},\mathbf{w}\right)\mathrm{d}\mathbf{w}.
\end{align}
See Appendix~\ref{sec:AppB} for the detailed derivation of Eq.~\eqref{eq4.6}. Correspondingly, the minimum MSE goes down to zero, that is,
\begin{align}\label{eq4.8}
\mathop{\min}\limits_{H(\mathbf{z})}\sigma_{\mathrm{MSE}}^{2}=0.
\end{align}
See Appendix~\ref{sec:AppC} for the detailed derivation of Eq.~\eqref{eq4.8}.\\
\indent Substituting Eq.~\eqref{eq4.7} into Eq.~\eqref{eq3.2} gives the least-squares adaptive filter-based CMCD
\begin{align}\label{eq4.9} \mathrm{C}_{\mathrm{LSAF}}(\mathcal{M},\mathcal{M}_1,\mathcal{M}_2,\mathcal{M}_3,\mathcal{M}_4,\mathcal{M}_5,\mathcal{M}_6)g(\mathbf{x},\mathbf{u})=\left(\mathrm{W}(\mathcal{M},\mathcal{M}_1,\mathcal{M}_2,\mathcal{M}_3)g\Theta_{\mathcal{M}_4,\mathcal{M}_5,\mathcal{M}_6}H_{\mathrm{opt}}\right)(\mathbf{x},\mathbf{u}).
\end{align}
\section{Simulations}\label{sec:5}
\indent In this section, we perform three examples to verify the correctness and effectiveness of the least-squares adaptive filter-based CMCD. We also compare the denoising effect of the proposed adaptive CMCD filtering method with some state-of-the-art techniques, including the Cohen's distribution filtering method with fixed kernel functions (e.g., the Margenau-Hill distribution filtering method, the Kirkwood-Rihaczek distribution filtering method, the Born-Jordan distribution filtering method and the Page distribution filtering method), the classical Wiener filter, the adaptive Cohen's distribution filtering method, the adaptive metaplectic Wigner distribution-based Cohen's distribution filtering method, and the adaptive generalized metaplectic convolution-based Cohen's distribution filtering method.\\
\indent For simplicity, we call the filtering methods using the Margenau-Hill distribution, the Kirkwood-Rihaczek distribution, the Born-Jordan distribution, the Page distribution, the adaptive Cohen's distribution, the adaptive metaplectic Wigner distribution-based Cohen's distribution and the adaptive generalized metaplectic convolution-based Cohen's distribution as the Margenau-Hill, the Kirkwood-Rihaczek, the Born-Jordan, the Page, the adaptive CD, the adaptive MWD-CD and the adaptive GMC-CD, respectively.\\
\indent \emph{Example~1 (Linear frequency-modulated signal):} The
polluted signal is selected for the linear frequency-modulated signal $\mathrm{e}^{2\pi\mathrm{i}\left(x+\frac{x^{2}}{2}\right)}$ added with a complex white Gaussian noise with SNRs ranging from $-4\mathrm{dB}$ to $6\mathrm{dB}$. The symplectic matrices are chosen as $\mathcal{M}=\begin{pmatrix}0&1\\-1&0\end{pmatrix}$, $\mathcal{M}_1=\mathcal{M}_2=\begin{pmatrix}0&1\\-1&2\end{pmatrix}$, $\mathcal{M}_3=\begin{pmatrix}1&1\\-\frac{1}{2}&\frac{1}{2}\end{pmatrix}$, $\mathcal{M}_4=\mathcal{M}_5=\mathcal{M}_6=\begin{pmatrix}-5&0&1&0\\0&5&0&1\\0&0&-\frac{1}{5}&0\\0&0&0&\frac{1}{5}\end{pmatrix}$.

\begin{figure}[htbp]
	\centering
	\subfigure[\textit{}]
	{\label{1a}\includegraphics[width=1\textwidth,height=0.4\textheight]{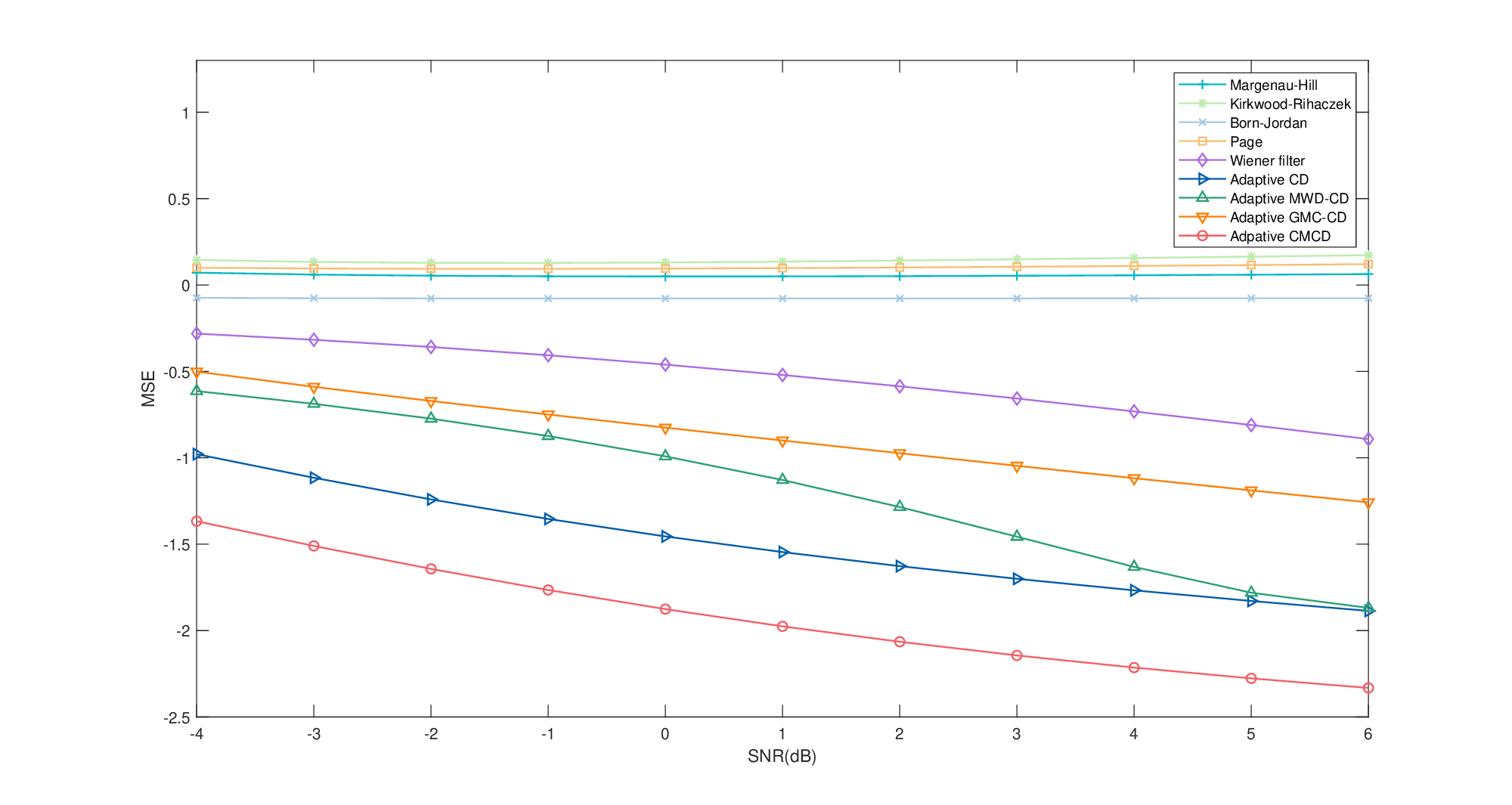}}
	\subfigure[\textit{}]
	{\label{1b}\includegraphics[width=1\textwidth,height=0.4\textheight]{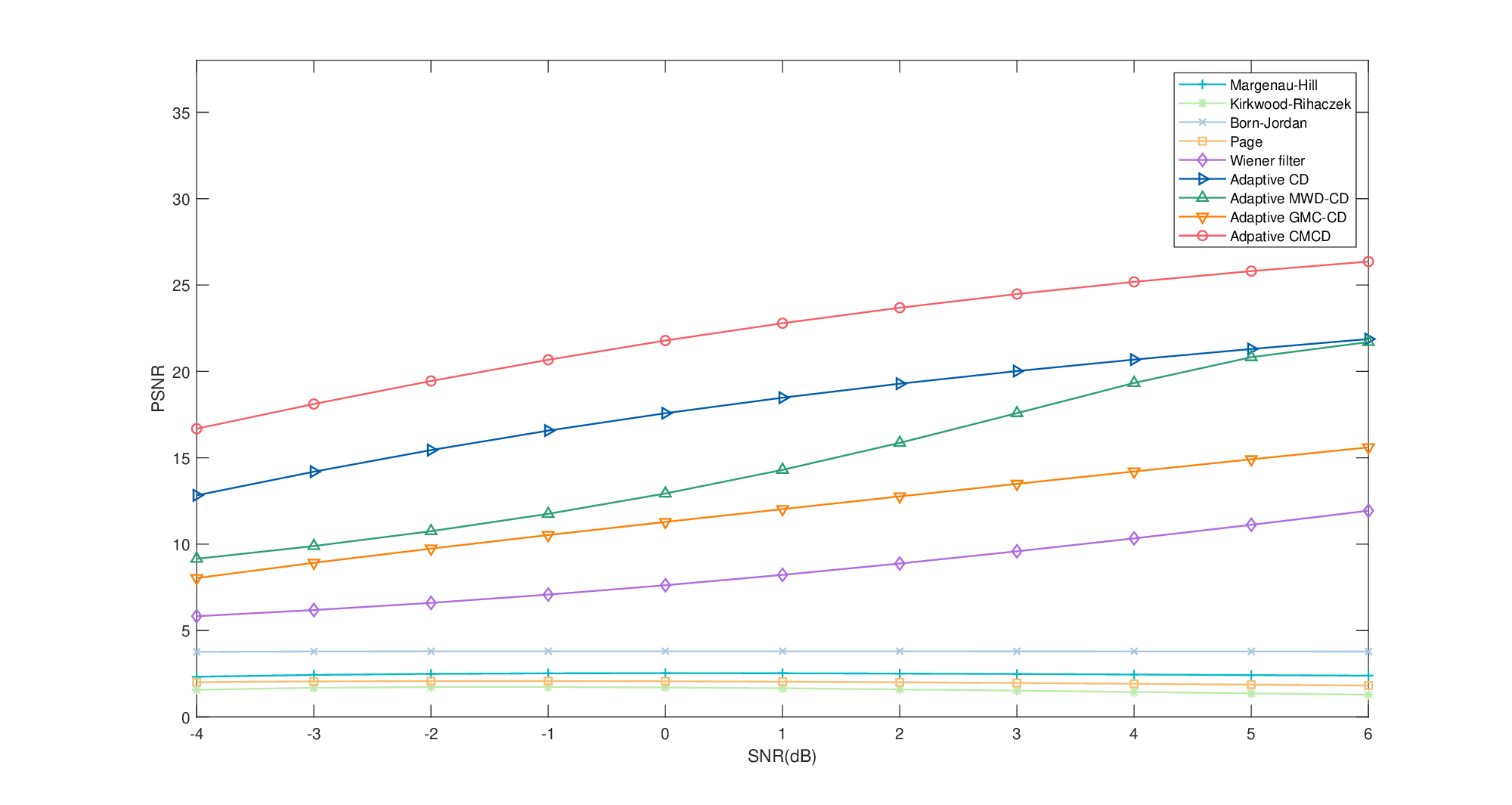}}
	\caption{The MSE and PSNR performance indexes of the estimate linear frequency-modulated signals using nine filtering methods. (a) The SNR-MSE diagram of the estimate linear frequency-modulated signals; (b) The SNR-PSNR diagram of the estimate linear frequency-modulated signals.}
	\label{fig1}
\end{figure}

\indent \emph{Example~2 (Gaussian enveloped linear frequency-modulated signal):} The polluted signal is selected for the Gaussian enveloped linear frequency-modulated signal $\mathrm{e}^{-\frac{1}{8}\left(x+1\right)^{2}}\mathrm{e}^{2\pi\mathrm{i}x^{2}}$ added with a complex white Gaussian noise with SNRs ranging from $-4\mathrm{dB}$ to $6\mathrm{dB}$. The symplectic matrices are chosen as $\mathcal{M}=\begin{pmatrix}0&1\\-1&0\end{pmatrix}$, $\mathcal{M}_1=\mathcal{M}_2=\begin{pmatrix}0&1\\-1&\frac{5}{2}\end{pmatrix}$, $\mathcal{M}_3=\begin{pmatrix}1&1\\-\frac{1}{2}&\frac{1}{2}\end{pmatrix}$, $\mathcal{M}_4=\mathcal{M}_5=\mathcal{M}_6=\begin{pmatrix}1&0&4&0\\0&1&0&1\\1&0&5&0\\0&1&0&2\end{pmatrix}$.

\begin{figure}[htbp]
	\centering
	\subfigure[\textit{}]
	{\label{2a}\includegraphics[width=1\textwidth,height=0.4\textheight]{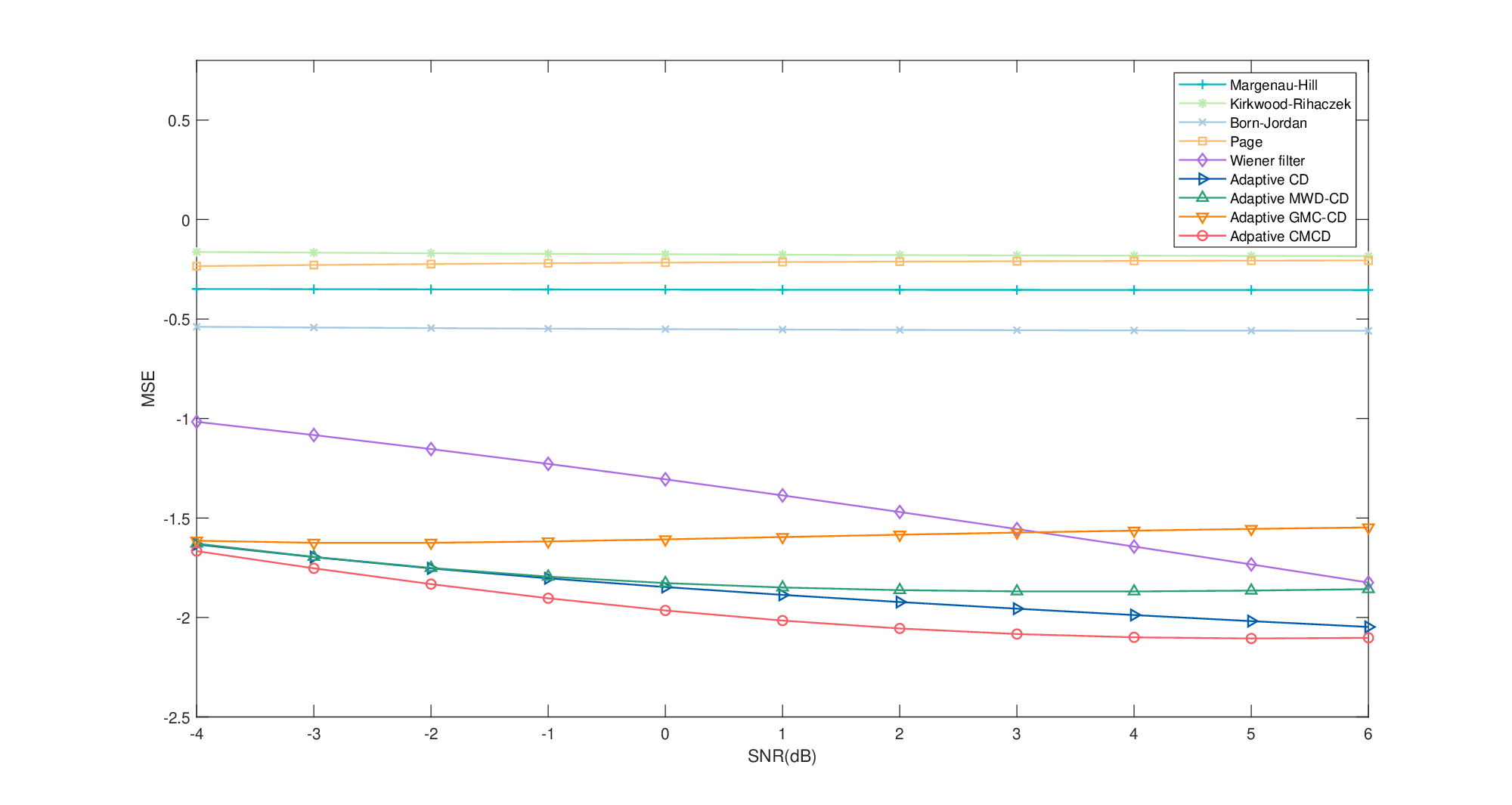}}
	\subfigure[\textit{}]
	{\label{2b}\includegraphics[width=1\textwidth,height=0.4\textheight]{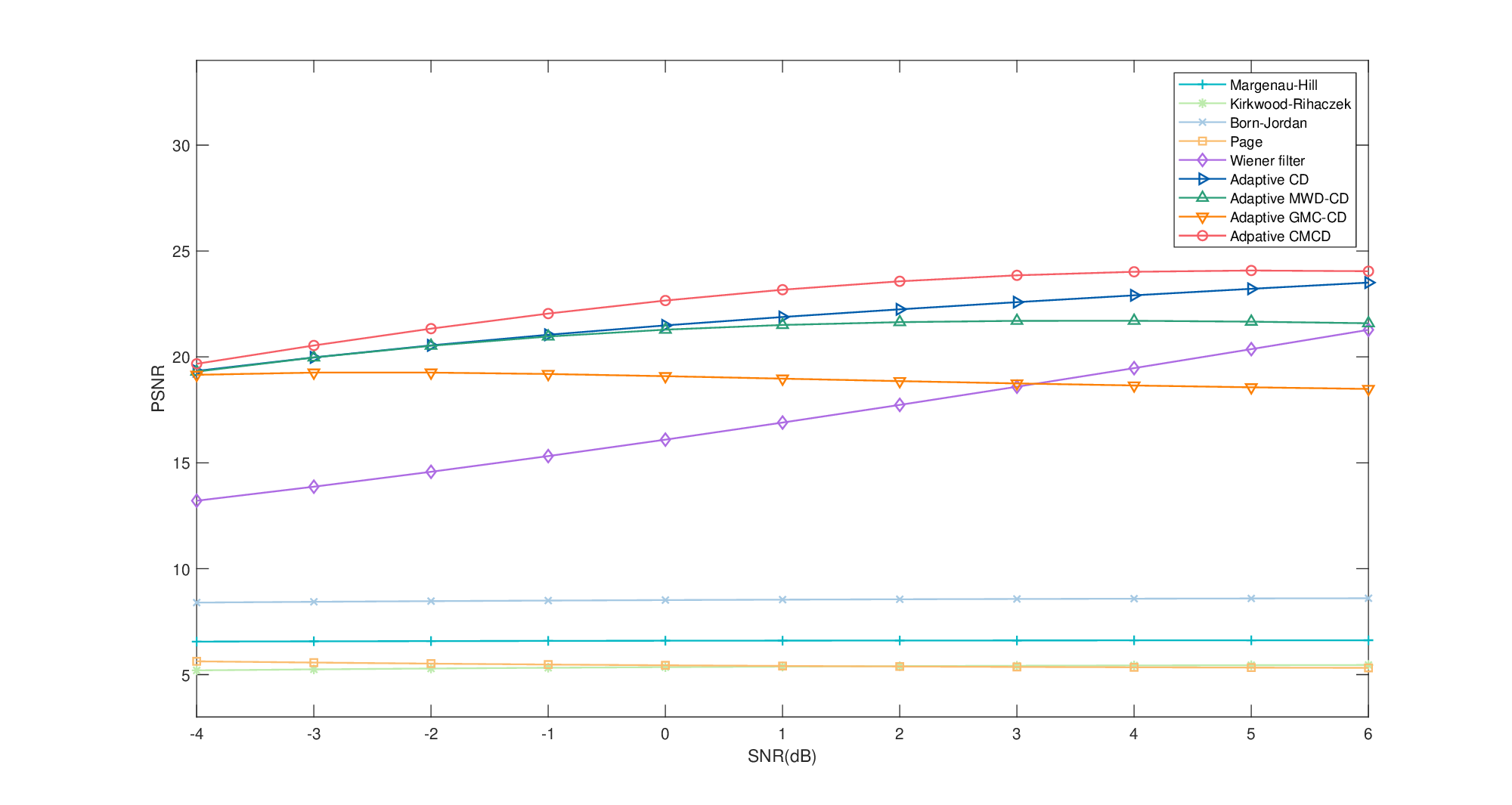}}
	\caption{The MSE and PSNR performance indexes of the estimate Gaussian enveloped linear frequency-modulated signals using nine filtering methods. (a) The SNR-MSE diagram of the estimate Gaussian enveloped linear frequency-modulated signals; (b) The SNR-PSNR diagram of the estimate Gaussian enveloped linear frequency-modulated signals.}
	\label{fig2}
\end{figure}

\emph{Example~3 (Complex exponential signal):} The polluted signal is selected for the complex exponential signal $\mathrm{e}^{\pi\mathrm{i}x}$ added with a complex white Gaussian noise with SNRs ranging from $-4\mathrm{dB}$ to $6\mathrm{dB}$. The symplectic matrices are chosen as $\mathcal{M}=\begin{pmatrix}0&\frac{10}{21}\\-\frac{21}{10}&\frac{10}{7}\end{pmatrix}$, $\mathcal{M}_1=\mathcal{M}_2=\begin{pmatrix}0&1\\-1&0\end{pmatrix}$, $\mathcal{M}_3=\begin{pmatrix}1&1\\-\frac{1}{2}&\frac{1}{2}\end{pmatrix}$, $\mathcal{M}_4=\mathcal{M}_5=\mathcal{M}_6=\begin{pmatrix}0&0&1&0\\0&3&0&-2\\-1&0&2&0\\0&-1&0&1\end{pmatrix}$.

\begin{figure}[htbp]
	\centering
	\subfigure[\textit{}]
	{\label{3a}\includegraphics[width=1\textwidth,height=0.4\textheight]{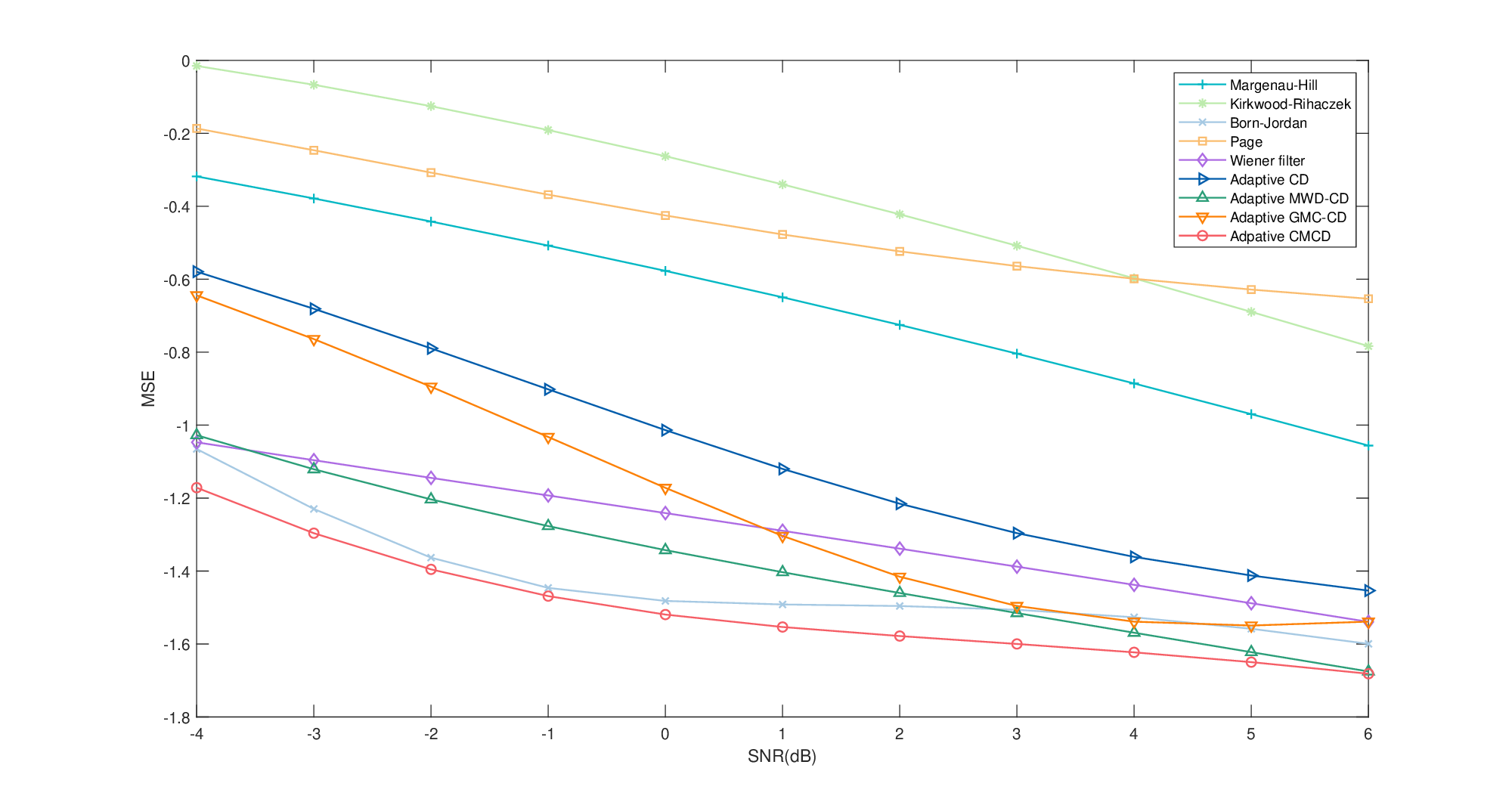}}
	\subfigure[\textit{}]
	{\label{3b}\includegraphics[width=1\textwidth,height=0.4\textheight]{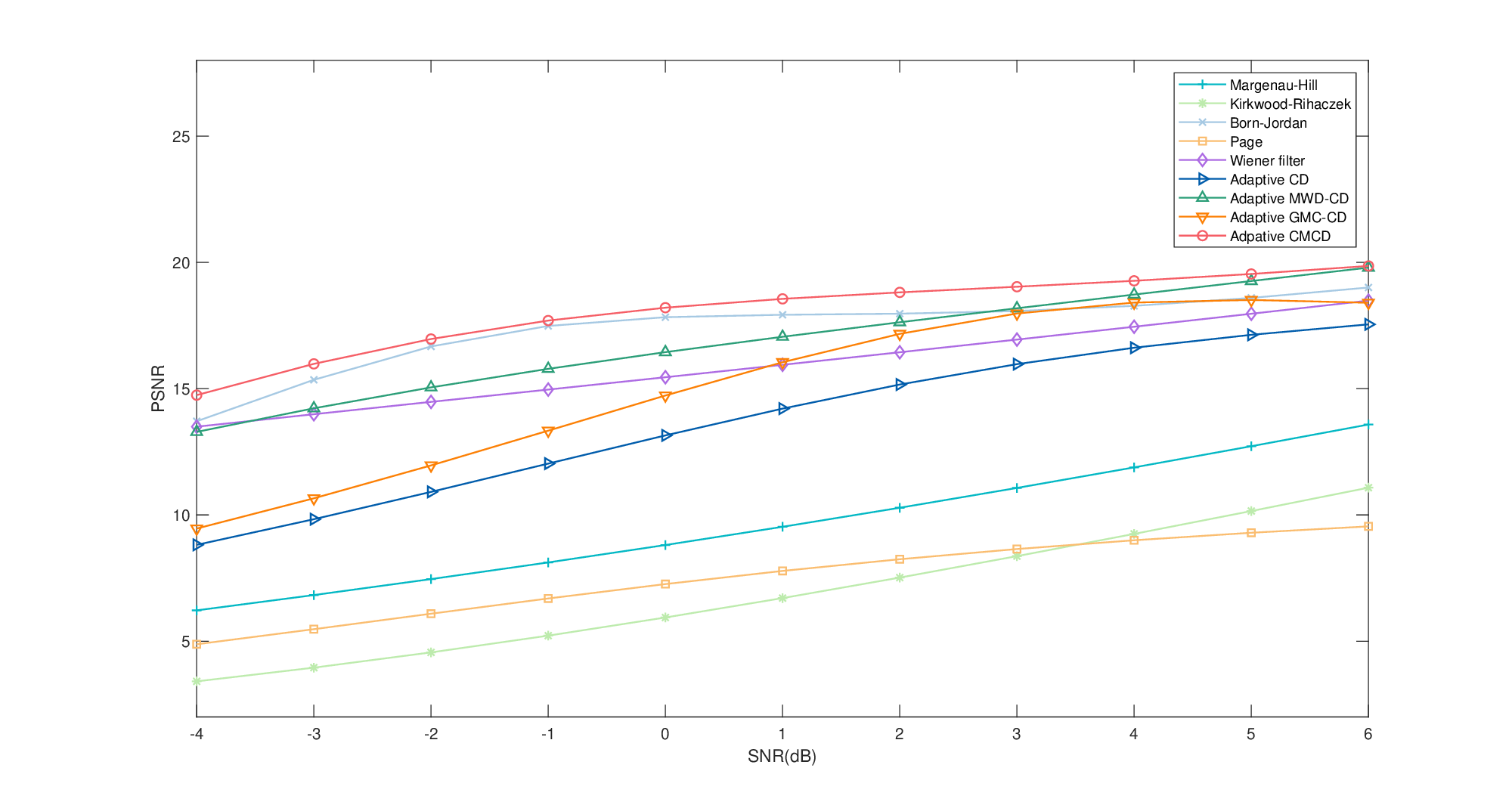}}
	\caption{The MSE and PSNR performance indexes of the estimate complex exponential signals using nine filtering methods. (a) The SNR-MSE diagram of the estimate complex exponential signals; (b) The SNR-PSNR diagram of the estimate complex exponential signals.}
	\label{fig3}
\end{figure}

\indent Note that the observing interval is set to $[-5\mathrm{s},5\mathrm{s}]$ for examples 1--3, and the sampling frequency is set to $30\mathrm{Hz}$ for example 1 and to $50\mathrm{Hz}$ for examples 2 and 3. Moreover, the logarithm of MSE is taken, and the PSNR is calculated by the average of the real and imaginary parts.\\
\indent Fig.~\ref{fig1} presents the MSE and PSNR performance indexes of the estimate linear frequency-modulated signals using nine filtering methods including the Margenau-Hill, the Kirkwood-Rihaczek, the Born-Jordan, the Page, the Wiener filter, the adaptive CD, the adaptive MWD-CD, the adaptive GMC-CD and the proposed adaptive CMCD. To be specific, Figs.~\ref{fig1}(a) and (b) plot the SNR-MSE and the SNR-PSNR diagrams of the estimate linear frequency-modulated signals, respectively. Fig.~\ref{fig2} presents the MSE and PSNR performance indexes of the estimate Gaussian enveloped linear frequency-modulated signals using these nine filtering methods. To be specific, Figs.~\ref{fig2}(a) and (b) plot the SNR-MSE and the SNR-PSNR diagrams of the estimate Gaussian enveloped linear frequency-modulated signals, respectively. Fig.~\ref{fig3} presents the MSE and PSNR performance indexes of the estimate complex exponential signals using these nine filtering methods. To be specific, Figs.~\ref{fig3}(a) and (b) plot the SNR-MSE and the SNR-PSNR diagrams of the estimate complex exponential signals, respectively.\\
\indent A smaller MSE value indicates higher accuracy and a larger PSNR value suggests less distortion. As seen in Figs~\ref{fig1}--\ref{fig3}, the proposed adaptive CMCD achieves better noise suppression performance than the Margenau-Hill, the Kirkwood-Rihaczek, the Born-Jordan, the Page, the Wiener filter, the adaptive CD, the adaptive MWD-CD and the adaptive GMC-CD.
\section{Discussions on the optimal symplectic matrices}\label{sec:6}
\indent In examples 1--3, the denoising effect of the proposed adaptive CMCD depends on the symplectic matrices $\mathcal{M},\mathcal{M}_1,\mathcal{M}_2,\mathcal{M}_3,\mathcal{M}_4,\mathcal{M}_5,\mathcal{M}_6$. It becomes therefore interesting and meaningful to explore the optimal symplectic matrices that reach the best denoising performance. In this section, we provide a theoretical analysis on how to choose symplectic matrices to achieve the highest-performance denoising.\\
\indent It should be recalled here that there exist chirp-stationary assumptions on $\mathrm{W}f$ and $\mathrm{W}(\mathcal{M},\mathcal{M}_1,\mathcal{M}_2,\mathcal{M}_3)g$ to simplify the Wiener-Hopf equation \eqref{eqB.3} into \eqref{eqB.4}. Unfortunately, often that is not the case in practical engineering applications. Indeed, either $\mathrm{W}f$ and $\mathrm{W}(\mathcal{M},\mathcal{M}_1,\mathcal{M}_2,\mathcal{M}_3)g$ are usually not chirp-stationary, or it is difficult to verify their chirp-stationarity. Even worse, it seems impossible to verify the chirp-stationarity of $\mathrm{W}f$ when the ideal $f(\mathbf{x})$ is unknown. Consequently, we are unable to solve explicitly the Wiener-Hopf equation~\eqref{eqB.3}. Instead we can apply the classical numerical algorithms to obtain the approximate solution of this equation. Since there is only a numerical solution regarding to the least-squares adaptive filter $H_{\mathrm{opt}}$ in the metaplectic Wigner distribution domain, the minimum MSE $\mathop{\min}\limits_{H(\mathbf{z})}\sigma_{\mathrm{MSE}}^2$ does not go down to zero. It can be seen as the minimum MSE function whose arguments are the symplectic matrices $\mathcal{M},\mathcal{M}_1,\mathcal{M}_2,\mathcal{M}_3,\mathcal{M}_4,\mathcal{M}_5,\mathcal{M}_6$. We denote the minimum MSE function as
\begin{align}\label{eq5.1}
\sigma_{\mathcal{M},\mathcal{M}_1,\mathcal{M}_2,\mathcal{M}_3,\mathcal{M}_4,\mathcal{M}_5,\mathcal{M}_6}^2.
\end{align}
By taking the symplectic matrices $\mathcal{M},\mathcal{M}_1,\mathcal{M}_2,\mathcal{M}_3,\mathcal{M}_4,\mathcal{M}_5,\mathcal{M}_6$ and the minimum MSE function $\sigma_{\mathcal{M},\mathcal{M}_1,\mathcal{M}_2,\mathcal{M}_3,\mathcal{M}_4,\mathcal{M}_5,\mathcal{M}_6}^2$ as decision variables and objective function, respectively, the minimum MSE minimization model follows
\begin{align}\label{eq5.2}
\mathop{\min}\limits_{\mathcal{M},\mathcal{M}_1,\mathcal{M}_2,\mathcal{M}_3\in Sp(N,\mathbb{R}),\mathcal{M}_4,\mathcal{M}_5,\mathcal{M}_6\in Sp(2N,\mathbb{R})}\sigma_{\mathcal{M},\mathcal{M}_1,\mathcal{M}_2,\mathcal{M}_3,\mathcal{M}_4,\mathcal{M}_5,\mathcal{M}_6}^2.
\end{align}
Obviously, the solution of the optimization model~\eqref{eq5.2} is none other than the optimal symplectic matrices selection strategy that minimizes the minimum MSE function, giving birth to the best denoising performance of the least-squares adaptive filter in the metaplectic Wigner distribution domain. Two points deserve to be underlined regarding how to solve the optimization model~\eqref{eq5.2} to come out the optimal symplectic matrices selection strategy.
\begin{itemize}

    \item By investigating the structural feature of the symplectic matrices optimization problem~\eqref{eq5.2}, we translate it equivalently into one of the common matrix optimization problems \cite{Abs08}, for example, the linear semi-definite programming problem, the robust principle component analysis problem, the matrix rank optimization problem, the inverse semi-definite quadratic programming problem, the inverse damped gyroscopic eigenvalue problem, the nonnegative matrix weighted/tri/multi-factorization problem, the matrix convex feasible problem, and the optimal matrix approximation problem.

    \item By employing numerical optimization algorithms frequently used in signal processing and machine learning \cite{Noc06,Bot18}, such as the peak detection algorithm, the uniform design, the meta-heuristic algorithm, the alternating direction method of multipliers, the spectral projection gradient algorithm, the majorization-minimization algorithm, the potential reduction algorithm, the variational inequality algorithm, and the zeroth order gradient method, in solving the derived matrix optimization problem, we obtain the optimal symplectic matrices selection strategy of the proposed adaptive CMCD.

\end{itemize}
\section{Conclusion}\label{sec:7}
\indent The time-frequency analysis theory, method and technology systems of the CMCD, which can meet the requirement of additive noises jamming signals adaptive optimal filter denoising under the condition of low SNR, have been established.\\
\indent The CMCD exhibits the strongest generalization capability in terms of Wigner operator fractionization and convolution operator fractionization. It includes particular cases the generalized metaplectic convolution-based Cohen's distribution, the metaplectic Wigner distribution-based Cohen's distribution, the Cohen's distribution, the metaplectic Wigner distribution, and the joint metaplectic Wigner and I/II/III-type metaplectic convolution operators Cohen's distribution. It essential properties, including marginal distribution, energy conservation, unique reconstruction, Moyal formula, complex conjugate symmetry, time reversal symmetry, scaling property, time translation property, frequency modulation property, and metaplectic invariance, are generalizations of those of the Cohen's distribution and hold for some constraints on the symplectic matrices and the kernel function.\\
\indent The proposed adaptive CMCD can automatically adjust its kernel function which minimizes the MSE in Wigner distribution domain according to the change of signal to adapt to different signal characteristics. It turns out that its denoising effect is better that of the Margenau-Hill, the Kirkwood-Rihaczek, the Born-Jordan, the Page, the Wiener filter, the adaptive CD, the adaptive MWD-CD and the adaptive GMC-CD.\\
\indent Without chirp-stationary assumptions, the denoising effect of the proposed adaptive CMCD depends on the symplectic matrices. The optimal symplectic matrices selection strategy can be generated by first converting the minimum MSE minimization model to the matrix optimization problem and then solving it through numerical optimization algorithms frequently used in signal processing and machine learning.
\section*{Appendix}
\indent In this section, we prove all the theoretical results in the paper.
\subsection{Proof of essential properties of the CMCD}\label{sec:AppA}
\indent (1) \emph{Proof of the time marginal distribution, i.e., Eq.~\eqref{Time marginal distribution}:} When $\mathbf{F}_{j2}=\mathbf{F}_{j3}=\mathbf{0}_N$, $j=4,5$, $\mathbf{F}_{44}=-\mathbf{D}\mathbf{B}^{-1}$ and $\mathbf{F}_{54}=\mathbf{0}_N$, we have
\begin{align}\label{eqA.1} &\int_{\mathbb{R}^N}\mathrm{C}(\mathcal{M},\mathcal{M}_1,\mathcal{M}_2,\mathcal{M}_3,\mathcal{M}_4,\mathcal{M}_5,\mathcal{M}_6)f(\mathbf{x},\mathbf{u})\mathrm{e}^{\pi\mathrm{i}\left(\mathbf{x},\mathbf{u}\right)\mathbf{B}_6^{-1}\mathbf{A}_6\left(\mathbf{x},\mathbf{u}\right)^{\mathrm{T}}}\mathrm{d}\mathbf{u}\nonumber\\
=&\frac{1}{\sqrt{-\mathrm{det}(\mathbf{B})}}\overbrace{\idotsint}^{5}_{\mathbb{R}^{N\times N\times N\times N\times N}}\mu(\mathcal{M}_1)f\left(\mathbf{p}\mathbf{B}_3+\mathbf{t}\mathbf{D}_3\right)\overline{\mu(\mathcal{M}_2)f\left(\mathbf{p}\mathbf{A}_3+\mathbf{t}\mathbf{C}_3\right)}\phi\left(\mathbf{v},\mathbf{z}\right)\nonumber\\ &\times\mathrm{e}^{-2\pi\mathrm{i}\left(\mathbf{v}(\mathbf{x}-\mathbf{p})^{\mathrm{T}}-\mathbf{z}\mathbf{q}^{\mathrm{T}}\right)}\mathrm{e}^{\pi\mathrm{i}\left(\mathbf{t}\mathbf{B}^{-1}\mathbf{A}\mathbf{t}^{\mathrm{T}}+\mathbf{p}\mathbf{F}_{41}\mathbf{p}^{\mathrm{T}}\right)}\mathrm{e}^{-2\pi\mathrm{i}\mathbf{t}\mathbf{B}^{-1}\mathbf{q}^{\mathrm{T}}}\mathrm{e}^{\pi\mathrm{i}(\mathbf{x}-\mathbf{p})\mathbf{F}_{51}(\mathbf{x}-\mathbf{p})^{\mathrm{T}}}\nonumber\\ &\times\int_{\mathbb{R}^N}\mathrm{e}^{-2\pi\mathrm{i}\mathbf{z}\mathbf{u}^{\mathrm{T}}}\mathrm{d}\mathbf{u}\mathrm{d}\mathbf{v}\mathrm{d}\mathbf{z}\mathrm{d}\mathbf{t}\mathrm{d}\mathbf{p}\mathrm{d}\mathbf{q}.
\tag{A.1}
\end{align}
When the kernel function satisfies $\phi(\mathbf{v},\mathbf{0})=1$, it follows that
\begin{align}\label{eqA.2}	&\int_{\mathbb{R}^N}\mathrm{C}(\mathcal{M},\mathcal{M}_1,\mathcal{M}_2,\mathcal{M}_3,\mathcal{M}_4,\mathcal{M}_5,\mathcal{M}_6)f(\mathbf{x},\mathbf{u})\mathrm{e}^{\pi\mathrm{i}\left(\mathbf{x},\mathbf{u}\right)\mathbf{B}_6^{-1}\mathbf{A}_6\left(\mathbf{x},\mathbf{u}\right)^{\mathrm{T}}}\mathrm{d}\mathbf{u}\nonumber\\
=&\frac{1}{\sqrt{-\mathrm{det}(\mathbf{B})}}\iiint_{\mathbb{R}^{N\times N\times N}}\mu(\mathcal{M}_1)f\left(\mathbf{p}\mathbf{B}_3+\mathbf{t}\mathbf{D}_3\right)\overline{\mu(\mathcal{M}_2)f\left(\mathbf{p}\mathbf{A}_3+\mathbf{t}\mathbf{C}_3\right)}\mathrm{e}^{\pi\mathrm{i}\left(\mathbf{t}\mathbf{B}^{-1}\mathbf{A}\mathbf{t}^{\mathrm{T}}+\mathbf{p}\mathbf{F}_{41}\mathbf{p}^{\mathrm{T}}\right)}\nonumber\\ &\times\mathrm{e}^{-2\pi\mathrm{i}\mathbf{t}\mathbf{B}^{-1}\mathbf{q}^{\mathrm{T}}}\mathrm{e}^{\pi\mathrm{i}(\mathbf{x}-\mathbf{p})\mathbf{F}_{51}(\mathbf{x}-\mathbf{p})^{\mathrm{T}}}\int_{\mathbb{R}^N}\mathrm{e}^{-2\pi\mathrm{i}\mathbf{v}(\mathbf{x}-\mathbf{p})^{\mathrm{T}}}\mathrm{d}\mathbf{v}\mathrm{d}\mathbf{t}\mathrm{d}\mathbf{p}\mathrm{d}\mathbf{q}\nonumber\\ =&\frac{1}{\sqrt{-\mathrm{det}(\mathbf{B})}}\int_{\mathbb{R}^N}\mu(\mathcal{M}_1)f\left(\mathbf{x}\mathbf{B}_3+\mathbf{t}\mathbf{D}_3\right)\overline{\mu(\mathcal{M}_2)f\left(\mathbf{x}\mathbf{A}_3+\mathbf{t}\mathbf{C}_3\right)}\mathrm{e}^{\pi\mathrm{i}\left(\mathbf{t}\mathbf{B}^{-1}\mathbf{A}\mathbf{t}^{\mathrm{T}}+\mathbf{x}\mathbf{F}_{41}\mathbf{x}^{\mathrm{T}}\right)}\nonumber\\ &\times\int_{\mathbb{R}^N}\mathrm{e}^{-2\pi\mathrm{i}\mathbf{t}\mathbf{B}^{-1}\mathbf{q}^{\mathrm{T}}}\mathrm{d}\mathbf{q}\mathrm{d}\mathbf{t}.
\tag{A.2}
\end{align}
Thus, we arrive the required result \eqref{Time marginal distribution}.$\hfill\blacksquare$\\	
\indent (2) \emph{Proof of the frequency marginal distribution, i.e., Eq.~\eqref{Frequency marginal distribution}:} When $\mathcal{M}_5=\mathbf{0}_{4N}$, $\mathbf{A}=\mathbf{0}_N$ and $\mathbf{F}_{41}=\mathbf{F}_{42}=\mathbf{F}_{43}=\mathbf{0}_N$, we have
\begin{align}\label{eqA.3} &\int_{\mathbb{R}^N}\mathrm{C}(\mathcal{M},\mathcal{M}_1,\mathcal{M}_2,\mathcal{M}_3,\mathcal{M}_4,\mathcal{M}_5,\mathcal{M}_6)f(\mathbf{x},\mathbf{u})\mathrm{e}^{\pi\mathrm{i}\left(\mathbf{x},\mathbf{u}\right)\mathbf{B}_6^{-1}\mathbf{A}_6\left(\mathbf{x},\mathbf{u}\right)^{\mathrm{T}}}\mathrm{d}\mathbf{x}\nonumber\\
=&\frac{1}{\sqrt{-\mathrm{det}(\mathbf{B})}}\iiiint_{\mathbb{R}^{N\times N\times N\times N}}\mu(\mathcal{M}_1)f\left(\mathbf{p}\mathbf{B}_3+\mathbf{t}\mathbf{D}_3\right)\overline{\mu(\mathcal{M}_2)f\left(\mathbf{p}\mathbf{A}_3+\mathbf{t}\mathbf{C}_3\right)}\phi(\mathbf{0},\mathbf{z})\mathrm{e}^{-2\pi\mathrm{i}\mathbf{z}(\mathbf{u}-\mathbf{q})^{\mathrm{T}}}\nonumber\\ &\times\mathrm{e}^{\pi\mathrm{i}\mathbf{q}\mathbf{D}\mathbf{B}^{-1}\mathbf{q}^{\mathrm{T}}}\mathrm{e}^{-2\pi\mathrm{i}\mathbf{t}\mathbf{B}^{-1}\mathbf{q}^{\mathrm{T}}}\mathrm{e}^{\pi\mathrm{i}\mathbf{q}\mathbf{F}_{44}\mathbf{q}^{\mathrm{T}}}\mathrm{d}\mathbf{p}\mathrm{d}\mathbf{t}\mathrm{d}\mathbf{z}\mathrm{d}\mathbf{q}.
\tag{A.3}
\end{align}
When $\phi(\mathbf{0},\mathbf{z})=1$, it follows that
\begin{align}\label{eqA.4} &\int_{\mathbb{R}^N}\mathrm{C}(\mathcal{M},\mathcal{M}_1,\mathcal{M}_2,\mathcal{M}_3,\mathcal{M}_4,\mathcal{M}_5,\mathcal{M}_6)f(\mathbf{x},\mathbf{u})\mathrm{e}^{\pi\mathrm{i}\left(\mathbf{x},\mathbf{u}\right)\mathbf{B}_6^{-1}\mathbf{A}_6\left(\mathbf{x},\mathbf{u}\right)^{\mathrm{T}}}\mathrm{d}\mathbf{x}\nonumber\\
=&\frac{1}{\sqrt{-\mathrm{det}(\mathbf{B})}}\iiint_{\mathbb{R}^{N\times N\times N}}\mu(\mathcal{M}_1)f\left(\mathbf{p}\mathbf{B}_3+\mathbf{t}\mathbf{D}_3\right)\overline{\mu(\mathcal{M}_2)f\left(\mathbf{p}\mathbf{A}_3+\mathbf{t}\mathbf{C}_3\right)}\mathrm{e}^{\pi\mathrm{i}\left(\mathbf{q}\mathbf{D}\mathbf{B}^{-1}\mathbf{q}^{\mathrm{T}}-2\mathbf{t}\mathbf{B}^{-1}\mathbf{q}^{\mathrm{T}}+\mathbf{q}\mathbf{F}_{44}\mathbf{q}^{\mathrm{T}}\right)}\nonumber\\ &\times\int_{\mathbb{R}^N}\mathrm{e}^{-2\pi\mathrm{i}\mathbf{z}(\mathbf{u}-\mathbf{q})^{\mathrm{T}}}\mathrm{d}\mathbf{z}\mathrm{d}\mathbf{q}\mathrm{d}\mathbf{p}\mathrm{d}\mathbf{t}\nonumber\\ =&\frac{1}{\sqrt{-\mathrm{det}(\mathbf{B})}}\mathrm{e}^{\pi\mathrm{i}\left(\mathbf{u}\mathbf{D}\mathbf{B}^{-1}\mathbf{u}^{\mathrm{T}}+\mathbf{u}\mathbf{F}_{44}\mathbf{u}^{\mathrm{T}}\right)}\iint_{\mathbb{R}^{N\times N}}\mu(\mathcal{M}_1)f\left(\mathbf{p}\mathbf{B}_3+\mathbf{t}\mathbf{D}_3\right)\overline{\mu(\mathcal{M}_2)f\left(\mathbf{p}\mathbf{A}_3+\mathbf{t}\mathbf{C}_3\right)}\nonumber\\ &\times\mathrm{e}^{-2\pi\mathrm{i}\mathbf{t}\mathbf{B}^{-1}\mathbf{u}^{\mathrm{T}}}\mathrm{d}\mathbf{p}\mathrm{d}\mathbf{t}.
\tag{A.4}
\end{align}
Taking the change of variables $\mathbf{x}=\mathbf{p}\mathbf{B}_3+\mathbf{t}\mathbf{D}_3$ and $\mathbf{y}=\mathbf{p}\mathbf{A}_3+\mathbf{t}\mathbf{C}_3$ yields
\begin{align}\label{eqA.5} &\int_{\mathbb{R}^N}\mathrm{C}(\mathcal{M},\mathcal{M}_1,\mathcal{M}_2,\mathcal{M}_3,\mathcal{M}_4,\mathcal{M}_5,\mathcal{M}_6)f(\mathbf{x},\mathbf{u})\mathrm{e}^{\pi\mathrm{i}\left(\mathbf{x},\mathbf{u}\right)\mathbf{B}_6^{-1}\mathbf{A}_6\left(\mathbf{x},\mathbf{u}\right)^{\mathrm{T}}}\mathrm{d}\mathbf{x}\nonumber\\ =&\frac{\left|\mathrm{det}(\mathbf{A}_3)\right|\left|\mathrm{det}(\mathbf{D}_3)\right|}{\sqrt{-\mathrm{det}(\mathbf{B})}}\mathrm{e}^{\pi\mathrm{i}\left(\mathbf{u}\mathbf{D}\mathbf{B}^{-1}\mathbf{u}^{\mathrm{T}}+\mathbf{u}\mathbf{F}_{44}\mathbf{u}^{\mathrm{T}}\right)}\int_{\mathbb{R}^N}\mu\left(\mathcal{M}_1\right)f(\mathbf{x})\mathcal{K}_{\begin{pmatrix}\mathbf{0}_N&\mathbf{B}{\left(\mathbf{A}_3^{-1}\right)}^{\mathrm{T}}\\-\left(\mathbf{B}^{-1}\right)^{\mathrm{T}}\mathbf{A}_3&\mathbf{I}_N\end{pmatrix}}\left(\mathbf{u},\mathbf{x}\right)\mathrm{d}\mathbf{x}\nonumber\\ &\times\overline{\int_{\mathbb{R}^N}\mu\left(\mathcal{M}_2\right)f(\mathbf{y})\mathcal{K}_{\begin{pmatrix}\mathbf{0}_N&\mathbf{B}{\left(\mathbf{B}_3^{-1}\right)}^{\mathrm{T}}\\-\mathbf{B}_3^{\mathrm{T}}\mathbf{B}^{-1}&\mathbf{B}_3^{-1}\mathbf{A}_3\end{pmatrix}}\left(\mathbf{u},\mathbf{y}\right)\mathrm{d}\mathbf{y}}, \tag{A.5}
\end{align}
where $\mathrm{det}(\mathbf{A}_3)\ne0$, and $\begin{pmatrix}\mathbf{0}_N&\mathbf{B}{\left(\mathbf{A}_3^{-1}\right)}^{\mathrm{T}}\\-\left(\mathbf{B}^{-1}\right)^{\mathrm{T}}\mathbf{A}_3&\mathbf{I}_N\end{pmatrix},\begin{pmatrix}\mathbf{0}_N&\mathbf{B}{\left(\mathbf{B}_3^{-1}\right)}^{\mathrm{T}}\\-\mathbf{B}_3^{\mathrm{T}}\mathbf{B}^{-1}&\mathbf{B}_3^{-1}\mathbf{A}_3\end{pmatrix}\in Sp(N,\mathbb{R})$. By using the cascadability of the metaplectic transform, we arrive the required result \eqref{Frequency marginal distribution}.$\hfill\blacksquare$\\
\indent (3) \emph{Proof of the time delay marginal distribution, i.e., Eq.~\eqref{Time delay marginal distribution}:} By setting $\mathbf{x}=\mathbf{0}$ in Eq.~\eqref{eq3.2}, we arrive the required result \eqref{Time delay marginal distribution}.$\hfill\blacksquare$\\
\indent	(4) \emph{Proof of the frequency shift marginal distribution, i.e., Eq.~\eqref{Frequency shift marginal distribution}:} By setting $\mathbf{u}=\mathbf{0}$ in Eq.~\eqref{eq3.2}, we arrive the required result \eqref{Frequency shift marginal distribution}.$\hfill\blacksquare$\\
\indent (5) \emph{Proof of the time marginal distribution based energy conservation, i.e., Eq.~\eqref{Time marginal distribution based energy conservation}:} When $\mathbf{A}_3=\mathbf{B}_3$ and $\mathcal{M}_1=\mathcal{M}_2$, by integrating with respect to the variables $\mathbf{x}$ on both sides of Eq.~\eqref{Time marginal distribution}, we arrive the required result \eqref{Time marginal distribution based energy conservation}.$\hfill\blacksquare$\\
\indent (6) \emph{Proof of the frequency marginal distribution based energy conservation, i.e., Eq.~\eqref{Frequency marginal distribution based energy conservation}:} When $\begin{pmatrix}\mathbf{0}_N&\mathbf{B}{\left(\mathbf{A}_3^{-1}\right)}^{\mathrm{T}}\\-\left(\mathbf{B}^{-1}\right)^{\mathrm{T}}\mathbf{A}_3&\mathbf{I}_N\end{pmatrix}\mathcal{M}_1=\begin{pmatrix}\mathbf{0}_N&\mathbf{B}{\left(\mathbf{B}_3^{-1}\right)}^{\mathrm{T}}\\-\mathbf{B}_3^{\mathrm{T}}\mathbf{B}^{-1}&\mathbf{B}_3^{-1}\mathbf{A}_3\end{pmatrix}\mathcal{M}_2$, by integrating with respect to the variables $\mathbf{u}$ on both sides of Eq.~\eqref{Frequency marginal distribution}, we arrive the required result \eqref{Frequency marginal distribution based energy conservation}.$\hfill\blacksquare$\\
\indent (7) \emph{Proof of the time delay or frequency shift marginal distribution based energy conservation, i.e., Eq.~\eqref{Time delay or frequency shift marginal distribution based energy conservation}:} Setting $\mathbf{x}=\mathbf{u}=\mathbf{0}$ in Eq.~\eqref{eq3.2} gives
\begin{align}\label{eqA.6} &\mathrm{C}(\mathcal{M},\mathcal{M}_1,\mathcal{M}_2,\mathcal{M}_3,\mathcal{M}_4,\mathcal{M}_5,\mathcal{M}_6)f(\mathbf{0},\mathbf{0})\nonumber\\
=&\iiint_{\mathbb{R}^{N\times N\times N}}\mu(\mathcal{M}_1)f\left(\mathbf{p}\mathbf{B}_3+\mathbf{t}\mathbf{D}_3\right)\overline{\mu(\mathcal{M}_2)f\left(\mathbf{p}\mathbf{A}_3+\mathbf{t}\mathbf{C}_3\right)}\mathcal{K}_{\mathcal{M}}(\mathbf{q},\mathbf{t})\mathrm{e}^{\pi\mathrm{i}(\mathbf{p},\mathbf{q})\left(\mathbf{B}_4^{-1}\mathbf{A}_4+\mathbf{B}_5^{-1}\mathbf{A}_5\right)(\mathbf{p},\mathbf{q})^{\mathrm{T}}}\nonumber\\
&\times\iint_{\mathbb{R}^{N\times N}}\phi\left(\mathbf{z},\mathbf{v}\right)\mathrm{e}^{2\pi\mathrm{i}\left(\mathbf{v}\mathbf{p}^{\mathrm{T}}+\mathbf{z}\mathbf{q}^{\mathrm{T}}\right)}\mathrm{d}\mathbf{z}\mathrm{d}\mathbf{v}\mathrm{d}\mathbf{p}\mathrm{d}\mathbf{q}\mathrm{d}\mathbf{t}.
\tag{A.6}
\end{align}
When $\mathbf{A}=\mathbf{0}_N$, $\mathbf{C}_3=\mathbf{D}_3$, $\mathcal{M}_1=\mathcal{M}_2$, and $\phi\left(\mathbf{z},\mathbf{v}\right)=1$, we arrive the required result \eqref{Time delay or frequency shift marginal distribution based energy conservation}.$\hfill\blacksquare$\\
\indent (8) \emph{Proof of the unique reconstruction, i.e., Eq.~\eqref{Reconstruction formula}:} The CMCD can be expressed as
\begin{align}\label{eqA.7}	&\mathrm{C}(\mathcal{M},\mathcal{M}_1,\mathcal{M}_2,\mathcal{M}_3,\mathcal{M}_4,\mathcal{M}_5,\mathcal{M}_6)f(\mathbf{x},\mathbf{u})\nonumber\\ =&\mathrm{e}^{-\pi\mathrm{i}(\mathbf{x},\mathbf{u})\mathbf{B}_6^{-1}\mathbf{A}_6(\mathbf{x,u})^{\mathrm{T}}}\iint_{\mathbb{R}^{N\times N}}\mathrm{W}(\mathcal{M},\mathcal{M}_1,\mathcal{M}_2,\mathcal{M}_3)f(\mathbf{p},\mathbf{q})\mathrm{e}^{\pi\mathrm{i}(\mathbf{p},\mathbf{q})\mathbf{B}_4^{-1}\mathbf{A}_4(\mathbf{p},\mathbf{q})^{\mathrm{T}}}\Pi(\mathbf{x}-\mathbf{p},\mathbf{u}-\mathbf{q})\nonumber\\ &\times\mathrm{e}^{\pi\mathrm{i}(\mathbf{x}-\mathbf{p},\mathbf{u}-\mathbf{q})\mathbf{B}_5^{-1}\mathbf{A}_5(\mathbf{x}-\mathbf{p},\mathbf{u}-\mathbf{q})^{\mathrm{T}}}\mathrm{d}\mathbf{p}\mathrm{d}\mathbf{q}.
\tag{A.7}
\end{align}
Thanks to the conventional convolution theorem, we have
\begin{align}\label{eqA.8}
&\mathrm{W}(\mathcal{M},\mathcal{M}_1,\mathcal{M}_2,\mathcal{M}_3)f(\mathbf{x,u})\nonumber\\
=&\mathrm{e}^{-\pi\mathrm{i}(\mathbf{x},\mathbf{u})\mathbf{B}_4^{-1}\mathbf{A}_4(\mathbf{x},\mathbf{u})^{\mathrm{T}}}\iint_{\mathbb{R}^{N\times N}}\mathrm{e}^{2\pi\mathrm{i}\left(\mathbf{p}\mathbf{x}^{\mathrm{T}}+\mathbf{q}\mathbf{u}^{\mathrm{T}}\right)}\nonumber\\
&\times\frac{\mathcal{F}\left[\mathrm{e}^{\pi\mathrm{i}\left(\mathbf{x},\mathbf{u}\right)\mathbf{B}_6^{-1}\mathbf{A}_6\left(\mathbf{x},\mathbf{u}\right)^{\mathrm{T}}}\mathrm{C}(\mathcal{M},\mathcal{M}_1,\mathcal{M}_2,\mathcal{M}_3,\mathcal{M}_4,\mathcal{M}_5,\mathcal{M}_6)f(\mathbf{x},\mathbf{u})\right]\left(\mathbf{p},\mathbf{q}\right)}{\mathcal{F}\left[\mathrm{e}^{\pi\mathrm{i}\left(\mathbf{x},\mathbf{u}\right)\mathbf{B}_5^{-1}\mathbf{A}_5\left(\mathbf{x},\mathbf{u}\right)^{\mathrm{T}}}\Pi(\mathbf{x,u})\right]\left(\mathbf{p},\mathbf{q}\right)}\mathrm{d}\mathbf{p}\mathrm{d}\mathbf{q}.
\tag{A.8}
\end{align}
Then, it follows that
\begin{align}\label{eqA.9}	&\int_{\mathbb{R}^N}\mu(\mathcal{M}_1)f\left(\mathbf{x}\mathbf{B}_3+\mathbf{t}\mathbf{D}_3\right)\overline{\mu(\mathcal{M}_2)f\left(\mathbf{x}\mathbf{A}_3+\mathbf{t}\mathbf{C}_3\right)}\mathrm{e}^{\pi\mathrm{i}\mathbf{t}\mathbf{B}^{-1}\mathbf{A}\mathbf{t}^{\mathrm{T}}}\int_{\mathbb{R}^N}\mathrm{e}^{-2\pi\mathrm{i}\mathbf{t}\mathbf{B}^{-1}\mathbf{u}^{\mathrm{T}}}\mathrm{d}\mathbf{u}\mathrm{d}\mathbf{t}\nonumber\\ =&\sqrt{-\mathrm{det}(\mathbf{B})}\int_{\mathbb{R}^N}\mathrm{e}^{-\pi\mathrm{i}\mathbf{u}\mathbf{D}\mathbf{B}^{-1}\mathbf{u}^{\mathrm{T}}}\mathrm{e}^{-\pi\mathrm{i}(\mathbf{x},\mathbf{u})\mathbf{B}_4^{-1}\mathbf{A}_4(\mathbf{x},\mathbf{u})^{\mathrm{T}}}\iint_{\mathbb{R}^{N\times N}}\mathrm{e}^{2\pi\mathrm{i}\left(\mathbf{p}\mathbf{x}^{\mathrm{T}}+\mathbf{q}\mathbf{u}^{\mathrm{T}}\right)}\nonumber\\
&\times\frac{\mathcal{F}\left[\mathrm{e}^{\pi\mathrm{i}\left(\mathbf{x},\mathbf{u}\right)\mathbf{B}_6^{-1}\mathbf{A}_6\left(\mathbf{x},\mathbf{u}\right)^{\mathrm{T}}}\mathrm{C}(\mathcal{M},\mathcal{M}_1,\mathcal{M}_2,\mathcal{M}_3,\mathcal{M}_4,\mathcal{M}_5,\mathcal{M}_6)f(\mathbf{x},\mathbf{u})\right]\left(\mathbf{p},\mathbf{q}\right)}{\mathcal{F}\left[\mathrm{e}^{\pi\mathrm{i}\left(\mathbf{x},\mathbf{u}\right)\mathbf{B}_5^{-1}\mathbf{A}_5\left(\mathbf{x},\mathbf{u}\right)^{\mathrm{T}}}\Pi(\mathbf{x,u})\right]\left(\mathbf{p},\mathbf{q}\right)}\mathrm{d}\mathbf{p}\mathrm{d}\mathbf{q}\mathrm{d}\mathbf{u}.
\tag{A.9}
\end{align}
Note that the left-hand-side of the above equation reads $\mu(\mathcal{M}_1)f\left(\mathbf{x}\mathbf{B}_3\right)\overline{\mu(\mathcal{M}_2)f\left(\mathbf{x}\mathbf{A}_3\right)}$. When $\mathbf{A}_{3}=\mathbf{0}$, we arrive the required result \eqref{Reconstruction formula}.$\hfill\blacksquare$\\
\indent (9) \emph{Proof of the Moyal formula, i.e., Eq.~\eqref{Moyal formula}:} Because of the generalized metaplectic convolution theorem, given by Eq.~\eqref{eq2.16}, the CMCD can be rewritten as
\begin{align}\label{eqA.10} &\mathrm{C}(\mathcal{M},\mathcal{M}_1,\mathcal{M}_2,\mathcal{M}_3,\mathcal{M}_4,\mathcal{M}_5,\mathcal{M}_6)f(\mathbf{x},\mathbf{u})\nonumber\\		=&\int_{\mathbb{R}^{2N}}\epsilon_{\mathcal{M}_4,\mathcal{M}_5,\mathcal{M}_6}(\mathbf{w})\mu(\mathcal{M}_4)\mathrm{W}(\mathcal{M},\mathcal{M}_1,\mathcal{M}_2,\mathcal{M}_3)f\left(\mathbf{w}\left(\mathbf{B}_6^{-1}\right)^{\mathrm{T}}\mathbf{B}_4^{\mathrm{T}}\right)\overline{\mu(\mathcal{M}_5)\Pi\left(\mathbf{w}\left(\mathbf{B}_6^{-1}\right)^{\mathrm{T}}\mathbf{B}_5^{\mathrm{T}}\right)}\nonumber\\
&\times\mathcal{K}_{\mathcal{M}_6^{-1}}\left(\left(\mathbf{x},\mathbf{u}\right),\mathbf{w}\right)\mathrm{d}\mathbf{w}.
\tag{A.10}
\end{align}
Then, it follows that
\begin{align}\label{eqA.11}
&\iint_{\mathbb{R}^{N\times N}}\mathrm{C}(\mathcal{M},\mathcal{M}_1,\mathcal{M}_2,\mathcal{M}_3,\mathcal{M}_4,\mathcal{M}_5,\mathcal{M}_6)f(\mathbf{x},\mathbf{u})\nonumber\\
&\times\overline{\mathrm{C}(\mathcal{M},\mathcal{M}_1,\mathcal{M}_2,\mathcal{M}_3,\mathcal{M}_4,\mathcal{M}_5,\mathcal{M}_6)g(\mathbf{x},\mathbf{u})}\mathrm{d}\mathbf{x}\mathrm{d}\mathbf{u}\nonumber\\ =&\int_{\mathbb{R}^{2N}}\int_{\mathbb{R}^{2N}}\epsilon_{\mathcal{M}_4,\mathcal{M}_5,\mathcal{M}_6}(\mathbf{w})\overline{\epsilon_{\mathcal{M}_4,\mathcal{M}_5,\mathcal{M}_6}(\mathbf{w'})}\mu(\mathcal{M}_4)\mathrm{W}(\mathcal{M},\mathcal{M}_1,\mathcal{M}_2,\mathcal{M}_3)f\left(\mathbf{w}\left(\mathbf{B}_6^{-1}\right)^{\mathrm{T}}\mathbf{B}_4^{\mathrm{T}}\right)\nonumber\\ &\times\overline{\mu(\mathcal{M}_4)\mathrm{W}(\mathcal{M},\mathcal{M}_1,\mathcal{M}_2,\mathcal{M}_3)g\left(\mathbf{w'}\left(\mathbf{B}_6^{-1}\right)^{\mathrm{T}}\mathbf{B}_4^{\mathrm{T}}\right)}\mu(\mathcal{M}_5)\Pi\left(\mathbf{w}\left(\mathbf{B}_6^{-1}\right)^{\mathrm{T}}\mathbf{B}_5^{\mathrm{T}}\right)\nonumber\\ &\times\overline{\mu(\mathcal{M}_5)\Pi\left(\mathbf{w'}\left(\mathbf{B}_6^{-1}\right)^{\mathrm{T}}\mathbf{B}_5^{\mathrm{T}}\right)}\iint_{\mathbb{R}^{N\times N}}\mathcal{K}_{\mathcal{M}_6^{-1}}\left(\left(\mathbf{x},\mathbf{u}\right),\mathbf{w}\right)\overline{\mathcal{K}_{\mathcal{M}_6^{-1}}\left(\left(\mathbf{x},\mathbf{u}\right),\mathbf{w'}\right)}\mathrm{d}\mathbf{x}\mathrm{d}\mathbf{u}\mathrm{d}\mathbf{w}\mathrm{d}\mathbf{w'}\nonumber\\ =&\frac{\left|\mathrm{det}(\mathbf{B}_{4})\right|\left|\mathrm{det}(\mathbf{B}_{5})\right|}{\left|\mathrm{det}(\mathbf{B}_{6})\right|}\int_{\mathbb{R}^{2N}}\left|\Pi\left(\mathbf{w}\left(\mathbf{B}_{6}^{-1}\right)^{\mathrm{T}}\mathbf{B}_{5}^{\mathrm{T}}\right)\right|^{2}\mu(\mathcal{M}_4)\mathrm{W}(\mathcal{M},\mathcal{M}_1,\mathcal{M}_2,\mathcal{M}_3)f\left(\mathbf{w}\left(\mathbf{B}_{6}^{-1}\right)^{\mathrm{T}}\mathbf{B}_{4}^{\mathrm{T}}\right)\nonumber\\ &\times\overline{\mu(\mathcal{M}_4)\mathrm{W}(\mathcal{M},\mathcal{M}_1,\mathcal{M}_2,\mathcal{M}_3)g\left(\mathbf{w}\left(\mathbf{B}_6^{-1}\right)^{\mathrm{T}}\mathbf{B}_4^{\mathrm{T}}\right)}\mathrm{d}\mathbf{w}.
\tag{A.11}
\end{align}
When $\left|\mu(\mathcal{M}_5)\Pi(\mathbf{w})\right|^2=1$ for all $\mathbf{w}\in\mathbb{R}^{2N}$, it derives from the Parseval's relation of the meteplectic transform that
\begin{align}\label{eqA.12}
&\iint_{\mathbb{R}^{N\times N}}\mathrm{C}(\mathcal{M},\mathcal{M}_1,\mathcal{M}_2,\mathcal{M}_3,\mathcal{M}_4,\mathcal{M}_5,\mathcal{M}_6)f(\mathbf{x},\mathbf{u})\nonumber\\
&\times\overline{\mathrm{C}(\mathcal{M},\mathcal{M}_1,\mathcal{M}_2,\mathcal{M}_3,\mathcal{M}_4,\mathcal{M}_5,\mathcal{M}_6)g(\mathbf{x},\mathbf{u})}\mathrm{d}\mathbf{x}\mathrm{d}\mathbf{u}\nonumber\\
=&\left|\mathrm{det}(\mathbf{B}_5)\right|\iint_{\mathbb{R}^{N\times N}}\mathrm{W}(\mathcal{M},\mathcal{M}_1,\mathcal{M}_2,\mathcal{M}_3)f(\mathbf{x},\mathbf{u})\overline{\mathrm{W}(\mathcal{M},\mathcal{M}_1,\mathcal{M}_2,\mathcal{M}_3)g(\mathbf{x},\mathbf{u})}\mathrm{d}\mathbf{x}\mathrm{d}\mathbf{u}.
\tag{A.12}
\end{align}
From Eq.~\eqref{eq3.1} and by applying the change of variables, we arrive the required result \eqref{Moyal formula}.$\hfill\blacksquare$\\
\indent (10) \emph{Proof of the complex conjugate symmetry, i.e., Eq.~\eqref{Complex conjugate symmetry}:} By substituting the complex conjugate function $\overline{f}$ into Eq.~\eqref{eq3.2}, and thanks to $\overline{\Pi}=\Pi$, we arrive the required result \eqref{Complex conjugate symmetry}.$\hfill\blacksquare$\\
\indent	(11) \emph{Proof of the time reversal symmetry, i.e., Eq.~\eqref{Time reversal symmetry}:} By substituting the time reversal function $\underline{f}$ into Eq.~\eqref{eq3.2}, we arrive the required result \eqref{Time reversal symmetry}.$\hfill\blacksquare$\\
\indent	(12) \emph{Proof of the scaling property, i.e., Eq.~\eqref{Scaling property}:} By substituting the scaling function $\mathcal{S}_\sigma f$ into Eq.~\eqref{eq3.2}, we arrive the required result \eqref{Scaling property}.$\hfill\blacksquare$\\
\indent (13) \emph{Proof of the time translation property, i.e., Eq.~\eqref{Time translation property}:} When $\mathcal{M}_3^{\mathrm{T}}$ is a symplectic matrix, we have
\begin{align}\label{eqA.13}	&\mathrm{W}(\mathcal{M},\mathcal{M}_1,\mathcal{M}_2,\mathcal{M}_3^{\mathrm{T}})T_{\boldsymbol{\tau}}f\left(\mathbf{x},\mathbf{u}\right)\nonumber\\
=&\mathrm{e}^{\pi\mathrm{i}\boldsymbol{\tau}\left(\mathbf{B}_2^{-1}\mathbf{A}_2-\mathbf{B}_2^{-1}\mathbf{A}_2\right)\boldsymbol{\tau}^{\mathrm{T}}}\mathrm{e}^{-\pi\mathrm{i}\boldsymbol{\tau}\left(\mathbf{A}_1^{\mathrm{T}}\mathbf{D}_1\mathbf{B}_1^{-1}\mathbf{A}_1-\mathbf{A}_2^{\mathrm{T}}\mathbf{D}_2\mathbf{B}_2^{-1}\mathbf{A}_2\right)\boldsymbol{\tau}^{\mathrm{T}}}\mathrm{e}^{2\pi\mathrm{i}\boldsymbol{\tau}\mathbf{B}_1^{\mathrm{T}}\mathbf{C}_1\mathbf{B}_1^{-1}\mathbf{C}_3\mathbf{x}^{\mathrm{T}}}\mathrm{e}^{-2\pi\mathrm{i}\boldsymbol{\tau}\mathbf{B}_2^{\mathrm{T}}\mathbf{C}_2\mathbf{B}_2^{-1}\mathbf{A}_3\mathbf{x}^{\mathrm{T}}}\nonumber\\
&\times\int_{\mathbb{R}^N}\mu(\mathcal{M}_1)f\left(\mathbf{x}\mathbf{C}_3^\mathrm{T}+\mathbf{y}\mathbf{D}_3^\mathrm{T}-\boldsymbol{\tau}\mathbf{A}_1^\mathrm{T}\right)\overline{\mu(\mathcal{M}_2)f\left(\mathbf{x}\mathbf{A}_3^\mathrm{T}+\mathbf{y}\mathbf{B}_3^\mathrm{T}-\boldsymbol{\tau}\mathbf{A}_2^\mathrm{T}\right)}\nonumber\\
&\times\mathrm{e}^{2\pi\mathrm{i}\boldsymbol{\tau}\mathbf{B}_1^{\mathrm{T}}\mathbf{C}_1\mathbf{B}_1^{-1}\mathbf{D}_3\mathbf{y}^{\mathrm{T}}}\mathrm{e}^{-2\pi\mathrm{i}\boldsymbol{\tau}\mathbf{B}_2^{\mathrm{T}}\mathbf{C}_2\mathbf{B}_2^{-1}\mathbf{B}_3\mathbf{y}^{\mathrm{T}}}\mathcal{K}_{\mathcal{M}}(\mathbf{u},\mathbf{y})\mathrm{d}\mathbf{y}\nonumber\\
=&\mathrm{e}^{\pi\mathrm{i}\boldsymbol{\tau}\left(\mathbf{B}_1^{-1}\mathbf{A}_1-\mathbf{B}_2^{-1}\mathbf{A}_2\right)\boldsymbol{\tau}^{\mathrm{T}}}\mathrm{e}^{-\pi\mathrm{i}\boldsymbol{\tau}\left(\mathbf{A}_1^{\mathrm{T}}\mathbf{D}_1\mathbf{B}_1^{-1}\mathbf{A}_1-\mathbf{A}_2^{\mathrm{T}}\mathbf{D}_2\mathbf{B}_2^{-1}\mathbf{A}_2\right)\boldsymbol{\tau}^{\mathrm{T}}}\mathrm{e}^{2\pi\mathrm{i}\boldsymbol{\tau}\mathbf{B}_1^{\mathrm{T}}\mathbf{C}_1\mathbf{B}_1^{-1}\mathbf{C}_3\mathbf{x}^{\mathrm{T}}}\mathrm{e}^{-2\pi\mathrm{i}\boldsymbol{\tau}\mathbf{B}_2^{\mathrm{T}}\mathbf{C}_2\mathbf{B}_2^{-1}\mathbf{A}_3\mathbf{x}^{\mathrm{T}}}\nonumber\\
&\times\int_{\mathbb{R}^N}\mu(\mathcal{M}_1)f\left(\left(\mathbf{x}-\boldsymbol{\tau}\mathbf{A}_1^{\mathrm{T}}(\mathbf{D}_3-\mathbf{B}_3)\right)\mathbf{C}_3^{\mathrm{T}}+\left(\mathbf{y}-\boldsymbol{\tau}\mathbf{A}_1^{\mathrm{T}}(\mathbf{A}_3-\mathbf{C}_3)\right)\mathbf{D}_3^{\mathrm{T}}\right)\nonumber\\
&\times\overline{\mu(\mathcal{M}_2)f\left(\left(\mathbf{x}-\boldsymbol{\tau}\mathbf{A}_2^{\mathrm{T}}(\mathbf{D}_3-\mathbf{B}_3)\right)\mathbf{A}_3^{\mathrm{T}}+\left(\mathbf{y}-\boldsymbol{\tau}\mathbf{A}_2^{\mathrm{T}}(\mathbf{A}_3-\mathbf{C}_3)\right)\mathbf{B}_3^{\mathrm{T}}\right)}\nonumber\\
&\times\mathrm{e}^{2\pi\mathrm{i}\boldsymbol{\tau}\mathbf{B}_1^{\mathrm{T}}\mathbf{C}_1\mathbf{B}_1^{-1}\mathbf{D}_3\mathbf{y}^{\mathrm{T}}}\mathrm{e}^{-2\pi\mathrm{i}\boldsymbol{\tau}\mathbf{B}_2^{\mathrm{T}}\mathbf{C}_2\mathbf{B}_2^{-1}\mathbf{B}_3\mathbf{y}^{\mathrm{T}}}\mathcal{K}_{\mathcal{M}}(\mathbf{u},\mathbf{y})\mathrm{d}\mathbf{y}.
\tag{A.13}
\end{align}
When $\mathbf{A}_1=\mathbf{A}_2$ and $\mathbf{B}_1^{\mathrm{T}}\mathbf{C}_1\mathbf{B}_1^{-1}\mathbf{D}_3-\mathbf{B}_2^{\mathrm{T}}\mathbf{C}_2\mathbf{B}_2^{-1}\mathbf{B}_3=\mathbf{0}_N$, it follows that
\begin{align}\label{eqA.14} &\mathrm{W}(\mathcal{M},\mathcal{M}_1,\mathcal{M}_2,\mathcal{M}_3^{\mathrm{T}})T_{\boldsymbol{\tau}}f\left(\mathbf{x}+\boldsymbol{\tau}\mathbf{A}_1^{\mathrm{T}}(\mathbf{D}_3-\mathbf{B}_3),\mathbf{u}\right)\nonumber\\
=&\mathrm{e}^{2\pi\mathrm{i}\boldsymbol{\tau}\mathbf{B}_1^{\mathrm{T}}\mathbf{C}_1\mathbf{B}_1^{-1}\mathbf{C}_3\left(\mathbf{x}+\boldsymbol{\tau}\mathbf{A}_1^{\mathrm{T}}(\mathbf{D}_3-\mathbf{B}_3)\right)^{\mathrm{T}}}\mathrm{e}^{-2\pi\mathrm{i}\boldsymbol{\tau}\mathbf{B}_2^{\mathrm{T}}\mathbf{C}_2\mathbf{B}_2^{-1}\mathbf{A}_3\left(\mathbf{x}+\boldsymbol{\tau}\mathbf{A}_1^{\mathrm{T}}(\mathbf{D}_3-\mathbf{B}_3)\right)^{\mathrm{T}}}\nonumber\\
&\times\mathrm{e}^{\pi\mathrm{i}\boldsymbol{\tau}\left(\mathbf{B}_1^{-1}-\mathbf{B}_2^{-1}\right)\mathbf{A}_1\boldsymbol{\tau}^{\mathrm{T}}}\mathrm{e}^{-\pi\mathrm{i}\boldsymbol{\tau}\mathbf{A}_1^{\mathrm{T}}\left(\mathbf{D}_1\mathbf{B}_1^{-1}-\mathbf{D}_2\mathbf{B}_2^{-1}\right)\mathbf{A}_1\boldsymbol{\tau}^{\mathrm{T}}}\nonumber\\
&\times\int_{\mathbb{R}^N}\mu(\mathcal{M}_1)f\left(\mathbf{x}\mathbf{C}_3^{\mathrm{T}}+\mathbf{y}\mathbf{D}_3^{\mathrm{T}}\right)\overline{\mu(\mathcal{M}_2)f\left(\mathbf{x}\mathbf{A}_3^{\mathrm{T}}+\mathbf{y}\mathbf{B}_3^{\mathrm{T}}\right)}\mathcal{K}_{\mathcal{M}}\left(\mathbf{u},\mathbf{y}+\boldsymbol{\tau}\mathbf{A}_1^{\mathrm{T}}(\mathbf{A}_3-\mathbf{C}_3)\right)\mathrm{d}\mathbf{y}.
\tag{A.14}
\end{align}
Thanks to the relation
\begin{align}\label{eqA.15} \mathcal{K}_{\mathcal{M}}\left(\mathbf{u},\mathbf{y}+\boldsymbol{\tau}\mathbf{A}_1^{\mathrm{T}}(\mathbf{A}_3-\mathbf{C}_3)\right)=&\mathrm{e}^{-\pi\mathrm{i}\boldsymbol{\tau}\mathbf{A}_1^{\mathrm{T}}\left(\mathbf{A}_3-\mathbf{C}_3\right)\mathbf{B}^{\mathrm{T}}\mathbf{C}\mathbf{B}^{-1}\mathbf{A}\left(\mathbf{A}_3-\mathbf{C}_3\right)^{\mathrm{T}}\mathbf{A}_1\boldsymbol{\tau}^{\mathrm{T}}}\nonumber\\ &\times\mathrm{e}^{2\pi\mathrm{i}\boldsymbol{\tau}\mathbf{A}_1^{\mathrm{T}}\left(\mathbf{A}_3-\mathbf{C}_3\right)\mathbf{B}^{\mathrm{T}}\mathbf{C}\mathbf{B}^{-1}\mathbf{u}^{\mathrm{T}}}\mathcal{K}_{\mathcal{M}}\left(\mathbf{u}-\boldsymbol{\tau}\mathbf{A}_1^{\mathrm{T}}(\mathbf{A}_3-\mathbf{C}_3)\mathbf{A}^{\mathrm{T}},\mathbf{y}\right),
\tag{A.15}
\end{align}
there is
\begin{align}\label{eqA.16} &\mathrm{W}(\mathcal{M},\mathcal{M}_1,\mathcal{M}_2,\mathcal{M}_3^{\mathrm{T}})T_{\boldsymbol{\tau}}f\left(\mathbf{x},\mathbf{u}\right)\nonumber\\ =&\mathrm{e}^{\pi\mathrm{i}\boldsymbol{\tau}\left(\mathbf{B}_1^{-1}-\mathbf{B}_2^{-1}\right)\mathbf{A}_1\boldsymbol{\tau}^{\mathrm{T}}} \mathrm{e}^{-\pi\mathrm{i}\boldsymbol{\tau}\mathbf{A}_1^{\mathrm{T}}\left(\mathbf{D}_1\mathbf{B}_1^{-1}-\mathbf{D}_2\mathbf{B}_2^{-1}\right)\mathbf{A}_1\boldsymbol{\tau}^{\mathrm{T}}}\mathrm{e}^{2\pi\mathrm{i}\boldsymbol{\tau}\mathbf{B}_1^{\mathrm{T}}\mathbf{C}_1\mathbf{B}_1^{-1}\mathbf{C}_3\mathbf{x}^{\mathrm{T}}}\mathrm{e}^{-2\pi\mathrm{i}\boldsymbol{\tau}\mathbf{B}_2^{\mathrm{T}}\mathbf{C}_2\mathbf{B}_2^{-1}\mathbf{A}_3\mathbf{x}^{\mathrm{T}}}\nonumber\\
&\times\mathrm{e}^{-\pi\mathrm{i}\boldsymbol{\tau}\mathbf{A}_1^{\mathrm{T}}\left(\mathbf{A}_3-\mathbf{C}_3\right)\mathbf{B}^{\mathrm{T}}\mathbf{C}\mathbf{B}^{-1}\mathbf{A}\left(\mathbf{A}_3-\mathbf{C}_3\right)^{\mathrm{T}}\mathbf{A}_1\boldsymbol{\tau}^{\mathrm{T}}} \mathrm{e}^{2\pi\mathrm{i}\boldsymbol{\tau}\mathbf{A}_1^{\mathrm{T}}\left(\mathbf{A}_3-\mathbf{C}_3\right)\mathbf{B}^{\mathrm{T}}\mathbf{C}\mathbf{B}^{-1}\mathbf{u}^{\mathrm{T}}}\nonumber\\ &\times\mathrm{W}(\mathcal{M},\mathcal{M}_1,\mathcal{M}_2,\mathcal{M}_3^{\mathrm{T}})f\left(\mathbf{x}-\boldsymbol{\tau}\mathbf{A}_1^{\mathrm{T}}(\mathbf{D}_3-\mathbf{B}_3),\mathbf{u}-\boldsymbol{\tau}\mathbf{A}_1^{\mathrm{T}}(\mathbf{A}_3-\mathbf{C}_3)\mathbf{A}^{\mathrm{T}}\right).
\tag{A.16}
\end{align}
By replacing $\mathcal{M}_3^{\mathrm{T}}$ with $\mathcal{M}_3$ in the above equation, and subsequently, substituting it into the CMCD of the time translation function $T_{\boldsymbol{\tau}}f$, we arrive the required result \eqref{Time translation property}.$\hfill\blacksquare$\\
\indent	(14) \emph{Proof of the frequency modulation property, i.e., Eq.~\eqref{Frequency modulation property}:} When $\mathcal{M}_3^{\mathrm{T}}$ is a symplectic matrix, we have
\begin{align}\label{eqA.17} &\mathrm{W}(\mathcal{M},\mathcal{M}_1,\mathcal{M}_2,\mathcal{M}_3^{\mathrm{T}})M_{\mathbf{w}}f\left(\mathbf{x},\mathbf{u}\right)\nonumber\\
=&\mathrm{e}^{-\pi\mathrm{i}\mathbf{w}\left(\mathbf{B}_1^{\mathrm{T}}\mathbf{D}_1-\mathbf{B}_2^{\mathrm{T}}\mathbf{D}_2\right)\mathbf{w}^{\mathrm{T}}} \mathrm{e}^{-\pi\mathrm{i}\mathbf{w}\left(\mathbf{D}_2^{\mathrm{T}}\mathbf{B}_3-\mathbf{D}_1^{\mathrm{T}}\mathbf{D}_3\right)\mathbf{B}^{\mathrm{T}}\mathbf{D}\left(\mathbf{D}_2^{\mathrm{T}}\mathbf{B}_3-\mathbf{D}_1^{\mathrm{T}}\mathbf{D}_3\right)^{\mathrm{T}}\mathbf{w}^{\mathrm{T}}}\mathrm{e}^{-2\pi\mathrm{i}\mathbf{u}\mathbf{D}\left(\mathbf{D}_2^{\mathrm{T}}\mathbf{B}_3-\mathbf{D}_1^{\mathrm{T}}\mathbf{D}_3\right)^{\mathrm{T}}\mathbf{w}^{\mathrm{T}}}\nonumber\\
&\times\mathrm{e}^{2\pi\mathrm{i}\mathbf{x}\left(\mathbf{D}_1^{\mathrm{T}}\mathbf{C}_3-\mathbf{D}_2^{\mathrm{T}}\mathbf{A}_3\right)^{\mathrm{T}}\mathbf{w}^{\mathrm{T}}}\int_{\mathbb{R}^N}\mu(\mathcal{M}_1)f\left(\mathbf{x}\mathbf{C}_3^\mathrm{T}+\mathbf{y}\mathbf{D}_3^\mathrm{T}-\mathbf{w}\mathbf{B}_1^\mathrm{T}\right)\overline{\mu(\mathcal{M}_2)f\left(\mathbf{x}\mathbf{A}_3^\mathrm{T}+\mathbf{y}\mathbf{B}_3^\mathrm{T}-\mathbf{w}\mathbf{B}_2^\mathrm{T}\right)}\nonumber\\
&\times\mathcal{K}_{\mathcal{M}}\left(\mathbf{u}+\mathbf{w}\mathbf{D}_2^{\mathrm{T}}\mathbf{B}_3\mathbf{B}^{\mathrm{T}}-\mathbf{w}\mathbf{D}_1^{\mathrm{T}}\mathbf{D}_3\mathbf{B}^{\mathrm{T}},\mathbf{y}\right)\mathrm{d}\mathbf{y}\nonumber\\
=&\mathrm{e}^{-\pi\mathrm{i}\mathbf{w}\left(\mathbf{B}_1^{\mathrm{T}}\mathbf{D}_1-\mathbf{B}_2^{\mathrm{T}}\mathbf{D}_2\right)\mathbf{w}^{\mathrm{T}}}\mathrm{e}^{-\pi\mathrm{i}\mathbf{w}\left(\mathbf{D}_2^{\mathrm{T}}\mathbf{B}_3-\mathbf{D}_1^{\mathrm{T}}\mathbf{D}_3\right)\mathbf{B}^{\mathrm{T}}\mathbf{D}\left(\mathbf{D}_2^{\mathrm{T}}\mathbf{B}_3-\mathbf{D}_1^{\mathrm{T}}\mathbf{D}_3\right)^{\mathrm{T}}\mathbf{w}^{\mathrm{T}}} \mathrm{e}^{-2\pi\mathrm{i}\mathbf{u}\mathbf{D}\left(\mathbf{D}_2^{\mathrm{T}}\mathbf{B}_3-\mathbf{D}_1^{\mathrm{T}}\mathbf{D}_3\right)^{\mathrm{T}}\mathbf{w}^{\mathrm{T}}}\nonumber\\
&\times\mathrm{e}^{2\pi\mathrm{i}\mathbf{x}\left(\mathbf{D}_1^{\mathrm{T}}\mathbf{C}_3-\mathbf{D}_2^{\mathrm{T}}\mathbf{A}_3\right)^{\mathrm{T}}\mathbf{w}^{\mathrm{T}}}\int_{\mathbb{R}^N}\mu(\mathcal{M}_1)f\left(\left(\mathbf{x}-\mathbf{w}\mathbf{B}_1^\mathrm{T}(\mathbf{D}_3-\mathbf{B}_3)\right)\mathbf{C}_3^{\mathrm{T}}+\left(\mathbf{y}-\mathbf{w}\mathbf{B}_1^\mathrm{T}(\mathbf{A}_3-\mathbf{C}_3)\right)\mathbf{D}_3^{\mathrm{T}}\right)\nonumber\\ &\times\overline{\mu(\mathcal{M}_2)f\left(\left(\mathbf{x}-\mathbf{w}\mathbf{B}_2^\mathrm{T}(\mathbf{D}_3-\mathbf{B}_3)\right)\mathbf{A}_3^{\mathrm{T}}+\left(\mathbf{y}-\mathbf{w}\mathbf{B}_2^\mathrm{T}(\mathbf{A}_3-\mathbf{C}_3)\right)\mathbf{B}_3^{\mathrm{T}}\right)}\nonumber\\
&\times\mathcal{K}_{\mathcal{M}}\left(\mathbf{u}+\mathbf{w}\mathbf{D}_2^{\mathrm{T}}\mathbf{B}_3\mathbf{B}^{\mathrm{T}}-\mathbf{w}\mathbf{D}_1^{\mathrm{T}}\mathbf{D}_3\mathbf{B}^{\mathrm{T}},\mathbf{y}\right)\mathrm{d}\mathbf{y}.
\tag{A.17}
\end{align}
When $\mathbf{B}_1=\mathbf{B}_2$, it follows that
\begin{align}\label{eqA.18} &\mathrm{W}(\mathcal{M},\mathcal{M}_1,\mathcal{M}_2,\mathcal{M}_3^{\mathrm{T}})M_{\mathbf{w}}f\left(\mathbf{x},\mathbf{u}\right)\nonumber\\
=&\mathrm{e}^{-\pi\mathrm{i}\mathbf{w}\left(\mathbf{B}_1^{\mathrm{T}}\mathbf{D}_1-\mathbf{B}_2^{\mathrm{T}}\mathbf{D}_2\right)\mathbf{w}^{\mathrm{T}}} \mathrm{e}^{-\pi\mathrm{i}\mathbf{w}\left(\mathbf{D}_2^{\mathrm{T}}\mathbf{B}_3-\mathbf{D}_1^{\mathrm{T}}\mathbf{D}_3\right)\mathbf{B}^{\mathrm{T}}\mathbf{D}\left(\mathbf{D}_2^{\mathrm{T}}\mathbf{B}_3-\mathbf{D}_1^{\mathrm{T}}\mathbf{D}_3\right)^{\mathrm{T}}\mathbf{w}^{\mathrm{T}}}\mathrm{e}^{-2\pi\mathrm{i}\mathbf{u}\mathbf{D}\left(\mathbf{D}_2^{\mathrm{T}}\mathbf{B}_3-\mathbf{D}_1^{\mathrm{T}}\mathbf{D}_3\right)^{\mathrm{T}}\mathbf{w}^{\mathrm{T}}}\nonumber\\
&\times\mathrm{e}^{2\pi\mathrm{i}\mathbf{x}\left(\mathbf{D}_1^{\mathrm{T}}\mathbf{C}_3-\mathbf{D}_2^{\mathrm{T}}\mathbf{A}_3\right)^{\mathrm{T}}\mathbf{w}^{\mathrm{T}}}\int_{\mathbb{R}^N}\mu(\mathcal{M}_1)f\left(\left(\mathbf{x}-\mathbf{w}\mathbf{B}_1^\mathrm{T}(\mathbf{D}_3-\mathbf{B}_3)\right)\mathbf{C}_3^{\mathrm{T}}+\mathbf{y}\mathbf{D}_3^{\mathrm{T}}\right)\nonumber\\ &\times\overline{\mu(\mathcal{M}_2)f\left(\left(\mathbf{x}-\mathbf{w}\mathbf{B}_1^\mathrm{T}(\mathbf{D}_3-\mathbf{B}_3)\right)\mathbf{A}_3^{\mathrm{T}}+\mathbf{y}\mathbf{B}_3^{\mathrm{T}}\right)}\nonumber\\
&\times\mathcal{K}_{\mathcal{M}}\left(\mathbf{u}+\mathbf{w}\mathbf{D}_2^{\mathrm{T}}\mathbf{B}_3\mathbf{B}^{\mathrm{T}}-\mathbf{w}\mathbf{D}_1^{\mathrm{T}}\mathbf{D}_3\mathbf{B}^{\mathrm{T}},\mathbf{y}+\mathbf{w}\mathbf{B}_1^\mathrm{T}(\mathbf{A}_3-\mathbf{C}_3)\right)\mathrm{d}\mathbf{y}\nonumber\\
=&\mathrm{e}^{-\pi\mathrm{i}\mathbf{w}\left(\mathbf{B}_1^{\mathrm{T}}\mathbf{D}_1-\mathbf{B}_2^{\mathrm{T}}\mathbf{D}_2\right)\mathbf{w}^{\mathrm{T}}}\mathrm{e}^{-\pi\mathrm{i}\mathbf{w}\left(\mathbf{D}_2^{\mathrm{T}}\mathbf{B}_3-\mathbf{D}_1^{\mathrm{T}}\mathbf{D}_3\right)\mathbf{B}^{\mathrm{T}}\mathbf{D}\left(\mathbf{D}_2^{\mathrm{T}}\mathbf{B}_3-\mathbf{D}_1^{\mathrm{T}}\mathbf{D}_3\right)^{\mathrm{T}}\mathbf{w}^{\mathrm{T}}}\nonumber\\
&\times\mathrm{e}^{-\pi\mathrm{i}\mathbf{w}\mathbf{B}_1^{\mathrm{T}}\left(\mathbf{A}_3-\mathbf{C}_3\right)\mathbf{B}^{\mathrm{T}}\mathbf{C}\mathbf{B}^{-1}\mathbf{A}\left(\mathbf{A}_3-\mathbf{C}_3\right)^{\mathrm{T}}\mathbf{B}_1\mathbf{w}^{\mathrm{T}}}\mathrm{e}^{2\pi\mathrm{i}\mathbf{w}\mathbf{B}_1^{\mathrm{T}}\left(\mathbf{A}_3-\mathbf{C}_3\right)\mathbf{B}^{\mathrm{T}}\mathbf{C}\left(\mathbf{D}_2^{\mathrm{T}}\mathbf{B}_3-\mathbf{D}_1^{\mathrm{T}}\mathbf{D}_3\right)^{\mathrm{T}}\mathbf{w}^{\mathrm{T}}}\nonumber\\
&\times\mathrm{e}^{2\pi\mathrm{i}\mathbf{x}\left(\mathbf{D}_1^{\mathrm{T}}\mathbf{C}_3-\mathbf{D}_2^{\mathrm{T}}\mathbf{A}_3\right)^{\mathrm{T}}\mathbf{w}^{\mathrm{T}}}\mathrm{e}^{-2\pi\mathrm{i}\mathbf{u}\mathbf{D}\left(\mathbf{D}_2^{\mathrm{T}}\mathbf{B}_3-\mathbf{D}_1^{\mathrm{T}}\mathbf{D}_3\right)^{\mathrm{T}}\mathbf{w}^{\mathrm{T}}}\mathrm{e}^{2\pi\mathrm{i}\mathbf{w}\mathbf{B}_1^{\mathrm{T}}\left(\mathbf{A}_3-\mathbf{C}_3\right)\mathbf{B}^{\mathrm{T}}\mathbf{C}\mathbf{B}^{-1}\mathbf{u}^{\mathrm{T}}}\mathrm{W}(\mathcal{M},\mathcal{M}_1,\mathcal{M}_2,\mathcal{M}_3^{\mathrm{T}})f\nonumber\\
&\left(\mathbf{x}-\mathbf{w}\mathbf{B}_1^\mathrm{T}(\mathbf{D}_3-\mathbf{B}_3),\mathbf{u}+\mathbf{w}\mathbf{D}_2^{\mathrm{T}}\mathbf{B}_3\mathbf{B}^{\mathrm{T}}-\mathbf{w}\mathbf{D}_1^{\mathrm{T}}\mathbf{D}_3\mathbf{B}^{\mathrm{T}}-\mathbf{w}\mathbf{B}_1^{\mathrm{T}}(\mathbf{A}_3-\mathbf{C}_3)\mathbf{A}^{\mathrm{T}}\right).
\tag{A.18}
\end{align}
By replacing $\mathcal{M}_3^{\mathrm{T}}$ with $\mathcal{M}_3$ in the above equation, and subsequently, substituting it into the CMCD of the frequency modulation function $M_{\mathbf{w}}f$, we arrive the required result \eqref{Frequency modulation property}.$\hfill\blacksquare$\\
\indent	(15) \emph{Proof of the metaplectic invariance, i.e., Eq.~\eqref{Metaplectic invariance}:} According to the cascadability of the metaplectic transform, we arrive the required result \eqref{Metaplectic invariance}.$\hfill\blacksquare$
\subsection{Proof of Eq.~\eqref{eq4.6}}\label{sec:AppB}
\indent The orthogonal principle implies that
\begin{align}\label{eqB.1}
\mathbb{E}\left\{\left(\mathrm{W}f\left(\mathbf{z}\right)-\mathrm{W}\widehat{f}\left(\mathbf{z}\right)\right)\overline{\mathrm{W}(\mathcal{M},\mathcal{M}_1,\mathcal{M}_2,\mathcal{M}_3)g\left(\mathbf{z'}\right)}\right\}=0,\mathbf{z'}\in\mathbb{R}^{2N}.
\tag{B.1}
\end{align}
By multiplying $\mathrm{e}^{\pi\mathrm{i}\left(\mathbf{z}\mathbf{B}_6^{-1}\mathbf{A}_6\mathbf{z}^{\mathrm{T}}-\mathbf{z'}\mathbf{B}_4^{-1}\mathbf{A}_4\mathbf{z'}^{\mathrm{T}}\right)}$ on both sides of Eq.~\eqref{eqB.1}, we have
\begin{align}\label{eqB.2}
\mathbb{E}\left\{\left(\mathop{\widetilde{\mathrm{W}}}\limits^6f\left(\mathbf{z}\right)-\mathop{\widetilde{\mathrm{W}}}\limits^6\widehat{f}\left(\mathbf{z}\right)\right)\overline{\mathop{\widetilde{\mathrm{W}}}\limits^4(\mathcal{M},\mathcal{M}_1,\mathcal{M}_2,\mathcal{M}_3)g\left(\mathbf{z'}\right)}\right\}=0,
\tag{B.2}
\end{align}
based on which we establish the Wiener-Hopf equation
\begin{align}\label{eqB.3} R_{\mathop{\widetilde{\mathrm{W}}}\limits^6f,\mathop{\widetilde{\mathrm{W}}}\limits^4(\mathcal{M},\mathcal{M}_1,\mathcal{M}_2,\mathcal{M}_3)g}\left(\mathbf{z},\mathbf{z'}\right)-\int_{\mathbb{R}^{2N}}R_{\mathop{\widetilde{\mathrm{W}}}\limits^4(\mathcal{M},\mathcal{M}_1,\mathcal{M}_2,\mathcal{M}_3)g}\left(\mathbf{k},\mathbf{z'}\right)\mathop{\widetilde{H_{\mathrm{opt}}}}\limits^5\left(\mathbf{z}-\mathbf{k}\right)\mathrm{d}\mathbf{k}=0,
\tag{B.3}
\end{align}
where $R_{\mathop{\widetilde{\mathrm{W}}}\limits^6f,\mathop{\widetilde{\mathrm{W}}}\limits^4(\mathcal{M},\mathcal{M}_1,\mathcal{M}_2,\mathcal{M}_3)g}$ denotes the cross-correlation function between $\mathop{\widetilde{\mathrm{W}}}\limits^6f$ and $\mathop{\widetilde{\mathrm{W}}}\limits^4(\mathcal{M},\mathcal{M}_1,\mathcal{M}_2,\mathcal{M}_3)g$, and $R_{\mathop{\widetilde{\mathrm{W}}}\limits^4(\mathcal{M},\mathcal{M}_1,\mathcal{M}_2,\mathcal{M}_3)g}$ denotes the auto-correlation function of $\mathop{\widetilde{\mathrm{W}}}\limits^4(\mathcal{M},\mathcal{M}_1,\mathcal{M}_2,\mathcal{M}_3)g$. In general, we can obtain $H_{\mathrm{opt}}$ in the metaplectic Wigner distribution domain by solving Eq.~\eqref{eqB.3} numerically. Particularly, if $\mathrm{W}f$ and $\mathrm{W}(\mathcal{M},\mathcal{M}_1,\mathcal{M}_2,\mathcal{M}_3)g$ are chirp-stationary, Eq.~\eqref{eqB.3} simplifies to
\begin{align}\label{eqB.4} R_{\mathop{\widetilde{\mathrm{W}}}\limits^6f,\mathop{\widetilde{\mathrm{W}}}\limits^4(\mathcal{M},\mathcal{M}_1,\mathcal{M}_2,\mathcal{M}_3)g}\left(\mathbf{z}-\mathbf{z'}\right)-\int_{\mathbb{R}^{2N}}R_{\mathop{\widetilde{\mathrm{W}}}\limits^4(\mathcal{M},\mathcal{M}_1,\mathcal{M}_2,\mathcal{M}_3)g}\left(\mathbf{k}-\mathbf{z'}\right)\mathop{\widetilde{H_{\mathrm{opt}}}}\limits^5\left(\mathbf{z}-\mathbf{k}\right)\mathrm{d}\mathbf{k}=0.
\tag{B.4}
\end{align}
Taking the change of variables $\mathbf{z}- \mathbf{z'}=\mathbf{p}$ and $\mathbf{z}- \mathbf{k}=\mathbf{q}$ yields
\begin{align}\label{eqB.5} R_{\mathop{\widetilde{\mathrm{W}}}\limits^6f,\mathop{\widetilde{\mathrm{W}}}\limits^4(\mathcal{M},\mathcal{M}_1,\mathcal{M}_2,\mathcal{M}_3)g}\left(\mathbf{p}\right)-\int_{\mathbb{R}^{2N}}R_{\mathop{\widetilde{\mathrm{W}}}\limits^4(\mathcal{M},\mathcal{M}_1,\mathcal{M}_2,\mathcal{M}_3)g}\left(\mathbf{p}-\mathbf{q}\right)\mathop{\widetilde{H_{\mathrm{opt}}}}\limits^5\left(\mathbf{q}\right)\mathrm{d}\mathbf{q}=0.
\tag{B.5}
\end{align}
Thanks to the conventional convolution and correlation theorems, we solve Eq.~\eqref{eqB.5} to obtain
\begin{align}\label{eqB.6}
\mathcal{F}\left[\mathop{\widetilde{\mathrm{W}}}\limits^6f\right]\left(\mathbf{w}\right)\overline{\mathcal{F}\left[\mathop{\widetilde{\mathrm{W}}}\limits^4(\mathcal{M},\mathcal{M}_1,\mathcal{M}_2,\mathcal{M}_3)g\right]\left(\mathbf{w}\right)}=\mathcal{F}\left[\mathop{\widetilde{H_{\mathrm{opt}}}}\limits^5\right]\left(\mathbf{u}\right)\left|\mathcal{F}\left[\mathop{\widetilde{\mathrm{W}}}\limits^4(\mathcal{M},\mathcal{M}_1,\mathcal{M}_2,\mathcal{M}_3)g\right]\left(\mathbf{w}\right)\right|^2.
\tag{B.6}
\end{align}
It follows from the relationship between the metaplectic transform and Fourier transform that
\begin{align} \label{eqB.7}
&\sqrt{-{\mathrm{det}}\left(\mathbf{B}_6\right)}\mathrm{e}^{-\pi\mathrm{i}\mathbf{w}\mathbf{B}_6^{\mathrm{T}}\mathbf{D}_6\mathbf{u}^{\mathrm{T}}}\mu\left(\mathcal{M}_6\right)\mathrm{W}f\left(\mathbf{w}\mathbf{B}_6^{\mathrm{T}}\right)\nonumber\\
=&\sqrt{-{\mathrm{det}}\left(\mathbf{B}_4\right)}\sqrt{-{\mathrm{det}}\left(\mathbf{B}_5\right)}\mathrm{e}^{-\pi\mathrm{i}\mathbf{w}\mathbf{B}_4^{\mathrm{T}}\mathbf{D}_4\mathbf{u}^{\mathrm{T}}}\mathrm{e}^{-\pi\mathrm{i}\mathbf{w}\mathbf{B}_5^{\mathrm{T}}\mathbf{D}_5\mathbf{u}^{\mathrm{T}}}\nonumber\\
&\times\mu\left(\mathcal{M}_5\right)H_{\mathrm{opt}}\left(\mathbf{w} \mathbf{B}_5^{\mathrm{T}}\right)\mu\left(\mathcal{M}_4\right)\mathrm{W}(\mathcal{M},\mathcal{M}_1,\mathcal{M}_2,\mathcal{M}_3)g\left(\mathbf{w}\mathbf{B}_4^{\mathrm{T}}\right),
\tag{B.7}
\end{align}
and therefore, we arrive the required result \eqref{eq4.6}.$\hfill\blacksquare$
\subsection{Proof of Eq.~\eqref{eq4.8}}\label{sec:AppC}
\indent The minimize MSE takes
\begin{align}\label{eqC.1}
\mathop{\min}\limits_{H\left(\mathbf{z}\right)}\sigma_{\mathrm{MSE}}^2=&\mathbb{E}\left\{\left|\mathrm{W}f\left(\mathbf{z}\right)-\mathrm{W}\widehat{f}\left(\mathbf{z}\right)\right|^2\right\}\nonumber\\
=&\mathbb{E}\left\{\left|\mathop{\widetilde{\mathrm{W}}}\limits^6f\left(\mathbf{z}\right)-\mathop{\widetilde{\mathrm{W}}}\limits^6\widehat{f}\left(\mathbf{z}\right)\right|^2\right\}\nonumber\\
=&\mathbb{E}\left\{\left(\mathop{\widetilde{\mathrm{W}}}\limits^6f\left(\mathbf{z}\right)-\mathrm{e}^{\pi\mathrm{i}\mathbf{z}\mathbf{B}_6^{-1}\mathbf{A}_6\mathbf{z}^{\mathrm{T}}}\left(\mathrm{W}(\mathcal{M},\mathcal{M}_1,\mathcal{M}_2,\mathcal{M}_3)g\Theta_{\mathcal{M}_4,\mathcal{M}_5,\mathcal{M}_6}H\right)\left(\mathbf{z}\right)\right)\overline{\mathop{\widetilde{\mathrm{W}}}\limits^6f\left(\mathbf{z}\right)}\right\}.
\tag{C.1}
\end{align}
Similar to Eqs.~\eqref{eqB.3}--\eqref{eqB.5}, we have
\begin{align}\label{eqC.2}
\mathop{\min}\limits_{H\left(\mathbf{z}\right)}\sigma_{\mathrm{MSE}}^2=R_{\mathop{\widetilde{\mathrm{W}}}\limits^6f}\left(\mathbf{0}\right)-\int_{\mathbb{R}^{2N}}R_{\mathop{\widetilde{\mathrm{W}}}\limits^4(\mathcal{M},\mathcal{M}_1,\mathcal{M}_2,\mathcal{M}_3)g,\mathop{\widetilde{\mathrm{W}}}\limits^6f}\left(-\mathbf{k}\right)\mathop{\widetilde{H_{\mathrm{opt}}}}\limits^5\left(\mathbf{k}\right)\mathrm{d}\mathbf{k}. \tag{C.2}
\end{align}
Because of the Parseval's relation of the Wigner distribution, it follows that
\begin{align}\label{eqC.3}
R_{\mathop{\widetilde {\mathrm{W}}}\limits^6f}\left(\mathbf{0}\right)=\int_{\mathbb{R}^{2N}}\left|\mathop{\widetilde{\mathrm{W}}}\limits^6f(\mathbf{z})\right|^2\mathrm{d}\mathbf{z}=\int_{\mathbb{R}^{2N}}\left|\mathrm{W}f(\mathbf{z})\right|^2\mathrm{d}\mathbf{z}=\left\|f\right\|_2^4.
\tag{C.3}
\end{align}
Thanks to the conventional convolution and correction theorems, there is
\begin{align}\label{eqC.4}
&\int_{\mathbb{R}^{2N}}R_{\mathop{\widetilde{\mathrm{W}}}\limits^4(\mathcal{M},\mathcal{M}_1,\mathcal{M}_2,\mathcal{M}_3)g,\mathop{\widetilde{\mathrm{W}}}\limits^6f}\left(-\mathbf{k}\right)\mathop{\widetilde{H_{\mathrm{opt}}}}\limits^5\left(\mathbf{k}\right)\mathrm{d}\mathbf{k}\nonumber\\
=&\left(\left(\mathop{\widetilde{H_{\mathrm{opt}}}}\limits^5(\mathbf{z})\right)\ast\left(R_{\mathop{\widetilde{\mathrm{W}}}\limits^4(\mathcal{M},\mathcal{M}_1,\mathcal{M}_2,\mathcal{M}_3)g,\mathop{\widetilde{\mathrm{W}}}\limits^6f}(\mathbf{z})\right)\right)\Bigg|_{\mathbf{z=0}}\nonumber\\
=&\int_{\mathbb{R}^{2N}}\mathcal{F}\left[\mathop{\widetilde{H_{\mathrm{opt}}}}\limits^5\right](\mathbf{w})\mathcal{F}\left[\mathop{\widetilde{\mathrm{W}}}\limits^4(\mathcal{M},\mathcal{M}_1,\mathcal{M}_2,\mathcal{M}_3)g\right](\mathbf{w})\overline{\mathcal{F}\left[\mathop{\widetilde{\mathrm{W}}}\limits^6f\right](\mathbf{w})}\mathrm{d}\mathbf{w}\nonumber\\
=&\sqrt{-\mathrm{det}(\mathbf{B}_4)}\sqrt{-\mathrm{det}(\mathbf{B}_5)}\sqrt{-\mathrm{det}(\mathbf{B}_6)}\int_{\mathbb{R}^{2N}}\mathrm{e}^{-\pi\mathrm{i}\mathbf{w}\mathbf{B}_4^{\mathrm{T}}\mathbf{D}_4\mathbf{w}^{\mathrm{T}}}\mathrm{e}^{-\pi\mathrm{i}\mathbf{w}\mathbf{B}_5^{\mathrm{T}}\mathbf{D}_5\mathbf{w}^{\mathrm{T}}}\mathrm{e}^{\pi\mathrm{i}\mathbf{w}\mathbf{B}_6^{\mathrm{T}}\mathbf{D}_6\mathbf{w}^{\mathrm{T}}}\nonumber\\ &\times\mu\left(\mathcal{M}_4\right)\mathrm{W}(\mathcal{M},\mathcal{M}_1,\mathcal{M}_2,\mathcal{M}_3)g\left(\mathbf{w}\mathbf{B}_4^{\mathrm{T}}\right)\mu\left(\mathcal{M}_6\right)H_{\mathrm{opt}}\left(\mathbf{w}\mathbf{B}_5^{\mathrm{T}}\right)\overline{\mu\left(\mathcal{M}_6\right)\mathrm{W}f\left(\mathbf{w}\mathbf{B}_6^{\mathrm{T}}\right)}\mathrm{d}\mathbf{w}.
\tag{C.4}
\end{align}
By substituting Eq.~\eqref{eq4.6} into Eq.~\eqref{eqC.4}, and subsequently, using Eqs.~\eqref{eqC.2}--\eqref{eqC.4}, we arrive the required result \eqref{eq4.8}.$\hfill\blacksquare$
\bibliography{mybib}{}
\bibliographystyle{IEEEtran}
\end{document}